\documentstyle[12pt]{article}
\newcommand{\ret}{\nonumber \\}

\newcommand{\Section}[1]%
{\section{#1}\setcounter{equation}{0}%
\setcounter{theorem}{0}}
\newtheorem{theorem}{Theorem}
\newtheorem{lemma}[theorem]{Lemma}
\newtheorem{coro}[theorem]{Corollary}
\newtheorem{pro}[theorem]{Proposition}

\newtheorem{assumption}[theorem]{Assumption}

\newenvironment{proof}[1]%
{\par\noindent{\em #1:\ }}%
{~\rule{2mm}{2mm}\par\bigskip}
\begin{document}
%%%%%%%%%%%%%%%%%%%%%%%%%%%%%%%%%%%%%%%%%%%
\newpage\thispagestyle{empty}
{\topskip 2cm
\begin{center}
{\huge\bf Revisiting the Charge Transport\\  
\bigskip
in Quantum Hall Systems\\}
\bigskip\bigskip
{\Large Tohru Koma\\}
\bigskip\medskip
{\small Department of Physics, Gakushuin University, 
Mejiro, Toshima-ku, Tokyo 171-8588, JAPAN\\}
\smallskip
{\small\tt e-mail: tohru.koma@gakushuin.ac.jp}
\smallskip

\end{center}
%This is also for double spacing
%\newpage
\vfil
\noindent
We reexamine the charge transport induced by a weak electric field 
in two-dimensional quantum Hall systems in a finite, periodic box 
at very low temperatures. Our model covers random vector and electrostatic 
potentials and electron-electron interactions.
The resulting linear response coefficients consist of the time-independent 
term $\sigma_{xy}$ corresponding to the Hall conductance and  
the linearly time-dependent term $\gamma_{sy}\cdot t$ in 
the transverse and longitudinal directions $s=x,y$ in a slow 
switching limit for adiabatically applying the initial electric field.  
The latter terms $\gamma_{sy}\cdot t$ are due to the acceleration of 
the electrons by the uniform electric field in the finite and 
isolated system, and so the time-independent term $\sigma_{yy}$ 
corresponding  to the diagonal conductance which generates dissipation of heat 
always vanishes. The well known topological argument yields 
the integral and fractional quantization 
of the averaged Hall conductance $\overline{\sigma_{xy}}$ over gauge parameters
under the assumption on the existence of a spectral gap above the ground state. 
In addition to this fact, we show that the averaged acceleration 
coefficients $\overline{\gamma_{sy}}$ are vanishing under the same assumption. 
In the non-interacting case, the spectral gap between the neighbouring Landau 
levels persists if the vector and the electrostatic potentials together 
satisfy a certain condition, and then the Hall conductance $\sigma_{xy}$ 
without averaging exhibits the exact integral quantization 
in the infinite volume limit with the vanishing acceleration coefficients. 
We also estimate their finite size corrections.
In the interacting case, the averaged Hall conductance $\overline{\sigma_{xy}}$ 
for a non-integer filling of the electrons 
is quantized to a fraction not equal to an integer under the assumption that 
the potentials satisfy certain conditions in addition to the gap assumption. 
We also discuss the relation between the fractional quantum Hall effect and 
Atiyah-Singer index theorem for non-Abelian gauge fields. 
\par\noindent
\bigskip
\hrule
\bigskip
\noindent
{\bf KEY WORDS:} Charge transport; linear response theory; 
quantum Hall effect; geometric invariants; non-Abelian gauge fields. 
\hfill
\bigskip 

%\noindent
%---$\!$---$\!$---$\!$---
%\medskip

\vfil}\newpage
%%%%%%%%%%%%%%%%%%%%%%%%%%%%%%%%%%%%%%%%%%%
%%%%%%%%%%%%%%%%%%%%%%%%%%%%%%%%%%%%%%%%%%%
\tableofcontents
\newpage
%%%%%%%%%%%%%%%%%%%%%%%%%%%%%%%%%%%%%%%%%%%
\Section{Introduction}

The linear response theory \cite{HNakano,Kubo} for charge transport 
successfully elucidates some aspects of the quantum Hall effect observed 
experimentally \cite{KDP,KawajiWakabayashi} in two-dimensional electron gases 
in a strong magnetic field. In particular, it was found out that 
the integral quantization of the Hall conductance is a consequence 
of a topological nature of the Hall conductance \cite{TKNN,Kohmoto}. 
However, the derivation of the linear response formulas for conductance 
from the first principle is still an unsolved problem \cite{AG,BVS}. 
Actually it is very hard to take into account the effect of the reservoir 
explicitly. It is needless to say that there have been many, various arguments, 
each employing some simplifying feature, so far. For example, 
the infinite volume formalism without taking the infinite volume 
limit from a sequence of finite volumes and the adiabatic (slowly varying) 
switching of the external electric field to avoid accelerating the electrons 
have been often used instead of coupling the corresponding finite system to 
a reservoir.\footnote{For recent attempts 
to justify the linear response formulas, see refs.~\cite{BVS,ElSch}.}  
Thus the issue is still left somewhat hanging there although it has been debated 
again and again so far.   

Apart from the problem of the validity of the linear response formulas, 
the quantized Hall conductance was identified with a topological invariant 
of a certain fiber bundle \cite{DN} by using the resulting linear response 
formulas as mentioned above. 
The topological argument for the Hall conductance 
was first introduced into a quantum Hall system of non-interacting 
electron gas in a periodic potential \cite{TKNN,Kohmoto}. As a result, 
it was shown that the Hall conductance is quantized to an integer 
under the assumption of the existence of a spectral gap 
above the unique ground state. The integer of the quantization is 
equal to the filling factor of the Landau levels when the periodic potential 
is weak. However, one cannot expect 
the appearance of the conductance plateaus for varying the filling 
of the electrons because of the absence of disorder.\footnote{
The appearance of the Hall conductance plateaus was discussed 
with localization estimates in refs.~\cite{AG,BVS,Kunz,ASS2}. 
We will discuss this issue in the next paper \cite{Koma3}.}

Soon after these articles, this topological argument\footnote{See also 
related articles \cite{ASS,Simon,DAZ,Ishikawa}.} was extended 
to a quantum Hall system with disorder and with electron-electron 
interactions \cite{NT-1,NTW,AvronSeiler,ASY}. Instead of the crude 
Hall conductance with fixed gauge parameters, they treated 
the Hall conductance averaged over gauge parameters, and showed that 
the averaged Hall conductance is quantized to an integer 
under the assumption on the existence of a spectral gap above the ground state. 
The integer of the quantization is equal to the filling factor of the Landau levels 
in the case of the non-interacting electron gas 
with weak single-body potentials \cite{Kunz} as well as the above case 
with a periodic potential. Surprisingly the averaged  Hall 
conductance does not have any finite size correction to the exact integral 
quantization even though the system has disorder. 
Clearly one cannot expect that the crude Hall conductance 
is exactly quantized for any finite system. 

In order to explain the fractional quantization \cite{TSG,SCTHG,WESTGE} 
of the Hall conductance, 
this topological argument needs some ad hoc assumptions on the degeneracy 
of the ground state in addition to the assumption on averaging 
the Hall conductance \cite{NTW,TW,TH,AY}. In fact, if the ground state is 
nondegenerate with an excitation gap, then the averaged Hall conductance 
always shows integral quantization. The explicit value of the fraction of 
the quantized Hall conductance is not determined by the topological 
argument alone even if the dimension of the sector of the degenerate ground state 
is given. 
A degenerate ground state with an excitation gap is expected to appear 
only for a fractional filling $p/q$ of the electrons \cite{YHL,Yoshioka,Su}. 
Actually, the existence of a spectral gap for a non-integer filling of the electrons 
implies a degenerate ground state for a certain model \cite{Koma1}. 
In the real experiments, the fraction of the quantized Hall conductance is 
observed to equal to the fractional filling $p/q$. 
Under the assumption on the spectral gap, it was 
proved without relying on the topological argument that 
the Hall conductance is proportional to the filling factor of the electrons 
for certain interacting electron gases \cite{Koma2}. 
However, the existence of the spectral gap itself has not yet been proved 
for any interacting electron gas. Besides, the relation between 
the fractionally quantized Hall conductance and the filling factor 
of the electrons is still unclear. We should remark that a set of 
the possible quantized values of the Hall conductance can be derived 
from a mathematical argument relying on the universality. 
See ref.~\cite{Froehlich}.

We should also remark another topological approach by Bellissard \cite{BVS}.  
For an infinite volume quantum Hall system of a non-interacting electron gas 
with disorder, the quantized Hall conductance is identified with 
a Fredholm index of a certain operator that arises in Connes theory of 
noncommutative geometry \cite{Connes}. 
(See also refs.~\cite{AG,ASS2,Xia,FNakano}.) In comparison to 
the above topological approach, Bellissard's framework has the advantage 
that it does not need the assumption on averaging the Hall conductance 
over gauge parameters. However, it has not yet been extended to interacting 
quantum Hall electron gases. 

In this paper, we study a two-dimensional $N$ electrons system in a uniform 
magnetic field perpendicular to the two-dimensional plane 
in which the electrons are confined. For simplicity we assume that 
the electrons do not have the spin degrees of freedom, although we can 
treat a similar system with both the spin degrees of freedom and multiple 
layers in the same way. The explicit form of the Hamiltonian of the system 
is given by (\ref{Ham2dgeneral0}) in the next section. The model covers 
a wide class of potentials including a random vector potential, 
a random electrostatic potential and an electron-electron interaction. 
In order to measure an induced current as a response to an external electric field, 
we apply a time-dependent vector potential ${\bf A}_{\rm ex}(t)=(0,\alpha(t))$, 
where the function $\alpha(t)$ of time $t$ is given by (\ref{alphat}) in the 
next section. For $t\in [-T,0]$ with a large positive $T$, 
the corresponding electric field is adiabatically switched on, 
and for $t\ge 0$, the electric field becomes $(0,F)$ with 
the constant strength $F$.  
We consider the finite, isolated system of an $L_x\times L_y$ rectangular box, 
and impose periodic boundary conditions. Thus we will not consider 
a reservoir, and clearly the system does not exhibit any dissipation of heat.  
In this sense, we cannot measure the conductance. 
But the quantum Hall systems in the real experiments show 
negligibly small dissipation. 
Correspondingly the system we consider in this paper 
is expected to show weak acceleration of the electrons. 
Actually we will show that the acceleration is weak in a certain sense, 
and the constant Hall current flow is dominant. From these results, 
if the system is connected with a reservoir, then the acceleration 
of the electrons is expected to be further suppressed, and 
the linear response coefficient corresponding to the Hall conductance 
can be identified with the realistic one. 
Thus we believe that it is useful in future studies to reexamine the charge 
transport in such a finite, isolated quantum Hall system. 

Let us describe our results. The precise statement of the main results will 
be given in the next section. 
The linear response coefficients are given by 
\begin{equation}
\sigma_{{\rm tot},sy}=\lim_{F\rightarrow 0}\frac{j_{{\rm ind},s}}{F},
\end{equation}
where $j_{{\rm ind},s}$ are the induced current 
in the transverse direction $s=x$ and the longitudinal direction $s=y$ to 
the electric field.  
We obtain the generic forms of the coefficients for time $t\ge 0$ as follows: 
\begin{equation}
\sigma_{{\rm tot},xy}=\sigma_{xy}+\gamma_{xy}\cdot t +\delta\sigma_{xy}
\label{LRCx}
\end{equation}
and
\begin{equation}
\sigma_{{\rm tot},yy}=\gamma_{yy}\cdot t +\delta\sigma_{yy}.
\end{equation}
The first term $\sigma_{xy}$ in the right-hand side of (\ref{LRCx}) 
is constant in time $t$, 
and corresponds to the Hall conductance. As we mentioned above, 
the time-independent term $\sigma_{yy}$ corresponding to 
the diagonal conductance is absent. Instead of that, 
the linearly time-dependent terms $\gamma_{sy}\cdot t$ appear, i.e., 
there exist terms corresponding to the acceleration of the electrons by 
the external electric field. The appearance of the acceleration term 
$\gamma_{xy}\cdot t$ in the transverse direction is due to 
the disorder scattering of the electrons accelerated in the longitudinal direction. 
When the system couples to a reservoir, 
we can expect that the acceleration terms $\gamma_{sy}\cdot t$ disappear, 
and instead of them, the diagonal conductance $\sigma_{yy}$ appears . 
The rest of two terms $\delta\sigma_{sy}$ are the corrections depending on  
the initial switching process for adiabatically applying the external electric 
field. We prove that these two terms are negligibly small for 
slow switching. 
In the case without a uniform magnetic field, we demonstrate that 
the velocity of the electrons increases in time for an interacting electron gas 
with translation invariance in Section~\ref{Sec:translationinv}. 
More precisely, we have  
\begin{equation}
\sigma_{xy}=0,\quad \gamma_{xy}=0,\quad \mbox{and}\quad  
\gamma_{yy}=\frac{ne^2}{m_e},
\end{equation}
where $n$ is the density of the electrons, and 
$-e$ and $m_e$ are, respectively, the charge and the mass of the electron. 
For an interacting electron gas with translation invariance in 
a uniform magnetic field \cite{Wallet}, we also demonstrate   
\begin{equation}
\sigma_{xy}=-\frac{e^2}{h}\nu, \quad \mbox{and} \quad\gamma_{sy}=0\quad \mbox{for}\ 
s=x,y,
\end{equation}
where $h$ is the Planck constant, and $\nu$ is the filling factor of 
the Landau levels. 

Using the explicit expression of the linear response coefficients, we focus on 
the unsolved issues about the charge transport of the quantum Hall effect. 
For this purpose, we assume that there exists an excitation gap above the 
sector of the (quasi)degenerate ground state. 
Since our Hall conductance $\sigma_{xy}$ is the same as the standard one, 
the well-known topological argument yields the fractional 
quantization of the Hall conductance \cite{NTW,TW,TH,AY} as  
\begin{equation}
\overline{\sigma_{xy}}=-\frac{e^2}{h}\frac{p}{q},
\label{FQHC}
\end{equation}
under the assumption on the gap, where $\overline{\cdots}$ denotes 
the average over gauge parameters, the integer $p$ is given by 
the geometrical invariant\footnote{\label{geometric} In the following, 
we will use the term ``geometrical" instead of ``topological" because 
we will not consider any deformation for a manifold nor a change of 
the local coordinate.}  called the first Chern number, and the integer $q$ 
is the dimension of the sector of the ground state.  
Under the same assumption, we prove 
\begin{equation}
\overline{\gamma_{sy}}=0\quad\mbox{for}\ s=x,y.
\label{vanishinggamma} 
\end{equation}
Thus the acceleration of the electrons is absent in the sense of the average, 
and so we can expect that, for fixed gauge parameters, 
the acceleration of the electrons is weak. 

In the non-interacting case, assume an integer filling $\nu=\ell$ of 
the Landau levels. Then we prove that a spectral gap exists above 
the unique ground state for certain weak potentials, 
and that the Hall conductance $\sigma_{xy}$ and the acceleration 
coefficients $\gamma_{sy}$ satisfy 
\begin{equation}
\left|\sigma_{xy}+\frac{e^2}{h}\ell\right|\le{\rm const.}\times 
\max\{L_x^{-1},L_y^{-1}\},
\label{FSC}
\end{equation}
and 
\begin{equation}
\left|\gamma_{sy}\right|\le {\rm const.}\times\max\{L_x^{-1},L_y^{-1}\}
\quad\mbox{for}\ s=x,y,
\label{FSC2} 
\end{equation}
where $L_x,L_y$ are the system sizes. This Hall conductance $\sigma_{xy}$ 
is not averaged over the gauge parameters, and the result gives 
the upper bound for the 
finite size correction\footnote{The upper bound for the finite size 
correction would not be optimal \cite{NT}. But the inequality (\ref{FSC}) 
gives the mathematical rigorous upper bound!}
to the quantized value $-(e^2/h)\ell$. 
The second inequality for $\gamma_{sy}$ implies weak acceleration of 
the electrons. In particular, it vanishes in the infinite volume.  

In the general case of the interacting electron gas, we cannot remove 
both of the assumptions on the existence of a spectral gap above 
the sector of the ground state and on the average over the gauge parameters. 
About the assumption on the average, for general values of the gauge parameters, 
we cannot expect an exact fractional quantization 
of the Hall conductance as (\ref{FQHC}), 
and the finite-size corrections to the quantized Hall conductance 
should appear. Besides, the fraction $p/q$ cannot be determined 
by the geometrical argument alone. 
But we can get the following result: 
In addition to the assumption on the existence of the spectral gap, 
if the potentials satisfy certain technical 
assumptions, then the fraction $p/q$ must satisfy the bound, 
\begin{equation}
\nu(1-\delta)\le \frac{p}{q} \le \nu(1+\delta),
\label{fractionbound}
\end{equation}
where $\nu$ is the filling factor of the electrons, and $\delta$ is 
a positive number which is determined by certain norms of 
the single-body potentials.  
In order to clarify the meaning of 
the bound (\ref{fractionbound}), consider the situation that the interval 
$[\nu(1-\delta),\nu(1+\delta)]$ does not include any integer. This situation 
is indeed realized for a non-integer filling factor $\nu$ and for 
weak single-body potentials. Then the fraction $p/q$ must be a non-integer, 
and the degeneracy $q$ of the ground states must be 
greater than $1$. Thus the fractional quantization of the Hall conductance 
occurs with a degenerate ground state for a non-integer filling. 

The present paper is organized as follows: In Section~\ref{modelresults}, 
we give the precise definition of the model and describe our main theorems 
in a mathematical rigorous manner. In Section~\ref{LRT},  
the linear response coefficients are derived, starting from the basic 
Schr\"odinger equation with a time-dependent gauge field which gives 
a constant electric field for the time $t\ge 0$. 
We check that the linear response coefficients so obtained are physically reasonable 
ones for certain translationally invariant systems in Section~\ref{Sec:translationinv}. 
Section~\ref{Sec:QAHC} is devoted 
to the proofs of the fractional quantization (\ref{FQHC}) of 
the averaged Hall conductance $\overline{\sigma_{xy}}$ 
and of the vanishing (\ref{vanishinggamma}) of the averaged acceleration 
coefficients $\overline{\gamma_{sy}}$ in the most general setting.  
We also discuss the relation between the fractional 
quantum Hall effect and Atiyah-Singer index theorem for non-Abelian 
gauge fields \cite{Atiyah,CGKS,Gilkey,Weinberg}. 
The non-interacting electron gases with disorder is treated 
in Section~\ref{noninteracting}. As a result, 
we obtain the bound (\ref{FSC}) for the Hall conductance 
and the bound (\ref{FSC2}) for the acceleration coefficients.  
In Section~\ref{IntCase}, the interacting electron gases with disorder 
are treated, and we prove the bound (\ref{fractionbound}) 
for the fraction of the quantized Hall conductance (\ref{FQHC}). 
Appendices~\ref{appendix:prodiffPhi}-\ref{estEU} are devoted to 
the details of technical calculations and proofs of propositions and theorems. 

%%%%%%%%%%%%%%%%%%%%%%%%%%%%%%%%%%%%%%%%%%%%%%%%%%%%%%%%%%%%%%%%%%%%%%%%
\Section{The model and the main results}
\label{modelresults}

Consider a two-dimensional $N$ electrons system in a uniform 
magnetic field $(0,0,B)$ perpendicular to the $x$-$y$ plane 
in which the electrons are confined. For simplicity we assume that 
the electrons do not have the spin degrees of freedom, although we can 
treat a similar system with the spin degrees of freedom or with multiple 
layers in the same way. The Hamiltonian is given by 
\begin{equation}
H_0^{(N)}=\sum_{j=1}^N\left\{\frac{1}{2m_e}
\left[{\bf p}_j+e{\bf A}({\bf r}_j)+\mbox{\boldmath $\phi$}\right]^2+W({\bf r}_j)
\right\}+
\sum_{1\le i<j\le N}W^{(2)}({\bf r}_i-{\bf r}_j), 
\label{Ham2dgeneral0}
\end{equation}
where $-e$ and $m_e$ are, respectively, the charge of electron and 
the mass of electron, and ${\bf r}_j=(x_j,y_j)$ is the $j$ th Cartesian 
coordinate of the $N$ electrons. As usual, we define 
\begin{equation}
p_{x,j}=-i\hbar\frac{\partial}{\partial x_j}\quad \mbox{and}
\quad p_{y,j}=-i\hbar\frac{\partial}{\partial y_j}
\end{equation}
with the Planck constant $\hbar$. The system is defined on a rectangular box, 
\begin{equation}
{\cal T}:=[-L_x/2,L_x/2]\times[-L_y/2,L_y/2],
\end{equation}
with the periodic boundary conditions. The vector potential 
${\bf A}=(A_x,A_y)$ consists of two parts as ${\bf A}={\bf A}_{\rm P}+{\bf A}_0$,
where ${\bf A}_0({\bf r})=(-By,0)$ which gives the uniform magnetic 
field and the vector potential ${\bf A}_{\rm P}$ satisfies 
the periodic boundary conditions 
\begin{equation}
{\bf A}_{\rm P}(x,y)={\bf A}_{\rm P}(x+L_x,y)={\bf A}_{\rm P}(x,y+L_y). 
\end{equation}
We have also introduced the gauge parameters 
$\mbox{\boldmath $\phi$}=(\phi_x,\phi_y)\in{\cal T}_{\rm g}$, where 
the space ${\cal T}_{\rm g}\subset{\bf R}^2$ of the gauge 
parameters $\mbox{\boldmath $\phi$}$ is defined as 
\begin{equation}
{\cal T}_{\rm g}:=[0,\Delta\phi_x]\times[0,\Delta\phi_y]
\quad \mbox{with} \ \Delta\phi_s:=\frac{2\pi \hbar}{L_s}, \ s=x,y. 
\label{defTg}
\end{equation}
We call ${\cal T}_{\rm g}$ the gauge torus. As we will see in 
Section~\ref{Sec:QAHC}, 
the Hall conductance $\sigma_{xy}$ of the present system can be expressed in 
a geometric invariant on the gauge torus ${\cal T}_{\rm g}$. 
We assume ${\bf A}_{\rm P}\in C^1({\bf R}^2,{\bf R}^2)$, i.e., 
the components are continuously differentiable on ${\bf R}^2$. 
Further we assume that the single-body potential $W$ and 
the electron-electron interaction $W^{(2)}$ satisfy the following conditions: 
the periodic boundary conditions 
\begin{equation}
W(x+L_x,y)=W(x,y+L_y)=W(x,y)
\end{equation}
and 
\begin{equation}
W^{(2)}(x+L_x,y)=W^{(2)}(x,y+L_y)=W^{(2)}(x,y),
\end{equation}
and the boundedness\footnote{Let $f$ be a complex-valued function on 
${\bf R}^2$, and 
let $|f(x,y)|\le{\cal C}$ for some ${\cal C}$ except for a subset 
of Lebesgue measure zero in ${\bf R}^2$. Then the norm 
$\Vert f\Vert_\infty$ is given by the smallest such ${\cal C}$. If $f$ 
is a continuous function, then $\Vert f\Vert_\infty=\max|f(x,y)|$.}
\begin{equation}
\Vert W \Vert_\infty <w_0 <\infty \quad \mbox{and} \quad 
\Vert W^{(2)}\Vert_\infty <w_0^{(2)}<\infty
\end{equation}
with the positive constants $w_0$ and $w_0^{(2)}$ which are 
independent of the number $N$ of the electrons and of the system sizes 
$L_x,L_y$; the interaction $W^{(2)}$ is invariant under the interchange of 
two electron's coordinates as 
\begin{equation}
W^{(2)}(-x,-y)=W^{(2)}(x,y).
\end{equation}

For our purpose of accelerating the electrons on the torus ${\cal T}$ 
by the electric field, it is convenient to choose the periodic boundary 
conditions (\ref{PBCx}) and (\ref{PBCy}) below for the wavefunctions. 
Then the magnetic flux piercing the torus must be quantized \cite{Koma1,Koma2}
so that the Hamiltonian is self-adjoint, or 
we need additional conditions for the wavefunctions. 
See ref.~\cite{Wallet} for other choices of boundary conditions 
with a nonquantized flux. In the present paper,  
we choose the flux quantization condition, $L_xL_y=2\pi M\ell_B^2$, 
with a sufficiently large positive integer 
$M$, where $\ell_B$ is the so-called magnetic length defined as 
$\ell_B:=\sqrt{\hbar/(eB)}$. The number $M$ is exactly equal to 
the number of the states in a single Landau level of the single-electron 
Hamiltonian in the simple uniform magnetic field with no single-body potential 
and with no electric field. For simplicity we take $M$ even. 
We define by $\nu=N/M$ the filling factor. We assume $\nu<\nu_0$ with 
a positive constant $\nu_0$ which is independent of $L_x,L_y$ and $N$. 
In other words, the filling factor $\nu$ converges to 
a finite constant in an infinite volume limit. 
The above condition $L_xL_y=2\pi M\ell_B^2$ for the sizes $L_x,L_y$ is 
convenient for imposing the following periodic boundary conditions: 
For an $N$ electrons wavefunction $\Phi^{(N)}$, we impose 
periodic boundary conditions 
\begin{equation}
t_j^{(x)}(L_x)\Phi^{(N)}({\bf r}_1,{\bf r}_2,\ldots,{\bf r}_N)=
\Phi^{(N)}({\bf r}_1,{\bf r}_2,\ldots,{\bf r}_N)
\qquad\mbox{for}\ j=1,2,\ldots,N,
\label{PBCx}
\end{equation}
and
\begin{equation}
t_j^{(y)}(L_y)\Phi^{(N)}({\bf r}_1,{\bf r}_2,\ldots,{\bf r}_N)=
\Phi^{(N)}({\bf r}_1,{\bf r}_2,\ldots,{\bf r}_N)
\qquad\mbox{for}\ j=1,2,\ldots,N,
\label{PBCy}
\end{equation}
where $t^{(x)}(\cdots)$ and $t^{(y)}(\cdots)$ are magnetic translation 
operators \cite{Zak} defined as 
\begin{equation}
t^{(x)}(x')f(x,y)=f(x-x',y), \qquad t^{(y)}(y')f(x,y)
=\exp[iy'x/\ell_B^2]f(x,y-y')
\end{equation}
for a function $f$ on ${\bf R}^2$, and a subscript $j$ of an operator 
indicates that the operator acts on only the $j$-th coordinate 
of the function.\footnote{Throughout the present paper, we use 
this convention.} 

In order to get the expression of the current induced 
by an external electric field, 
we introduce a time-dependent vector potential \cite{Greenwood} 
into the Hamiltonian (\ref{Ham2dgeneral0}) as 
\begin{eqnarray}
H^{(N)}(t)&=&\sum_{j=1}^N\left\{\frac{1}{2m_e}
\left[p_{x,j}+eA_x({\bf r}_j)+\phi_x\right]^2+
\frac{1}{2m_e}\left[p_{y,j}+eA_y({\bf r}_j)+\phi_y+e\alpha(t)\right]^2
\right\}\ret
&+&\sum_{j=1}^NW({\bf r}_j)+\sum_{1\le i<j\le N}W^{(2)}({\bf r}_i-{\bf r}_j),
\label{Ham2dgeneral}
\end{eqnarray}
where the additional vector potential is given by 
${\bf A}_{\rm ex}(t)=(0,\alpha(t))$ with 
\begin{equation}
\alpha(t)=-Ft\times\cases{e^{\eta t}, & $t\le 0$;\cr 
                    1,           & $t>0$,\cr}
\label{alphat}
\end{equation}
with a small positive parameter $\eta$. The corresponding electric field 
is oriented along the $y$ direction with the constant strength $F$ for all $t\ge 0$. 
Namely we apply the electric field adiabatically from the initial 
time $t=t_0=-T$ with a large $T$, and observe the currents of the system 
for the time $t\ge 0$. 
Therefore we will consider those quantities for the time $t\ge 0$ only 
in the following. 

Throughout the present paper, we will consider the following two situations: 
(i) The ground state of the Hamiltonian $H_0^{(N)}$ of (\ref{Ham2dgeneral0}) is 
exactly $q$-fold degenerate, $(q=1,2,\ldots)$, with a small excitation energy gap  
which tends to zero in the infinite volume limit. 
(ii) The ground state is $q$-fold quasidegenerate with a uniform excitation 
energy gap which persists in the infinite volume limit in the sense of 
Assumption~\ref{gapassumption} below. In both of two situations, 
we take the initial state of the system at the time $t=t_0=-T$ as 
\begin{equation}
\omega_0(\cdots):=\frac{1}{q}\sum_{\mu=1}^q 
\left\langle\Phi_{0,\mu}^{(N)}(\mbox{\boldmath $\phi$}),
(\cdots)\Phi_{0,\mu}^{(N)}(\mbox{\boldmath $\phi$})
\right\rangle,
\label{omega0}
\end{equation}
where $\Phi_{0,\mu}^{(N)}(\mbox{\boldmath $\phi$})$ are the $q$ eigenvectors 
of the ground state  
of the Hamiltonian $H_0^{(N)}$ of (\ref{Ham2dgeneral0}). 
Namely we assume that the system is at a low temperature such that 
the corresponding inverse temperature $\beta$ satisfies the condition 
$\Delta{\cal E}\ll\beta^{-1}\ll\Delta E$, 
where $\Delta E$ is the excitation energy gap and 
$\Delta{\cal E}=\max_{\mu,\mu'}|E_{0,\mu}^{(N)}(\mbox{\boldmath $\phi$})
-E_{0,\mu'}^{(N)}(\mbox{\boldmath $\phi$})|$ 
with the energy eigenvalue $E_{0,\mu}^{(N)}(\mbox{\boldmath $\phi$})$ of 
the ground state eigenvector $\Phi_{0,\mu}^{(N)}(\mbox{\boldmath $\phi$})$. 
In the corresponding realistic situation, the transition between 
the degenerate ground states frequently occurs with finite probabilities 
owing to an external thermal perturbation. 
In consequence, all of the ground states are equally mixed as in the assumption 
of the initial state (\ref{omega0}).
But, when a symmetry breaking occurs at zero temperature,  
those transition probabilities become negligibly small 
in a large volume. In that situation, the assumption of the initial state 
(\ref{omega0}) may be physically unnatural. Instead of the mixed state (\ref{omega0}),  
we might have to take one of the symmetry breaking pure ground states 
as an initial state.  But we can expect that all of the pure ground states 
give the same current because the broken symmetry is the translational 
symmetry \cite{Koma1,Koma2}. Here we stress that we cannot justify this argument. 
To summarize, in both of the cases, we can expect that the assumption of the initial 
state (\ref{omega0}) leads to the realistic, correct current for 
the present quantum Hall system.

In general, it is believed that the existence of an energy gap 
above the ground state for an integral or a fractional filling of the electrons 
is essential to both integral and fractional quantization 
of the Hall conductance. In addition, the degeneracy \cite{NTW,AY} of the ground 
state is essential to the fractional quantization because the unique ground state 
with an energy gap always yields an integral quantization of the conductance 
by using the well-known topological argument \cite{NTW,AvronSeiler}.    
However, as Tao and Haldane \cite{TH} pointed out, one cannot expect 
the exact degeneracy of the ground state because the randomness of 
the potential(s) always lifts the degeneracy for a finite system.   
Therefore we require the following assumption 
on the quasidegeneracy of the ground state with the excitation energy gap  
in the quantum Hall case: 

\begin{assumption}
\label{gapassumption}
For any $\mbox{\boldmath $\phi$}\in {\cal T}_{\rm g}$, the ground state of the 
Hamiltonian $H_0^{(N)}$ of (\ref{Ham2dgeneral0}) is 
$q$-fold degenerate in the sense that 
\begin{equation}
\max_{\mu,\mu'\in\{1,2,\ldots,q\}}\left|E_{0,\mu}^{(N)}(\mbox{\boldmath $\phi$})
-E_{0,\mu'}^{(N)}(\mbox{\boldmath $\phi$})\right|\rightarrow 0 \ \mbox{as}\ 
L_x,L_y \rightarrow \infty,
\end{equation}
where $E_{0,\mu}^{(N)}(\mbox{\boldmath $\phi$})$, $\mu=1,2,\ldots,q$, are 
the energy eigenvalues 
of the ground state. Besides, there exists a uniform energy gap $\Delta E$ 
above the degenerate ground state in the sense that 
\begin{equation}
\inf_{\mbox{\boldmath $\phi$}\in{\cal T}_{\rm g}}E_1^{(N)}(\mbox{\boldmath $\phi$})
>\Delta E
+\sup_{\mbox{\boldmath $\phi$}\in{\cal T}_{\rm g}}\max_{\mu\in\{1,2,\ldots,q\}}
E_{0,\mu}^{(N)}(\mbox{\boldmath $\phi$}),
\label{gapcondition}
\end{equation}
where $E_1^{(N)}(\mbox{\boldmath $\phi$})$ is the energy of the first excited state, and 
$\Delta E$ is a positive constant which is independent of the number 
$N$ of the electrons and of the system sizes $L_x,L_y$. 
\end{assumption}
This assumption is justified for the non-interacting case 
with the potentials ${\bf A}_{\rm P}$ and $W$ satisfying 
the condition (\ref{nonintgapcondition}) in 
Theorem~\ref{theorem:noninteracting} below. 
Unfortunately, for the interacting case, we cannot justify the assumption 
of the gap. We call the subspace of the (quasi)degenerate ground state 
the sector of the ground state. We remark that the dimension $q$ of 
the sector of the ground state may depend on 
the system sizes $L_x,L_y$ and on the number $N$ of the electrons. 

The state of the system at the time $t\ge t_0$ is given by 
\begin{equation}
\omega(\cdots;t):=\omega_0([U^{(N)}(t,t_0)]^\dagger(\cdots)
U^{(N)}(t,t_0)).
\end{equation}
Here $U^{(N)}(t,t_0)$ is the time evolution operator for 
the Schr\"odinger equation,
\begin{equation}
i\hbar\frac{\partial}{\partial t}\Psi^{(N)}(t)=H^{(N)}(t)
\Psi^{(N)}(t),
\end{equation}
with the time dependent Hamiltonian $H^{(N)}(t)$ of (\ref{Ham2dgeneral}). 
Namely the solution of the equation is written as 
\begin{equation}
\Psi^{(N)}(t)=U^{(N)}(t,t_0)\Psi^{(N)}(t_0)
\end{equation}
in terms of the operator $U^{(N)}(t,t_0)$ with the initial 
vector $\Psi^{(N)}(t_0)$. 

Now we define the total velocity operator as 
\begin{equation}
v_{{\rm tot},i}(t):=\cases{\displaystyle{\sum_{j=1}^N 
\frac{1}{m_e}[p_{x,j}+eA_x({\bf r}_j)+\phi_x],} & $i=x$;\cr
\displaystyle{\sum_{j=1}^N\frac{1}{m_e}[p_{y,j}+eA_y({\bf r}_j)+\phi_y+
e\alpha(t)],} 
& $i=y$.\cr}
\label{totvelocityext}
\end{equation}
Then the total current density is given by 
\begin{equation}
{\bf j}_{\rm tot}(t):=-\frac{e}{L_xL_y}
\omega({\bf v}_{\rm tot}(t);t),
\end{equation}
where ${\bf v}_{\rm tot}(t)=(v_{{\rm tot},x}(t),v_{{\rm tot},y}(t))$. 
This total current density consists of the initial current 
density ${\bf j}_0$ and the induced current density 
${\bf j}_{\rm ind}(t)$ due to the external electric field as 
\begin{equation}
{\bf j}_{\rm tot}(t)={\bf j}_0+{\bf j}_{\rm ind}(t),
\end{equation}
where 
\begin{equation}
{\bf j}_0=-\frac{e}{L_xL_y}\omega_0({\bf v}_{\rm tot}^{(0)})
\end{equation}
with the velocity operator, 
\begin{equation}
{\bf v}_{\rm tot}^{(0)}=\sum_{j=1}^N\frac{1}{m_e}[{\bf p}_j+e{\bf A}
({\bf r}_j)+\mbox{\boldmath $\phi$}], 
\label{veloctynoF}
\end{equation}
without the vector potential $\alpha(t)$ giving the external 
electric field. The initial current density ${\bf j}_0$ is not 
necessarily vanishing because the persistent current may exist 
owing to the presence of the vector potentials. 
The linear response coefficients are given by  
\begin{equation}
\sigma_{{\rm tot},sy}(t;\mbox{\boldmath $\phi$},\eta,T)
:=\lim_{F\rightarrow 0}\frac{j_{{\rm ind},s}(t)}{F} 
\end{equation}
in the $s=x,y$ directions, where we have written ${\bf j}_{\rm ind}(t)=
(j_{{\rm ind},x}(t),j_{{\rm ind},y}(t))$.   
As we will show in the next Section~\ref{LRT}, these coefficients  
for the time $t\ge 0$ have the expressions, 
\begin{equation}
\sigma_{{\rm tot},xy}(t;\mbox{\boldmath $\phi$},\eta,T)
=\sigma_{xy}(\mbox{\boldmath $\phi$})+\gamma_{xy}(\mbox{\boldmath $\phi$})\cdot t
+\delta\sigma_{xy}(t;\mbox{\boldmath $\phi$},\eta,T),
\label{sigmatotxy0} 
\end{equation}
and 
\begin{equation}
\sigma_{{\rm tot},yy}(t;\mbox{\boldmath $\phi$},\eta,T)
=\gamma_{yy}(\mbox{\boldmath $\phi$})\cdot t+
\delta\sigma_{yy}(t;\mbox{\boldmath $\phi$},\eta,T),
\label{sigmatotyy0} 
\end{equation}
where $\sigma_{xy}(\mbox{\boldmath $\phi$})$, 
$\gamma_{xy}(\mbox{\boldmath $\phi$})$ and $\gamma_{yy}(\mbox{\boldmath $\phi$})$ are all 
independent of the time $t$. The rest of the two 
terms $\delta\sigma_{sy}(t;\mbox{\boldmath $\phi$},\eta,T)$ 
for $s=x,y$ are due to the initial switching process for adiabatically applying 
the electric field in the time $-T\le t\le 0$, and so the two terms are 
negligibly small for the slow switching condition $\eta T\gg 1$ and 
$\eta\ll \Delta E/\hbar$. In particular, they vanish in the slow switching limit 
$\eta T\rightarrow \infty$ and $\eta\rightarrow 0$.   
See the bound (\ref{deltasigmasybound}) and the equality (\ref{vanishB0}) 
in Section~\ref{Sec:translationinv} for 
certain translationally invariant systems, 
the bound (\ref{deltasigmasyboundnonint}) in Section~\ref{noninteracting} 
for the non-interacting electron gas 
and the bound (\ref{deltasigmasybound2}) in 
Section~\ref{IntCase} for the interacting electron gas.     
The term $\sigma_{xy}(\mbox{\boldmath $\phi$})$ corresponds to the Hall conductance. 
For simplicity, we will call $\sigma_{xy}(\mbox{\boldmath $\phi$})$ the Hall conductance. 
As mentioned in Introduction, the time-independent 
term $\sigma_{yy}(\mbox{\boldmath $\phi$})$ 
corresponding to the diagonal conductance does not appear \cite{AG,ElSch,NK}, 
and instead of that, there appear linearly time-dependent terms, 
$\gamma_{sy}(\mbox{\boldmath $\phi$})\cdot t$. 
In particular, if the system is translationally invariant 
in both $x$ and $y$ directions with no uniform magnetic field, 
then the total velocity of the electrons is proportional to the time $t$ 
owing to the uniform electric field. (See Section~\ref{Sec:translationinv}.)  
Therefore we will call $\gamma_{sy}(\mbox{\boldmath $\phi$})$ 
the acceleration coefficient.  

On the other hand, when a finite energy gap appears above the sector 
of the ground state as in a band insulator or in the quantum Hall case, 
we can expect that both of the acceleration coefficients 
$\gamma_{sy}(\mbox{\boldmath $\phi$})$ vanish in the infinite volume limit. 
Actually this statement holds for the non-interacting case 
as we will see in Theorem~\ref{theorem:noninteracting} below. 
However, we could not obtain a similar theorem for the interacting case 
except for a trivial case without disorder in 
Section~\ref{Sec:translationinv}. Of course, we cannot expect that 
the acceleration coefficients $\gamma_{sy}(\mbox{\boldmath $\phi$})$ are exactly 
vanishing for a finite volume in a generic situation with disorder 
because there may exist non-vanishing current due to the scattering of 
the electrons by the disorder. 

If the Hall conductance $\sigma_{xy}(\mbox{\boldmath $\phi$})$ alone is needed, 
then it is enough to measure the linear response coefficients 
at the time $t=0$ because 
the effect of the Lorentz force alone persists at $t=0$ without any 
acceleration of the electrons by the potentials of the system. 

We define the averaged Hall conductance and the averaged acceleration 
coefficients over the gauge parameters $\mbox{\boldmath $\phi$}$ on the gauge 
torus ${\cal T}_{\rm g}$ as 
\begin{equation}
\overline{\sigma_{xy}(\mbox{\boldmath $\phi$})}=\frac{1}{\Delta\phi_x\Delta\phi_y}
\int_{{\cal T}_{\rm g}}d\phi_x d\phi_y \ \sigma_{xy}(\mbox{\boldmath $\phi$}),
\label{avsigmaxy}
\end{equation}
and 
\begin{equation}
\overline{\gamma_{sy}(\mbox{\boldmath $\phi$})}=\frac{1}{\Delta\phi_x\Delta\phi_y}
\int_{{\cal T}_{\rm g}}d\phi_x d\phi_y \ \gamma_{sy}(\mbox{\boldmath $\phi$})
\quad \mbox{for } s=x,y.
\label{avgammasy} 
\end{equation}
Under Assumption~\ref{gapassumption}, the averaged Hall conductance 
$\overline{\sigma_{xy}(\mbox{\boldmath $\phi$})}$ can be written as  
\begin{equation}
\overline{\sigma_{xy}(\mbox{\boldmath $\phi$})}=\frac{e^2}{h}\frac{{\cal I}}{q}
\label{avsigmaFQG}
\end{equation}
in terms of the geometric invariant \cite{TKNN,Kohmoto,BVS,NTW,AvronSeiler,AY}, 
\begin{equation}
{\cal I}=\frac{1}{2\pi i}\int_{{\cal T}_{\rm g}}d\phi_x d\phi_y 
\ {\rm tr}\ {\cal F}(\mbox{\boldmath $\phi$}),
\label{GI}
\end{equation}
with the curvature ${\cal F}(\mbox{\boldmath $\phi$})$ given by 
\begin{equation}
{\cal F}(\mbox{\boldmath $\phi$})=\frac{\partial}{\partial \phi_x}
{\cal A}_y(\mbox{\boldmath $\phi$})
-\frac{\partial}{\partial \phi_y}{\cal A}_x(\mbox{\boldmath $\phi$})
+[{\cal A}_x(\mbox{\boldmath $\phi$}),{\cal A}_y(\mbox{\boldmath $\phi$})],
\label{CF}
\end{equation}
where ${\rm tr}$ stands for the trace of the matrix, 
and ${\cal A}_s(\mbox{\boldmath $\phi$})$ for $s=x,y$ are the connections on 
the gauge torus ${\cal T}_{\rm g}$, i.e., 
each ${\cal A}_s(\mbox{\boldmath $\phi$})$ takes 
a $q \times q$ matrix value in the Lie algebra of the unitary group $U(q)$ of 
$q\times q$ matrices \cite{Atiyah}. As we will see in Section~\ref{Sec:QAHC}, 
the connections ${\cal A}_s(\mbox{\boldmath $\phi$})$ are written in terms of 
the ground state wavefunctions. 
In the language of the non-Abelian gauge theory \cite{Weinberg}, 
the curvature ${\cal F}(\mbox{\boldmath $\phi$})$ corresponds to the field strength 
tensor or the ``electromagnetic" field, and the connections 
${\cal A}_s(\mbox{\boldmath $\phi$})$ 
correspond to the gauge fields on the torus ${\cal T}_{\rm g}$. 
Thus the averaged Hall conductance can be written in terms of the geometric invariant 
for the non-Abelian gauge fields on the torus. 
Since the geometric invariant ${\cal I}$ called the first Chern number takes 
an integer value, the averaged Hall conductance 
$\overline{\sigma_{xy}(\mbox{\boldmath $\phi$})}$ is integrally or fractionally 
quantized. 
 
On the other hand, we can prove that the averaged acceleration coefficients 
$\overline{\gamma_{sy}(\mbox{\boldmath $\phi$})}$ are exactly vanishing.  
We summarize the well-known result of the averaged Hall 
conductance $\overline{\sigma_{xy}(\mbox{\boldmath $\phi$})}$ and our result about 
the averaged acceleration coefficients $\overline{\gamma_{sy}(\mbox{\boldmath $\phi$})}$ 
as the following theorem in the most general setting for the present model: 

\begin{theorem}
\label{theorem:avconductance}
Suppose that, for a finite volume and a filling factor $\nu$ of 
the Landau levels, there exists a uniform gap above the sector of  
the ground state in the sense of Assumption~\ref{gapassumption}. 
Then there exists an integer $p$ such that the averaged Hall conductance 
for the finite volume is quantized as 
\begin{equation}
\overline{\sigma_{xy}(\mbox{\boldmath $\phi$})}=-\frac{e^2}{h}\frac{p}{q}
\label{wellknownTIresult}
\end{equation}
with the dimension $q$ of the sector of the ground state, 
and the averaged acceleration coefficients are vanishing as 
\begin{equation}
\overline{\gamma_{sy}(\mbox{\boldmath $\phi$})}=0 \quad \mbox{for both $s=x,y$ directions}.
\end{equation}
\end{theorem}
The proof will be given in Section~\ref{Sec:QAHC}. 
In such a general setting, we cannot determine the fraction $p/q$, 
which is expected to be equal to the filling factor $\nu$ 
of the Landau levels \cite{Koma2}. We also remark that, for general fixed 
values of the gauge parameters $\mbox{\boldmath $\phi$}\in{\cal T}_{\rm g}$, 
we cannot expect 
the same exact results about the fractional quantization of the Hall 
conductance and the vanishing acceleration coefficients 
without any finite size correction. 

For the quantum Hall case without the electron-electron interaction, 
we can obtain much stronger results as follows: 

\begin{theorem}
\label{theorem:noninteracting}
Assume $W^{(2)}=0$, i.e., no electron-electron interaction, 
and assume an integer filling factor $\nu=\ell$ of the Landau levels 
with $\ell\in{\bf N}$. 
Further we assume that the vector potential ${\bf A}_{\rm P}$ and 
the electrostatic potential $W$ satisfy the condition, 
\begin{equation}
\hbar\omega_c>\sqrt{\frac{2\hbar\omega_c}{m_e}}e
\left\Vert|{\bf A}_{\rm P}|\right\Vert_\infty
\left[\sqrt{\ell+{1}/{2}}+\sqrt{\ell-{1}/{2}}\right]
+\frac{e^2}{2m_e}\left(\left\Vert|{\bf A}_{\rm P}|\right\Vert_\infty\right)^2
+\Vert W^+\Vert_\infty+\Vert W^-\Vert_\infty,
\label{nonintgapcondition}
\end{equation}
where $\omega_c$ is the cyclotron frequency given by 
$\omega_c=eB/m_e$, $|{\bf A}_{\rm P}|=\sqrt{{\bf A}_{\rm P}\cdot{\bf A}_{\rm P}}$, 
and $W^\pm=\max\{\pm W,0\}$. 
Then there exists a uniform gap above the unique ground state 
in the sense of Assumption~\ref{gapassumption}, and the following bounds 
are valid:  
\begin{equation}
\left|\sigma_{xy}(\mbox{\boldmath $\phi$}_0)+\frac{e^2}{h}\ell\right|\le{\cal C}
\max\{L_x^{-1},L_y^{-1}\},
\label{fscsigmaxy}
\end{equation}
and 
\begin{equation}
\left|\gamma_{sy}(\mbox{\boldmath $\phi$}_0)\right|\le {\cal C}'
\max\{L_x^{-1},L_y^{-1}\}
\label{fscgammasy}
\end{equation}
for any gauge parameters $\mbox{\boldmath $\phi$}_0\in{\cal T}_{\rm g}$, 
and for $s=x,y$, 
where the positive constants ${\cal C}$ and ${\cal C}'$ are independent 
of the number $N$ of the electrons and of the system sizes $L_x,L_y$. 
\end{theorem}
The proof will be given in Section~\ref{noninteracting}. 
Thus the Hall conductance shows the integral quantization with the 
finite size correction, and the acceleration coefficients are vanishing in 
the infinite volume limit. However, the bounds for the finite size corrections 
would not be optimal \cite{NT} in comparison with the precision of the quantization 
and the weakness of the dissipation of heat in realistic quantum Hall systems. 
Perhaps, if possible, we should take account of the effect of the self-averaging 
about the disorder in the realistic systems. 

In order to explain the appearance of the Hall conductance plateaus, 
the gap condition in Theorem~\ref{theorem:noninteracting} 
must be replaced with localization estimates. 
Namely we must show that 
the quantized Hall conductance does not change when varying the Fermi level 
within the localization regime. For infinite volume systems,  
the Hall conductance formula by Bellissard shows the plateaus \cite{AG,BVS,ASS2}.
Kunz \cite{Kunz} also discussed the plateaus in an infinite volume system 
by combining the topological argument with certain assumptions on localization. 
This issue for finite volume systems will be discussed in ref.~\cite{Koma3}.  

For the interacting case, we could not estimate the corresponding finite size 
corrections. But we obtain the following theorem about the relation between 
the fraction $p/q$ of the quantization and the filling factor $\nu$ of the electrons: 

\begin{theorem}
\label{intFQE}
Assume ${\bf A}_{\rm P}=0$, and 
that the electrostatic potential $W$ and the electron-electron 
interaction $W^{(2)}$ satisfy $W\in C^3({\bf R}^2)$ and 
$W^{(2)}\in C^1({\bf R}^2)$, respectively.\footnote{The set $C^k(S)$ denotes 
$k$ times continuously differentiable functions on the set $S$.} 
Further we assume that, 
for a finite volume and a filling factor $\nu$ of 
the Landau levels, there exists a uniform gap $\Delta E$ 
above the sector of the ground state with the degeneracy $q$ in the sense of 
Assumption~\ref{gapassumption}. Then the fraction $p/q$ of the averaged Hall conductance 
$\overline{\sigma_{xy}(\mbox{\boldmath $\phi$})}$ of (\ref{wellknownTIresult}) satisfies  
\begin{equation}
\nu(1-\delta)\le \frac{p}{q}\le \nu(1+\delta),
\label{fractionbound0}
\end{equation}
where the positive number $\delta$ is given by 
\begin{equation}
\delta=2\ell_B^4\frac{\hbar\omega_c}{(\Delta E)^3}
\mathop{\max_{m,n\ge 0;}}_{m+n=2}\left\Vert\frac{\partial^{m+n}W}{\partial x^m 
\partial y^n}\right\Vert_\infty^2. 
\end{equation}
\end{theorem}
The proof will be given in Section~\ref{FQHE}. As we mentioned in 
Introduction, if the interval 
$[\nu(1-\delta),\nu(1+\delta)]$ does not include any integer, then 
the number $p/q$ must be equal to purely a fractional number, i.e., 
a non-integer. In addition, the dimension $q$ of the sector of 
the ground state must be greater than $1$. 
In the weak potential limit $\delta\rightarrow 0$, 
we obtain the desired result, 
\begin{equation}
\overline{\sigma_{xy}(\mbox{\boldmath $\phi$})}=-\frac{e^2}{h}\nu, \quad \mbox{with }
\nu=p/q
\end{equation}
from the result (\ref{wellknownTIresult}) with the inequality 
(\ref{fractionbound0}). Therefore one can expect the fractional quantization 
of the Hall conductance \cite{Koma1,Koma2} 
when a spectral gap appears above the sector of the ground state  
for {\em the fractional filling factor} $\nu=p/q$. 
But the Hall conductance would vanish in the exceptional 
case with a very strong periodic potential.\footnote{When a spectral gap 
exists owing to a strong periodic potential irrespective of 
the uniform magnetic field, 
the Hall conductance must vanish in the weak limit of 
the magnetic field. 
In such a situation, the integer $p$ in the fractionally quantized 
Hall conductance must be equal to zero because the Hall conductance is 
a continuous function of the uniform magnetic field. Thus the Hall conductance 
vanishes owing to the strong periodic potential \cite{NB}.} 

In the case with ${\bf A}_{\rm P}\ne 0$, we can also obtain very similar 
results to Theorem~\ref{intFQE}. But we need stronger assumptions on 
the potentials ${\bf A}_{\rm P}$, $W$ and $W^{(2)}$. See 
Section~\ref{FQHE} for the details. 

%%%%%%%%%%%%%%%%%%%%%%%%%%%%%%%%%%%%%%%%%%%%%%%%%%%%%%%%%%
\Section{Derivation of the linear response coefficients}
\label{LRT}

In this section, we derive the expressions for the linear response  
coefficients by using the perturbation theory \cite{Kato} 
in a mathematically rigorous manner. 
We denote by $\Phi_{0,\mu}^{(N)}(\mbox{\boldmath $\phi$})$ the ground state 
eigenvectors of 
the Hamiltonian $H_0^{(N)}$ of (\ref{Ham2dgeneral0}) with the eigenvalue 
$E_{0,\mu}^{(N)}(\mbox{\boldmath $\phi$})$, 
and denote by $\Phi_{n,\mu}^{(N)}(\mbox{\boldmath $\phi$})$ with $n\ge 1$ 
the eigenvectors of the excited states with the energy eigenvalue 
$E_n^{(N)}(\mbox{\boldmath $\phi$})$ and with the subscript $\mu$ for the degeneracy. 
In the following, we will often use the abbreviations, 
$\Phi_{n,\mu}^{(N)}$, $E_{0,\mu}^{(N)}$, $E_n^{(N)}$, 
by dropping the $\mbox{\boldmath $\phi$}$ dependence if there is no confusion. 

To begin with, we rewrite the Hamiltonian $H^{(N)}(t)$ of 
(\ref{Ham2dgeneral}) as 
\begin{equation}
H^{(N)}(t)=H_0^{(N)}+\Delta H_0^{(N)}(t)+H_{\rm per}^{(N)}(t) 
\end{equation}
with the diagonal part,
\begin{equation} 
\Delta H_0^{(N)}(t)=
\sum_nQ(E_n^{(N)})H_{\rm min}^{(N)}(t)Q(E_n^{(N)})
+N\frac{e^2}{2m_e}[\alpha(t)]^2,
\end{equation}
of the perturbation and the off-diagonal part,  
\begin{equation}
H_{\rm per}^{(N)}(t)=H_{\rm min}^{(N)}(t)-
\sum_nQ(E_n^{(N)})H_{\rm min}^{(N)}(t)Q(E_n^{(N)}),
\end{equation}
where $H_{\rm min}^{(N)}(t)$ is the minimal coupling with the external 
electric field, i.e., 
\begin{equation}
H_{\rm min}^{(N)}(t)=\frac{e}{m_e}\sum_{j=1}^N\alpha(t)
[p_{y,j}+eA_y({\bf r}_j)], 
\end{equation}
and $Q(E_0^{(N)})$ is the projection operator onto the subspace 
spanned by the ground state eigenvectors 
$\Phi_{0,\mu}^{(N)}$ of the unperturbed Hamiltonian $H_0^{(N)}$, 
and $Q(E_n^{(N)})$ for $n\ge 1$ is the projection operator onto 
the eigenspace spanned by the excited state eigenvector(s) of 
the Hamiltonian $H_0^{(N)}$ with the eigenvalue $E_n^{(N)}$. 

Consider the time-dependent Schr\"odinger equation 
\begin{equation}
i\hbar\frac{\partial}{\partial t}\Psi^{(N)}(t)
=H^{(N)}(t)\Psi^{(N)}(t)
\label{tdSchrodinger}
\end{equation}
with the Hamiltonian $H^{(N)}(t)$ of (\ref{Ham2dgeneral}). 
The solution $\Psi^{(N)}(t)$ can be written as 
$\Psi^{(N)}(t)=U^{(N)}(t,t_0)\Psi^{(N)}(t_0)$ by using the time evolution 
operator $U^{(N)}(t,s)$, with an initial vector $\Psi^{(N)}(t_0)$ at 
the initial time $t=t_0$. We denote by $U_0^{(N)}(t,s)$ 
the time evolution operator 
for the Hamiltonian $H_0^{(N)}+\Delta H_0^{(N)}(t)$. 

Let $\Psi^{(N)}(t)$ be a solution of the Schr\"odinger equation 
(\ref{tdSchrodinger}). Note that 
\begin{eqnarray}
& &\frac{\partial}{\partial s}U_0^{(N)}(t,s)\Psi^{(N)}(s)\ret
&=&
\frac{i}{\hbar}U_0^{(N)}(t,s)
\left\{H_0^{(N)}+\Delta H_0^{(N)}(s)\right\}
\Psi^{(N)}(s)+U_0^{(N)}(t,s)\frac{\partial}{\partial s}\Psi^{(N)}(s)\ret
&=&
\frac{i}{\hbar}U_0^{(N)}(t,s)\left\{H_0^{(N)}+
\Delta H_0^{(N)}(s)-H^{(N)}(s)\right\}
\Psi^{(N)}(s)\ret
&=&-\frac{i}{\hbar}U_0^{(N)}(t,s)H_{\rm per}^{(N)}(s)\Psi^{(N)}(s).
\end{eqnarray}
Integrating this on the time $s$ from $t_0$ to $t$, one obtains 
\begin{equation}
\Psi^{(N)}(t)-U_0^{(N)}(t,t_0)\Psi^{(N)}(t_0)= 
-\frac{i}{\hbar}\int_{t_0}^tds \ 
U_0^{(N)}(t,s)H_{\rm per}^{(N)}(s)\Psi^{(N)}(s).
\label{iteratePhi(t)}
\end{equation}
{From} the definition of $U^{(N)}(t,s)$, this can be rewritten as 
\begin{equation}
\left[U^{(N)}(t,t_0)-U_0^{(N)}(t,t_0)\right]\Psi^{(N)}(t_0)= 
-\frac{i}{\hbar}\int_{t_0}^tds \ 
U_0^{(N)}(t,s)H_{\rm per}^{(N)}(s)\Psi^{(N)}(s).
\end{equation}
Since both $U^{(N)}(t,s)$ and $U_0^{(N)}(t,s)$ are bounded, 
one has the following lemma \cite{Kato}:

\begin{lemma}
\label{lemma:diffU}
In the strong sense, $U^{(N)}(t,s)\rightarrow U_0^{(N)}(t,s)$ 
as $F\rightarrow 0$, uniformly in $t,s$ in any finite interval. 
\end{lemma}

In the equation (\ref{iteratePhi(t)}), we take the initial state at $t_0=-T$ 
as 
\begin{equation}
\Psi^{(N)}(t_0=-T)=U_0^{(N)}(-T,0)\Phi^{(N)}
\label{initialstate}
\end{equation}
with a vector $\Phi^{(N)}$ in the domain of the Hamiltonian $H_0^{(N)}$ of 
(\ref{Ham2dgeneral0}). Then the equation (\ref{iteratePhi(t)}) becomes 
\begin{equation}
\Psi^{(N)}(t)=U_0^{(N)}(t,0)\Phi^{(N)} 
-\frac{i}{\hbar}\int_{-T}^tds \ 
U_0^{(N)}(t,s)H_{\rm per}^{(N)}(s)\Psi^{(N)}(s).
\end{equation}
The following theorem can be obtained in the same way as in Theorem~2.19 
in Sec.~2 in Chap. IX of ref.~\cite{Kato}: 

\begin{theorem}
\label{theorem:Phi(t)expand}
Let $\Phi^{(N)}$ be a vector in the domain of the Hamiltonian $H_0^{(N)}$ of 
(\ref{Ham2dgeneral0}). Then 
\begin{equation}
\Psi^{(N)}(t)=U_0^{(N)}(t,0)\Phi^{(N)} 
-\frac{i}{\hbar}\int_{-T}^tds \ 
U_0^{(N)}(t,s)H_{\rm per}^{(N)}(s)U_0^{(N)}(s,0)\Phi^{(N)}+o(F),
\end{equation}
where $o(F)$ denotes a vector $\Psi_{\rm R}$ with the norm 
$\Vert\Psi_{\rm R}\Vert$ 
satisfying $\Vert\Psi_{\rm R}\Vert/F\rightarrow 0$ as $F\rightarrow 0$. 
\end{theorem}
\begin{proof}{Proof}
Since 
\begin{equation}
\Psi^{(N)}(s)=U^{(N)}(s,-T)\Psi^{(N)}(-T)
\end{equation}
and
\begin{equation}
U_0^{(N)}(s,0)\Phi^{(N)}=U_0^{(N)}(s,-T)U_0^{(N)}(-T,0)\Phi^{(N)}
=U_0^{(N)}(s,-T)\Psi^{(N)}(-T),
\end{equation}
it is sufficient to show that
\begin{equation}
\left\Vert H_{\rm per}^{(N)}(s)\left[U^{(N)}(s,-T)-U_0^{(N)}(s,-T)\right]
\Psi^{(N)}(-T)\right\Vert=o(F).
\end{equation}
Note that
\begin{equation}
H_{\rm per}^{(N)}(s)U^{(N)}(s,-T)\Psi^{(N)}(-T)=
H_{\rm per}^{(N)}(s)[S(s)]^{-1}{\tilde U}^{(N)}(s,-T)
S(-T)\Psi^{(N)}(-T)
\label{HperUPsi}
\end{equation}
and
\begin{equation}
H_{\rm per}^{(N)}(s)U_0^{(N)}(s,-T)\Psi^{(N)}(-T)=
H_{\rm per}^{(N)}(s)[{\tilde S}_0(s)]^{-1}{\tilde U}_0^{(N)}(s,-T)
{\tilde S}_0(-T)\Psi^{(N)}(-T), 
\label{HperU0Psi}
\end{equation}
where 
\begin{equation}
{\tilde U}(t,s)=S(t)U(t,s)[S(s)]^{-1},
\end{equation}
\begin{equation}
S(t)=\frac{i}{\hbar}H^{(N)}(t)+\lambda_0,
\end{equation}
\begin{equation}
{\tilde U}_0(t,s)={\tilde S}_0(t)U_0(t,s)[{\tilde S}_0(s)]^{-1},
\end{equation}
and
\begin{equation}
{\tilde S}_0(t)=\frac{i}{\hbar}\left[H_0^{(N)}+\Delta H_0^{(N)}(t)
\right]+\lambda_0.
\end{equation}
Here $\lambda_0$ is some real constant so that both $[S(t)]^{-1}$ and 
$[{\tilde S}_0]^{-1}$ exist, and the operators ${\tilde U}(t,s)$ and 
${\tilde U}_0(t,s)$ are well defined and bounded \cite{Kato2}. 
Formally one has 
\begin{equation}
\frac{d}{dr}{\tilde U}(t,r)U(r,s)=-{\tilde U}(t,r)
\left[\frac{d}{dr}S(r)\right][S(r)]^{-1}U(r,s).
\end{equation}
Integrating this on the time $r$ from $s$ to $t$, one 
obtain\footnote{For simplicity we have given the formal derivation here 
although the resulting integral equation (\ref{inteqtildeU}) is justified 
\cite{Kato2}.} 
\begin{equation}
{\tilde U}(t,s)=U(t,s)+\int_s^t dr {\tilde U}(t,r)
\left[\frac{d}{dr}S(r)\right][S(r)]^{-1}U(r,s).
\label{inteqtildeU}
\end{equation}
Further one has 
\begin{equation}
[S(s)]^{-1}=[{\tilde S}_0(s)]^{-1}-[S(s)]^{-1}\frac{i}{\hbar}
H_{\rm per}^{(N)}(s)[{\tilde S}_0(s)]^{-1}.
\end{equation}
Using these two formulas, the right-hand side of (\ref{HperUPsi}) can be 
evaluated as 
\begin{eqnarray}
& & H_{\rm per}^{(N)}(s)[S(s)]^{-1}{\tilde U}^{(N)}(s,-T)
S(-T)\Psi^{(N)}(-T)\ret&=&
 H_{\rm per}^{(N)}(s)[{\tilde S}_0(s)]^{-1}U^{(N)}(s,-T){\tilde S}_0(-T)
\Psi^{(N)}(-T)+o(F).
\label{HperSUSPsi}
\end{eqnarray}

Similarly one has 
\begin{equation}
{\tilde U}_0(t,s)=U_0(t,s)+\int_s^t dr {\tilde U}_0(t,r)
\left[\frac{d}{dr}{\tilde S}_0(r)\right][{\tilde S}_0(r)]^{-1}U_0(r,s). 
\end{equation}
Hence the right-hand side of (\ref{HperU0Psi}) can be evaluated as 
\begin{eqnarray}
& & H_{\rm per}^{(N)}(s)[{\tilde S}_0(s)]^{-1}{\tilde U}_0^{(N)}(s,-T)
{\tilde S}_0(-T)\Psi^{(N)}(-T)\ret
&=& H_{\rm per}^{(N)}(s)[{\tilde S}_0(s)]^{-1}U_0^{(N)}(s,-T)
{\tilde S}_0(-T)\Psi^{(N)}(-T)+o(F).
\label{HperSU0SPsi}
\end{eqnarray}
Combining (\ref{HperUPsi}), (\ref{HperU0Psi}), (\ref{HperSUSPsi}) and 
(\ref{HperSU0SPsi}), one has 
\begin{eqnarray}
& &\left\Vert H_{\rm per}^{(N)}(s)\left[U^{(N)}(s,-T)-U_0^{(N)}(s,-T)\right]
\Psi^{(N)}(-T)\right\Vert\ret
&=&\left\Vert H_{\rm per}^{(N)}(s)[{\tilde S}_0(s)]^{-1}
\left[U^{(N)}(s,-T)-U_0^{(N)}(s,-T)\right]{\tilde S}_0(-T)
\Psi^{(N)}(-T)\right\Vert+o(F).\ret 
\end{eqnarray}
This right-hand side is of $o(F)$ from Lemma~\ref{lemma:diffU} because 
the operator $H_{\rm per}^{(N)}(s)[{\tilde S}_0(s)]^{-1}$ is bounded and 
already of order $F$. 
\end{proof}

We take the initial state $\omega_0$ at the time $t=t_0=-T$ as 
\begin{equation}
\omega_0(\cdots)=\frac{1}{q}\sum_{\mu=1}^q
\left\langle\Phi_{0,\mu}^{(N)},(U_0^{(N)}(-T,0))^\dagger
(\cdots)U_0^{(N)}(-T,0),
\Phi_{0,\mu}^{(N)}\right\rangle,
\label{defomega0}
\end{equation}
where the ground state vectors $\Phi_{0,\mu}^{(N)}$ are normalized. 
Using the projection $Q(E_0^{(N)})$ onto the sector of the ground state, 
we have 
\begin{eqnarray}
\omega_0(a)&=&\frac{1}{q}\sum_{\mu=1}^q
\left\langle\Phi_{0,\mu}^{(N)},(U_0^{(N)}(-T,0))^\dagger a U_0^{(N)}(-T,0),
\Phi_{0,\mu}^{(N)}\right\rangle\ret
&=&\frac{1}{q}\ {\rm Tr}\ Q(E_0^{(N)})(U_0^{(N)}(-T,0))^\dagger a 
U_0^{(N)}(-T,0)
\ret
&=&\frac{1}{q}\ {\rm Tr}\ Q(E_0^{(N)}) a 
\end{eqnarray}
for any observable $a$ in the domain. 
Here Tr stands for the trace on the Hilbert space, and 
we have used the fact that $Q(E_0^{(N)})$ commutes with 
the unitary operator $U_0^{(N)}(t,s)$. 
Starting from this initial state is physically natural 
as we discussed in the preceding section. 

With this initial condition, the total current at time $t$ is given by 
\begin{equation}
{\bf j}_{\rm tot}(t)=
-\frac{e}{L_xL_y}\frac{1}{q}\sum_{\mu=1}^q 
\left\langle\Psi_{0,\mu}^{(N)}(t),{\bf v}_{\rm tot}(t)
\Psi_{0,\mu}^{(N)}(t)\right\rangle,
\label{currentjtott}
\end{equation}
where $\Psi_{0,\mu}^{(N)}(t)$ is the solution of the time-dependent 
Schr\"odinger equation (\ref{tdSchrodinger}) with the initial state 
$U_0^{(N)}(-T,0)\Phi_{0,\mu}^{(N)}$ for $\mu=1,2,\ldots,q$, and the total 
velocity operator ${\bf v}_{\rm tot}(t)$ is given by (\ref{totvelocityext}). 
This total current ${\bf j}_{\rm tot}(t)$ 
can be decomposed into two parts as 
\begin{equation}
{\bf j}_{\rm tot}(t)={\bf j}_0+{\bf j}_{\rm ind}(t)
\end{equation}
with the current, 
\begin{equation}
{\bf j}_0=-\frac{e}{L_xL_y}\frac{1}{q}\sum_{\mu=1}^q 
\left\langle\Phi_{0,\mu}^{(N)},\left[U_0^{(N)}(t,0)\right]^\dagger
{\bf v}_{\rm tot}^{(0)}
U_0^{(N)}(t,0)\Phi_{0,\mu}^{(N)}\right\rangle, 
\label{currentj0}
\end{equation}
where the velocity operator ${\bf v}_{\rm tot}^{(0)}$ is given by 
(\ref{veloctynoF}). 
The current ${\bf j}_0$ of (\ref{currentj0}) is independent of the time 
$t$, and is equal to the initial current or 
the persistent current without the external electric field. 
Actually it can be rewritten as 
\begin{equation}
{\bf j}_0=-\frac{e}{L_xL_y}\frac{1}{q}\ 
{\rm Tr}\left[{\bf v}_{\rm tot}^{(0)}Q(E_0^{(N)})\right]
\end{equation}
in the same way as in the above. 

Clearly the rest of the current, 
${\bf j}_{\rm ind}(t)=(j_{{\rm ind},x}(t),j_{{\rm ind},y}(t))$, is 
the induced current by the external electric field. 
Using Theorem~\ref{theorem:Phi(t)expand}, 
the Hall current (the $x$ component of ${\bf j}_{\rm ind}(t)$) 
induced by the electric field is given by 
\begin{eqnarray}
j_{{\rm ind},x}(t)&=&\frac{e}{L_xL_y}\frac{i}{\hbar}
\frac{1}{q}\sum_{\mu=1}^q\int_{-T}^t ds \left\langle U_0^{(N)}(t,0)
\Phi_{0,\mu}^{(N)},
v_{{\rm tot},x}^{(0)}\ U_0^{(N)}(t,s)H_{\rm per}^{(N)}(s)
U_0^{(N)}(s,0)\Phi_{0,\mu}^{(N)}\right\rangle\ret
&+&{\rm c.c.}+o(F)\ret
&=&\frac{e^2}{L_xL_y}\frac{i}{\hbar}\frac{1}{q}\sum_{\mu=1}^q\sum_{n\ge 1,\mu'}
\left\langle \Phi_{0,\mu}^{(N)},v_{{\rm tot},x}^{(0)}\Phi_{n,\mu'}^{(N)}
\right\rangle \left\langle\Phi_{n,\mu'}^{(N)},v_{{\rm tot},y}^{(0)}
\Phi_{0,\mu}^{(N)}
\right\rangle\ret
&\times&\exp\left[\frac{i}{\hbar}(E_{0,\mu}^{(N)}-E_n^{(N)})t\right]
\int_{-T}^t ds\ \exp\left[-\frac{i}{\hbar}(E_{0,\mu}^{(N)}-E_n^{(N)})s
\right]\alpha(s)
+{\rm c.c.}+o(F),\ret
\label{jindHall}
\end{eqnarray}
where $\Phi_{n,\mu'}^{(N)}$ is the normalized eigenvector of the Hamiltonian 
$H_0^{(N)}$ with the energy eigenvalue $E_n^{(N)}$ for $n\ge 1$, and 
${\rm c.c.}$ stands for the complex conjugate of the first part. 
Note that
\begin{eqnarray}
& &\frac{1}{F}\int_{-T}^t ds\ 
\exp\left[-\frac{i}{\hbar}(E_{0,\mu}^{(N)}-E_n^{(N)})s
\right]\alpha(s)\ret
&=&
\left[\frac{i\hbar T}{E_{0,\mu}^{(N)}-E_n^{(N)}+i\hbar\eta}
-\frac{\hbar^2}{\left(E_{0,\mu}^{(N)}-E_n^{(N)}+i\hbar\eta\right)^2}\right]
e^{-\eta T}\exp\left[i(E_{0,\mu}^{(N)}-E_n^{(N)})T/\hbar\right]\ret
&+&\hbar^2\left[
\frac{1}{\left(E_{0,\mu}^{(N)}-E_n^{(N)}\right)^2}
-\frac{1}{\left(E_{0,\mu}^{(N)}-E_n^{(N)}+i\hbar\eta\right)^2}\right]\ret
&-&\left[\frac{i\hbar t}{E_{0,\mu}^{(N)}-E_n^{(N)}}
+\frac{\hbar^2}{\left(E_{0,\mu}^{(N)}-E_n^{(N)}\right)^2}\right]
\exp\left[-{i}(E_{0,\mu}^{(N)}-E_n^{(N)})t/\hbar\right]
\label{tintegral}
\end{eqnarray}
for the time $t\ge 0$. In the following, we will consider only the time 
$t\ge 0$. Substituting (\ref{tintegral}) into (\ref{jindHall}), we obtain 
the linear response coefficient, 
\begin{equation}
\sigma_{{\rm tot},{xy}}(t;\eta,T):=
\lim_{F\rightarrow 0}\frac{j_{{\rm ind},x}(t)}{F}
=\sigma_{xy}+\gamma_{xy}\cdot t+\delta\sigma_{xy}(t;\eta,T),
\end{equation}
in the $x$ direction, where the first term which we call the Hall conductance 
is given by 
\begin{equation}
\sigma_{xy}
=-\frac{i\hbar e^2}{L_xL_y}\frac{1}{q}\sum_{\mu=1}^q\sum_{n\ge 1,\mu'}\left[
\frac{\left\langle \Phi_{0,\mu}^{(N)},v_{{\rm tot},x}^{(0)}
\Phi_{n,\mu'}^{(N)}\right\rangle 
\left\langle\Phi_{n,\mu'}^{(N)},v_{{\rm tot},y}^{(0)}\Phi_{0,\mu}^{(N)}
\right\rangle}
{\left(E_{0,\mu}^{(N)}-E_n^{(N)}\right)^2}-{\rm c.c.}\right], 
\label{sigmaxy}
\end{equation}
and the acceleration coefficient $\gamma_{xy}$ of the second term which is linear 
in the time $t$ is 
\begin{equation}
\gamma_{xy}=\frac{e^2}{L_xL_y}\frac{1}{q}\sum_{\mu=1}^q
\sum_{n\ge 1,\mu'}
\left[\left\langle \Phi_{0,\mu}^{(N)},v_{{\rm tot},x}^{(0)}
\Phi_{n,\mu'}^{(N)}\right\rangle 
\frac{1}{E_{0,\mu}^{(N)}-E_n^{(N)}}
\left\langle\Phi_{n,\mu'}^{(N)},v_{{\rm tot},y}^{(0)}\Phi_{0,\mu}^{(N)}
\right\rangle+{\rm c.c.}\right].
\label{gammaxy}
\end{equation}
The third term $\delta\sigma_{xy}(t;\eta,T)$ is given by 
\begin{equation}
\delta\sigma_{xy}(t;\eta,T)=\frac{ie^2}{L_xL_y}
\frac{1}{q}\sum_{\mu=1}^q\sum_{n\ge 1,\mu'}
\left\langle \Phi_{0,\mu}^{(N)},v_{{\rm tot},x}^{(0)}\Phi_{n,\mu'}^{(N)}
\right\rangle \left\langle\Phi_{n,\mu'}^{(N)},v_{{\rm tot},y}^{(0)}
\Phi_{0,\mu}^{(N)}
\right\rangle{\cal M}(t,{\cal E}_{0,\mu}^n;\eta,T)+{\rm c.c.}
\end{equation}
with ${\cal E}_{0,\mu}^n=E_{0,\mu}^{(N)}-E_n^{(N)}$ and 
\begin{equation}
{\cal M}(t,{\cal E};\eta,T)=\left\{\left[\frac{i T}{{\cal E}+i\hbar\eta}
-\frac{\hbar}{\left({\cal E}+i\hbar\eta\right)^2}\right]
e^{-\eta T}e^{i{\cal E}T/\hbar}
+\left[
\frac{\hbar}{{\cal E}^2}
-\frac{\hbar}{\left({\cal E}+i\hbar\eta\right)^2}\right]\right\}e^{i{\cal E}t/\hbar}.
\end{equation}
Similarly the induced current in the $y$ direction is 
\begin{eqnarray}
j_{{\rm ind},y}(t)&=&\frac{N}{L_xL_y}\frac{e^2}{m_e}Ft\ret
&+&\frac{e}{L_xL_y}\frac{i}{\hbar}\frac{1}{q}\sum_{\mu=1}^q
\int_{-T}^t ds \left\langle U_0^{(N)}(t,0)\Phi_{0,\mu}^{(N)},
v_{{\rm tot},y}^{(0)}\ U_0^{(N)}(t,s)H_{\rm per}^{(N)}(s)
U_0^{(N)}(s,0)\Phi_{0,\mu}^{(N)}\right\rangle\ret
&+&{\rm c.c.}+o(F),
\end{eqnarray}
and the linear response coefficient is given by 
\begin{equation}
\sigma_{{\rm tot},yy}(t;\eta,T):=
\lim_{F\rightarrow 0}\frac{j_{{\rm ind},y}(t)}{F}
=\gamma_{yy}\cdot t+\delta\sigma_{yy}(t;\eta,T),
\label{sigmayy}
\end{equation}
where 
\begin{equation}
\gamma_{yy}=\frac{e^2}{L_xL_y}\left[\frac{N}{m_e}+
\frac{2}{q}\sum_{\mu=1}^q\sum_{n\ge 1,\mu'}
\left\langle \Phi_{0,\mu}^{(N)},v_{{\rm tot},y}^{(0)}\Phi_{n,\mu'}^{(N)}
\right\rangle 
\frac{1}{E_{0,\mu}^{(N)}-E_n^{(N)}}
\left\langle\Phi_{n,\mu'}^{(N)},v_{{\rm tot},y}^{(0)}\Phi_{0,\mu}^{(N)}
\right\rangle\right]
\label{gammayy}
\end{equation}
and $\delta\sigma_{yy}(t;\eta,T)$ is given by replacing the velocity 
operator $v_{{\rm tot},x}^{(0)}$ with $v_{{\rm tot},y}^{(0)}$ in 
$\delta\sigma_{xy}(t;\eta,T)$. 

%%%%%%%%%%%%%%%%%%%%%%%%%%%%%%%%%%%%%%%%%%%%%%%%%%%%%%
\Section{The system with translation invariance}
\label{Sec:translationinv}

In this section, we check the validity of our linear response formulas 
in the special case without the vector potential 
${\bf A}_{\rm P}$ and without the electrostatic potential $W$. 

The Hamiltonian without the external electric field is given by 
\begin{equation}
H_0^{(N)}=\sum_{j=1}^N\frac{1}{2m_e}
\left\{(p_{x,j}-eBy_j+\phi_x)^2+(p_{y,j}+\phi_y)^2\right\}
+\sum_{1\le i<j\le N}W^{(2)}({\bf r}_i-{\bf r}_j).
\end{equation}
Clearly the system has translation invariance \cite{Wallet} in both $x$ and $y$ 
directions because the electron-electron interaction $W^{(2)}$ 
is a function of the relative coordinate only. We assume 
$W^{(2)}\in C^1({\bf R}^2)$. 
In this situation with $B\ne 0$, the well-known results are obtained as    
\begin{equation}
\sigma_{xy}=-\frac{e^2}{h}\nu,
\label{trinvsxy}
\end{equation}
and 
\begin{equation}
\gamma_{xy}=\gamma_{yy}=0,
\label{trinvtdsxy}
\end{equation}
where $\nu$ is the filling factor of the Landau level, i.e., 
$\nu=N/M$ with the number $N$ of the electrons and with the number $M$ of 
the single electron states in a single Landau level without 
the electron-electron interaction. Further we prove the bounds, 
\begin{equation}
\left|\delta\sigma_{sy}\right|\le      
\frac{e^2}{h}\nu\left(1+\omega_cT\right)e^{-\eta T}+
\frac{e^2}{h}\nu\left(2+\frac{\eta}{\omega_c}\right)\frac{\eta}{\omega_c}
\quad \mbox{for } s=x,y.
\label{deltasigmasybound} 
\end{equation}
Clearly the first term in the right-hand side vanishes in the large limit 
of $T$, and the second term also vanishes in the small limit of $\eta$. 
Further we also stress that the right-hand side is independent of the 
system sizes, $L_x,L_y$, and of the number $N$ of the electrons 
for a fixed filling factor $\nu$. 

In order to give proofs of (\ref{trinvsxy}) and (\ref{trinvtdsxy}), 
we note that 
\begin{equation}
v_{{\rm tot},x}^{(0)}=-\frac{im_e}{\hbar eB}
\left[v_{{\rm tot},y}^{(0)},H_0^{(N)}\right],
\label{vytotcommuH0}
\end{equation}
\begin{equation}
v_{{\rm tot},y}^{(0)}=\frac{im_e}{\hbar eB}
\left[v_{{\rm tot},x}^{(0)},H_0^{(N)}\right]
\label{vxtotcommuH0}
\end{equation}
and 
\begin{equation}
\frac{im_e}{\hbar eB}\left[v_{{\rm tot},x}^{(0)},v_{{\rm tot},y}^{(0)}\right]
=\frac{N}{m_e}.
\label{vvcommu}
\end{equation}
Let $\Phi$ be a vector in the domain of $H_0^{(N)}$, i.e., 
$\Vert H_0^{(N)}\Phi\Vert<\infty$. Then one has 
\begin{eqnarray}
\left\langle(E_{0,\mu}^{(N)}-H_0^{(N)})\Phi,v_{{\rm tot},x}^{(0)}
\Phi_{0,\mu}^{(N)}\right\rangle&=&
\left\langle v_{{\rm tot},x}^{(0)}\Phi,H_0^{(N)}
\Phi_{0,\mu}^{(N)}\right\rangle
-\left\langle H_0^{(N)}\Phi,v_{{\rm tot},x}^{(0)}\Phi_{0,\mu}^{(N)}\right\rangle\ret
&=&\lim_{n\rightarrow\infty}\left(
\left\langle v_{{\rm tot},x}^{(0)}\Phi, H_0^{(N)}\Psi_\mu^{(n)}\right\rangle
-\left\langle H_0^{(N)}\Phi,v_{{\rm tot},x}^{(0)}\Psi_\mu^{(n)}\right\rangle\right)\ret
&=&\lim_{n\rightarrow\infty}\left\langle\Phi,[v_{{\rm tot},x}^{(0)},H_0^{(N)}]
\Psi_\mu^{(n)}\right\rangle\ret
&=&\frac{\hbar eB}{im_e}
\left\langle\Phi,v_{{\rm tot},y}^{(0)}\Phi_{0,\mu}^{(N)}\right\rangle,
\label{matrixelementcommu} 
\end{eqnarray}
where $\Psi_\mu^{(n)}\in C^\infty({\cal T})$ is an approximate vector\footnote{One can 
easily find such a vector $\Psi_\mu^{(n)}$ by using the Fourier expansion in terms of 
the eigenvectors (\ref{varphiP}) of the single-electron Landau Hamiltonian 
in Section~\ref{singleLandau} below.} 
such that 
$\Vert(H_0^{(N)}+\lambda)(\Psi_\mu^{(n)}-\Phi_{0,\mu}^{(N)})\Vert\rightarrow 0$ 
as $n\rightarrow\infty$ with a positive constant $\lambda$ satisfying 
$H_0^{(N)}+\lambda>0$, and we have used the commutation 
relation (\ref{vxtotcommuH0}). 
Using this and the commutation relation (\ref{vvcommu}), one has 
\begin{eqnarray}
& &\frac{2}{q}\sum_{\mu=1}^q
\left\langle\Phi_{0,\mu}^{(N)},v_{{\rm tot},y}^{(0)}
\frac{1-Q(E_0^{(N)})}{E_{0,\mu}^{(N)}-H_0^{(N)}}v_{{\rm tot},y}^{(0)}
\Phi_{0,\mu}^{(N)}\right\rangle\ret
&=&\frac{1}{q}\sum_{\mu=1}^q
\frac{im_e}{\hbar eB}\left\langle\Phi_{0,\mu}^{(N)},v_{{\rm tot},y}^{(0)}
\left[1-Q(E_0^{(N)})\right]v_{{\rm tot},x}^{(0)}
\Phi_{0,\mu}^{(N)}\right\rangle+{\rm c.c.}\ret
&=&\frac{1}{q}\sum_{\mu=1}^q
\frac{im_e}{\hbar eB}\left\langle\Phi_{0,\mu}^{(N)},
\left[v_{{\rm tot},y}^{(0)},v_{{\rm tot},x}^{(0)}\right]
\Phi_{0,\mu}^{(N)}\right\rangle=-\frac{N}{m_e}.
\end{eqnarray}
Substituting this into the expression (\ref{gammayy}) of $\gamma_{yy}$, 
one gets $\gamma_{yy}=0$. 
In the same way, one can easily obtain $\gamma_{xy}=0$ by using 
the expression (\ref{gammaxy}) for $\gamma_{xy}$. 
For the rest of the Hall conductance $\sigma_{xy}$ of (\ref{trinvsxy}), one has 
\begin{eqnarray}
& &\frac{1}{q}\sum_{\mu=1}^q\left\langle\Phi_{0,\mu}^{(N)},
v_{{\rm tot},x}^{(0)}\frac{1-Q(E_0^{(N)})}{\left(E_{0,\mu}^{(N)}-H_0^{(N)}
\right)^2}v_{{\rm tot},y}^{(0)}\Phi_{0,\mu}^{(N)}\right\rangle-{\rm c.c.}\ret
&=&-\frac{1}{q}\sum_{\mu=1}^q\left(\frac{m_e}{\hbar eB}\right)^2
\left\langle\Phi_{0,\mu}^{(N)},v_{{\rm tot},y}^{(0)}
\left[1-Q(E_0^{(N)})\right]v_{{\rm tot},x}^{(0)}\Phi_{0,\mu}^{(N)}
\right\rangle+{\rm c.c.}\ret
&=&\frac{1}{q}\sum_{\mu=1}^q\left(\frac{m_e}{\hbar eB}\right)^2
\left\langle\Phi_{0,\mu}^{(N)},\left[v_{{\rm tot},x}^{(0)},
v_{{\rm tot},y}^{(0)}\right]\Phi_{0,\mu}^{(N)}\right\rangle
=-\frac{iN}{\hbar eB}.
\end{eqnarray}
Substituting this into the expression (\ref{sigmaxy}) of $\sigma_{xy}$, 
the desired result (\ref{trinvsxy}) is obtained. 

Next let us give a proof of the bounds (\ref{deltasigmasybound}). 
To this end, we recall the expression of $\delta\sigma_{sy}(t;\eta,T)$ as  
\begin{equation}
\delta\sigma_{sy}(t;\eta,T)=\frac{ie^2}{L_xL_y}
\frac{1}{q}\sum_{\mu=1}^q\sum_{n\ge 1,\mu'}
\left\langle \Phi_{0,\mu}^{(N)},v_{{\rm tot},s}^{(0)}\Phi_{n,\mu'}^{(N)}
\right\rangle \left\langle\Phi_{n,\mu'}^{(N)},v_{{\rm tot},y}^{(0)}
\Phi_{0,\mu}^{(N)}
\right\rangle{\cal M}(t,{\cal E}_{0,\mu}^n;\eta,T)+{\rm c.c.}
\label{exprdeltasigma}
\end{equation}
with ${\cal E}_{0,\mu}^n=E_{0,\mu}^{(N)}-E_n^{(N)}$ and 
\begin{equation}
{\cal M}(t,{\cal E};\eta,T)=\left\{\left[\frac{i T}{{\cal E}+i\hbar\eta}
-\frac{\hbar}{\left({\cal E}+i\hbar\eta\right)^2}\right]
e^{-\eta T}e^{i{\cal E}T/\hbar}
+\left[
\frac{\hbar}{{\cal E}^2}
-\frac{\hbar}{\left({\cal E}+i\hbar\eta\right)^2}\right]\right\}e^{i{\cal E}t/\hbar}.
\end{equation}
For a generic filling factor $\nu$, we cannot expect the existence of 
a system-size-independent 
energy gap above the sector of the ground state. Namely the gap might become 
small for a large volume of the system. Therefore we must treat carefully 
the denominators of the fractions appeared in the expression of 
${\cal M}(t,{\cal E};\eta,T)$.

To begin with, we note that 
\begin{equation}
\frac{1}{{\cal E}^2}-\frac{1}{\left({\cal E}+i\hbar\eta\right)^2}=
\frac{2i\hbar\eta}{{\cal E}({\cal E}+i\hbar\eta)^2}
-\frac{\hbar^2\eta^2}{{\cal E}^2({\cal E}+i\hbar\eta)^2}. 
\end{equation}
Using this identity, $L_xL_y=2\pi M\ell_B^2$ and $\nu=N/M$, we have 
\begin{equation}
\delta\sigma_{sy}(t;\eta,T)=i\frac{e^2}{h}\nu
\left[(A_1+A_2\omega_cT)e^{-\eta T}+
\left(2A_3+A_4{\eta}/{\omega_c}\right){\eta}/{\omega_c}\right]
+{\rm c.c.},
\label{deltasigmasy0}
\end{equation}
where 
\begin{equation} 
A_j=\frac{\hbar eB}{N}\frac{1}{q}\sum_{\mu=1}^q \sum_{n\ge 1;\mu'}
\left\langle\Phi_{0,\mu}^{(N)},v_{{\rm tot},s}^{(0)}\Phi_{n,\mu'}^{(N)}\right\rangle
{\cal M}_j({\cal E}_{0,\mu}^n)\left\langle\Phi_{n,\mu'}^{(N)},v_{{\rm tot},y}^{(0)}
\Phi_{0,\mu}^{(N)}\right\rangle
\end{equation}
with 
\begin{equation}
{\cal M}_1({\cal E})=-\frac{1}{({\cal E}+i\hbar\eta)^2}\exp[i{\cal E}(t+T)/\hbar],
\end{equation}
\begin{equation}
{\cal M}_2({\cal E})=\frac{i}{\hbar\omega_c}\frac{1}{{\cal E}+i\hbar\eta}
\exp[i{\cal E}(t+T)/\hbar],
\end{equation}
\begin{equation}
{\cal M}_3({\cal E})=\frac{i\hbar\omega_c}{{\cal E}({\cal E}+i\hbar\eta)^2}
e^{i{\cal E}t/\hbar}
\end{equation}
and 
\begin{equation}
{\cal M}_4({\cal E})=-\frac{\hbar^2\omega_c^2}{{\cal E}^2({\cal E}+i\hbar\eta)^2}
e^{i{\cal E}t/\hbar}.
\label{M4}
\end{equation}
We can prove $|A_j|\le 1/2$ for $j=1,2,3,4$. The desired 
bound (\ref{deltasigmasybound}) follows from these bounds for $A_j$. 
Since all of $A_j$ can be treated in the same way, we shall give a proof for 
$|A_1|\le 1/2$ only.
In the same way as the above, one has   
\begin{eqnarray}
A_1&=&-\frac{\hbar eB}{N}\frac{1}{q}\sum_{\mu=1}^q \left\langle
\Phi_{0\mu}^{(N)},v_{{\rm tot},s}^{(0)}e^{i{\hat \theta}}
\frac{1-Q(E_0^{(N)})}{(E_{0,\mu}^{(N)}-H_0^{(N)}+i\hbar\eta)^2}
v_{{\rm tot},y}^{(0)}\Phi_{0,\mu}^{(N)}\right\rangle\ret
&=&-\frac{im_e}{N}\frac{1}{q}\sum_{\mu=1}^q \left\langle
\Phi_{0\mu}^{(N)},v_{{\rm tot},s}^{(0)}e^{i{\hat \theta}}
\frac{1-Q(E_0^{(N)})}{(E_{0,\mu}^{(N)}-H_0^{(N)}+i\hbar\eta)^2}
(E_{0,\mu}^{(N)}-H_0^{(N)})
v_{{\rm tot},x}^{(0)}\Phi_{0,\mu}^{(N)}\right\rangle\ret
&=&-\frac{m_e^2}{\hbar eBN}\frac{1}{q}\sum_{\mu=1}^q \left\langle
\Phi_{0\mu}^{(N)},v_{{\rm tot},s}^{(0)}e^{i{\hat \theta}}
\frac{1-Q(E_0^{(N)})}{(E_{0,\mu}^{(N)}-H_0^{(N)}+i\hbar\eta)^2}
(E_{0,\mu}^{(N)}-H_0^{(N)})^2
v_{{\rm tot},y}^{(0)}\Phi_{0,\mu}^{(N)}\right\rangle,\ret
\label{A1equality}
\end{eqnarray}
where we have written ${\hat \theta}=(E_{0,\mu}^{(N)}-H_0^{(N)})(t+T)/\hbar$.
Let $\Phi$ be a vector in the domain of $v_{{\rm tot},x}^{(0)}$. Then one has 
\begin{eqnarray}
\left\langle\Phi,[1-Q(E_0^{(N)})]
v_{{\rm tot},y}^{(0)}\Phi_{0,\mu}^{(N)}\right\rangle
&=&-\frac{im_e}{\hbar eB}\lim_{n\rightarrow\infty}
\left\langle\Phi,[1-Q(E_0^{(N)})]
[H_0^{(N)},v_{{\rm tot},x}^{(0)}]\Psi_\mu^{(n)}\right\rangle\ret
&=&-\frac{im_e}{\hbar eB}\lim_{n\rightarrow\infty}\left\langle\Phi,
[1-Q(E_0^{(N)})](H_0^{(N)}-E_{0,\mu}^{(N)})v_{{\rm tot},x}^{(0)}
\Psi_\mu^{(n)}\right\rangle,\ret
\label{commuteidentity}
\end{eqnarray}
where $\Psi_\mu^{(n)}$ is the approximate vector which was introduced 
for proving (\ref{matrixelementcommu}),  
and we have used the commutation relation (\ref{vxtotcommuH0}).
Substituting this into the expression (\ref{A1equality}) of $A_1$ 
and then applying the Schwarz inequality, we obtain  
\begin{equation}
|A_1|\le \frac{m_e^3}{(\hbar eB)^2N}\max_s\lim_{n\rightarrow\infty}
\frac{1}{q}\sum_{\mu=1}^q
\left\langle v_{{\rm tot},s}^{(0)}\Psi_\mu^{(n)},
[1-Q(E_0^{(N)})](H_0^{(N)}-E_{0,\mu}^{(N)})
v_{{\rm tot},s}^{(0)}\Psi_\mu^{(n)}\right\rangle.
\label{A1bound}
\end{equation}
Note that, for a symmetric operator $A$, one has formally  
\begin{eqnarray}
& &\sum_{\mu=1}^q\left\langle A\Phi_{0,\mu}^{(N)},\left[1-Q(E_0^{(N)})\right]
(H_0^{(N)}-E_{0,\mu}^{(N)})A\Phi_{0,\mu}^{(N)}\right\rangle\ret 
&=&\sum_{\mu=1}^q\left\langle A\Phi_{0,\mu}^{(N)},
(H_0^{(N)}-E_{0,\mu}^{(N)})A\Phi_{0,\mu}^{(N)}\right\rangle\ret
&=&\frac{1}{2}\sum_{\mu=1}^q
\left\{\left\langle A\Phi_{0,\mu}^{(N)},
(H_0^{(N)}-E_{0,\mu}^{(N)})A\Phi_{0,\mu}^{(N)}\right\rangle
+\left\langle (H_0^{(N)}-E_{0,\mu}^{(N)})A\Phi_{0,\mu}^{(N)},
A\Phi_{0,\mu}^{(N)}\right\rangle\right\}\ret
&=&\frac{1}{2}\sum_{\mu=1}^q
\left\{\left\langle A\Phi_{0,\mu}^{(N)},
\left[H_0^{(N)},A\right]\Phi_{0,\mu}^{(N)}\right\rangle
+\left\langle \left[H_0^{(N)},A\right]\Phi_{0,\mu}^{(N)},
A\Phi_{0,\mu}^{(N)}\right\rangle\right\}\ret
&=&\frac{1}{2}\sum_{\mu=1}^q
\left\langle \Phi_{0,\mu}^{(N)},\left[A,
\left[H_0^{(N)},A\right]\right]\Phi_{0,\mu}^{(N)}\right\rangle,
\label{doublecommuEq}
\end{eqnarray}
where we have used 
\begin{equation}
\sum_{\mu=1}^q\sum_{\mu'=1}^q 
\left\langle A\Phi_{0,\mu}^{(N)},\Phi_{0,\mu'}^{(N)}\right\rangle 
(E_{0,\mu'}^{(N)}-E_{0,\mu}^{(N)})
\left\langle\Phi_{0,\mu'}^{(N)},A\Phi_{0,\mu}^{(N)}\right\rangle=0. 
\end{equation}
We can justify this formal identity (\ref{doublecommuEq}) for 
$A=v_{{\rm tot},s}^{(0)}$ by using the approximate vector $\Psi_\mu^{(n)}$. 
Combining this observation, the commutation relations, (\ref{vytotcommuH0}), 
(\ref{vxtotcommuH0}), (\ref{vvcommu}), and the bound (\ref{A1bound}), 
we obtain $|A_1|\le 1/2$. 

In the case with $B=0$, one has
\begin{equation}
\sigma_{xy}=\delta\sigma_{xy}(t;\eta,T)=\delta\sigma_{yy}(t;\eta,T)=0
\label{vanishB0} 
\end{equation} 
and 
\begin{equation}
\gamma_{xy}=0, \quad \gamma_{yy}=\frac{N}{L_xL_y}\frac{e^2}{m_e}
\end{equation}
under the same assumptions as in the above. Clearly these imply that 
the total velocity of the electrons is proportional to the time $t$. 
The derivation is not hard as follows: 
{From} the assumptions, we have 
\begin{equation}
\left[v_{{\rm tot},s}^{(0)},H_0^{(N)}\right]=0\quad\mbox{for}\ s=x,y,
\quad\mbox{and}\quad 
\left[v_{{\rm tot},x}^{(0)},v_{{\rm tot},y}^{(0)}\right]=0. 
\end{equation}
Namely all of these operators commute with each other. 
This implies that all of the matrix elements in the expressions 
(\ref{sigmaxy}), (\ref{gammaxy}), (\ref{gammayy}) and (\ref{exprdeltasigma}) 
of the coefficients vanishes. As a result, we get the desired results. 

%%%%%%%%%%%%%%%%%%%%%%%%%%%%%%%%%%%%%%%%%%%%%%%%%%%%%%
\Section{The linear response coefficients averaged over the gauge parameters}
\label{Sec:QAHC}

In this section, we treat the averaged Hall 
conductance $\overline{\sigma_{xy}(\mbox{\boldmath $\phi$})}$ 
of (\ref{avsigmaxy}) and the averaged acceleration coefficients 
$\overline{\gamma_{sy}(\mbox{\boldmath $\phi$})}$ of (\ref{avgammasy}). 
As is well known, the ``topological" argument 
\cite{TKNN,Kohmoto,BVS,Kunz,ASS,NTW,AvronSeiler,TH,AY} yields  
the integral and fractional quantization of the Hall conductance 
under the assumption on the excitation energy gap above the ground state. 
Following the argument, the fractional quantization (\ref{wellknownTIresult}) 
of the averaged Hall conductance $\overline{\sigma_{xy}(\mbox{\boldmath $\phi$})}$ 
will be proved in the most general setting of the present paper. 
In addition, we will prove that the averaged acceleration coefficients are exactly 
vanishing, i.e., $\overline{\gamma_{sy}(\mbox{\boldmath $\phi$})}=0$ for $s=x,y$.  
Thus we will give the proof of Theorem~\ref{theorem:avconductance} in 
Section~\ref{ProofQHC} below. In Section~\ref{AStheorem}, we will also discuss 
the geometric property of the averaged Hall conductance 
$\overline{\sigma_{xy}(\mbox{\boldmath $\phi$})}$ as a geometric invariant for 
non-Abelian gauge fields on the gauge torus ${\cal T}_{\rm g}$.
In particular, the integer ${\cal I}$ of the fractional quantized 
Hall conductance (\ref{avsigmaFQG}) is equal to an index of 
a Pauli-Dirac operator coupled to the gauge fields. In other words, 
the Hall conductance of the interacting electron gas is closely related to 
the ground state property of a single electron system coupled to 
the gauge fields.  

The Hamiltonian of the quantum Hall system without the external electric 
field is given by 
\begin{eqnarray}
H_0^{(N)}(\mbox{\boldmath $\phi$})&=&\sum_{j=1}^N\left\{\frac{1}{2m_e}
[{\bf p}_j+e{\bf A}({\bf r}_j)+\mbox{\boldmath $\phi$}]^2
+W({\bf r}_j)\right\}+\sum_{1\le i<j\le N}W^{(2)}({\bf r}_i-{\bf r}_j).
\label{Ham0phi}
\end{eqnarray}
Since the gauge parameters $\mbox{\boldmath $\phi$}=(\phi_x,\phi_y)$ play 
an important role 
in the following proof, we write the parameter dependence explicitly as 
$\Phi_{0,\mu}^{(N)}=\Phi_{0,\mu}^{(N)}(\mbox{\boldmath $\phi$})$ and $E_{0,\mu}^{(N)}
=E_{0,\mu}^{(N)}(\mbox{\boldmath $\phi$})$ for the (quasi)degenerate ground state, 
and $\Phi_{n,\mu}^{(N)}=\Phi_{n,\mu}^{(N)}(\mbox{\boldmath $\phi$})$ and 
$E_n^{(N)}=E_n^{(N)}(\mbox{\boldmath $\phi$})$ 
for the excited states throughout this section. 
We also write the velocity operator as 
\begin{equation}
{\bf v}_{\rm tot}^{(0)}(\mbox{\boldmath $\phi$})=\sum_{j=1}^N\frac{1}{m_e}
\left[{\bf p}_j+e{\bf A}({\bf r}_j)+\mbox{\boldmath $\phi$}\right].
\end{equation}
Further, throughout this section, we require Assumption~\ref{gapassumption} 
which we need for the proof of Theorem~\ref{theorem:avconductance}. 

%%%%%%%%%%%%%%%%%%%%%%%%%%%%%%%%%%%%%%%%%%%%%%%%%%%%%%%%%%%%%%%%%%%%%
\subsection{Proof of Theorem~\ref{theorem:avconductance}}
\label{ProofQHC}

To begin with, we prepare some tools to rewrite the expressions 
of $\sigma_{xy}(\mbox{\boldmath $\phi$})$ 
and $\gamma_{sy}(\mbox{\boldmath $\phi$})$. 
The derivative of the projection $Q(E_0^{(N)}(\mbox{\boldmath $\phi$}))$ onto 
the sector of the ground state becomes 
\begin{eqnarray}
Q_i(E_0^{(N)}(\mbox{\boldmath $\phi$})):=
\frac{\partial}{\partial\phi_i}
Q(E_0^{(N)}(\mbox{\boldmath $\phi$}))
&=&\frac{\partial}{\partial\phi_i}
\frac{1}{2\pi i}\int_\Gamma dz\frac{1}{z-H_0^{(N)}(\mbox{\boldmath $\phi$})}\ret
&=&
\frac{1}{2\pi i}\int_\Gamma dz\frac{1}{z-H_0^{(N)}(\mbox{\boldmath $\phi$})}
v_{{\rm tot},i}^{(0)}(\mbox{\boldmath $\phi$})
\frac{1}{z-H_0^{(N)}(\mbox{\boldmath $\phi$})}, 
\label{delQE}
\end{eqnarray}
where we have used the integral representation of the projection 
\begin{equation}
Q(E_0^{(N)}(\mbox{\boldmath $\phi$}))=
\frac{1}{2\pi i}\int_\Gamma dz \frac{1}{z-H_0^{(N)}(\mbox{\boldmath $\phi$})}
\label{intrepQphi}
\end{equation}
with the resolvent $[z-H_0^{(N)}(\mbox{\boldmath $\phi$})]^{-1}$. Here the closed path 
$\Gamma$ encircles all of the ground state energy eigenvalues 
$E_{0,\mu}^{(N)}(\mbox{\boldmath $\phi$})$ which are isolated 
from the rest of the spectrum. We can take the path $\Gamma$ to be 
independent of $\mbox{\boldmath $\phi$}$ because of Assumption~\ref{gapassumption}. 
Using the operator $Q_i(E_0^{(N)}(\mbox{\boldmath $\phi$}))$, the Hall conductance 
(\ref{sigmaxy}) can be rewritten as 
\begin{equation}
\sigma_{xy}(\mbox{\boldmath $\phi$})=-\frac{i\hbar e^2}{L_xL_y}\frac{1}{q}\ 
{\rm Tr}\ Q(E_0^{(N)}(\mbox{\boldmath $\phi$}))
[Q_x(E_0^{(N)}(\mbox{\boldmath $\phi$})),Q_y(E_0^{(N)}(\mbox{\boldmath $\phi$}))],
\label{sigmaxyphi}
\end{equation}
and for $\gamma_{xy}$ of (\ref{gammaxy}) and $\gamma_{yy}$ of (\ref{gammayy}), 
we have 
\begin{equation}
\gamma_{xy}(\mbox{\boldmath $\phi$})=
\frac{e^2}{L_xL_y}
\frac{1}{q}\ {\rm Tr}\ v_{{\rm tot},x}^{(0)}(\mbox{\boldmath $\phi$})
\left[Q_y(E_0^{(N)}(\mbox{\boldmath $\phi$}))
Q(E_0^{(N)}(\mbox{\boldmath $\phi$}))+Q(E_0^{(N)}(\mbox{\boldmath $\phi$}))
Q_y(E_0^{(N)}(\mbox{\boldmath $\phi$}))\right] 
\end{equation}
and 
$$
\gamma_{yy}(\mbox{\boldmath $\phi$})\hspace{14cm}
$$
\begin{equation}
=\frac{e^2}{L_xL_y}
\left\{\frac{N}{m_e}+\frac{1}{q}\ {\rm Tr}\ v_{{\rm tot},y}^{(0)}(\mbox{\boldmath $\phi$})
\left[Q_y(E_0^{(N)}(\mbox{\boldmath $\phi$}))
Q(E_0^{(N)}(\mbox{\boldmath $\phi$}))+Q(E_0^{(N)}(\mbox{\boldmath $\phi$}))
Q_y(E_0^{(N)}(\mbox{\boldmath $\phi$}))\right]
\right\},
\end{equation}
where Tr stands for the trace on the Hilbert space. 
The expression of the Hall conductance with the use of the derivative 
of a projection was used in refs.~\cite{ASS,AvronSeiler,AY}. 
Further $\gamma_{sy}(\mbox{\boldmath $\phi$})$ can be rewritten as 
\begin{equation}
\gamma_{sy}(\mbox{\boldmath $\phi$})=\frac{e^2}{L_xL_y}
\frac{1}{q}\frac{\partial}{\partial \phi_y}
{\rm Tr}\ v_{{\rm tot},s}^{(0)}(\mbox{\boldmath $\phi$})
Q(E_0^{(N)}(\mbox{\boldmath $\phi$}))\quad\mbox{for}\ s=x,y,
\label{tdsigmasyphi}
\end{equation}
where we have used the identity,
\begin{equation}
Q_s(E_0^{(N)}(\mbox{\boldmath $\phi$}))=
Q_s(E_0^{(N)}(\mbox{\boldmath $\phi$}))Q(E_0^{(N)}(\mbox{\boldmath $\phi$}))
+Q(E_0^{(N)}(\mbox{\boldmath $\phi$}))
Q_s(E_0^{(N)}(\mbox{\boldmath $\phi$})),\quad \mbox{for}\ s=x,y, 
\label{idQdiff}
\end{equation}
which is derived by differentiation of $Q(E_0^{(N)}(\mbox{\boldmath $\phi$}))=
Q(E_0^{(N)}(\mbox{\boldmath $\phi$}))^2$. We also note that 
\begin{equation}
Q(E_0^{(N)}(\mbox{\boldmath $\phi$}))Q_s(E_0^{(N)}(\mbox{\boldmath $\phi$}))
Q(E_0^{(N)}(\mbox{\boldmath $\phi$}))=0\quad \mbox{for}\ s=x,y, 
\label{QQsQ}
\end{equation}
which is easily derived from (\ref{idQdiff}).

\begin{pro}
\label{prodiffPhi}
Under Assumption~\ref{gapassumption}, there exist orthonormal vectors 
${\hat \Phi}_{0,\mu}^{(N)}(\mbox{\boldmath $\phi$})$, $\mu=1,2,\ldots,q$ such that 
the sector of the (quasi)degenerate ground state is spanned by 
the vectors ${\hat \Phi}_{0,\mu}^{(N)}(\mbox{\boldmath $\phi$})$, $\mu=1,2,\ldots,q$, 
and that all the vectors ${\hat \Phi}_{0,\mu}^{(N)}(\mbox{\boldmath $\phi$})$, 
$\mu=1,2,\ldots,q$, are infinitely differentiable with respect to 
the gauge parameters $\mbox{\boldmath $\phi$}$ on the gauge torus ${\cal T}_{\rm g}$. 
\end{pro}
This proposition is essentially due to T.~Kato. But, in his 
book \cite{Kato}, he treated only the case with a single variable. 
For reader's convenience, we give the proof of Proposition~\ref{prodiffPhi} 
in Appendix~\ref{appendix:prodiffPhi} along his line 
although the extension to two variables is not so difficult. 
In terms of these vectors ${\hat \Phi}_{0,\mu}^{(N)}(\mbox{\boldmath $\phi$})$, 
the acceleration coefficients $\gamma_{sy}(\mbox{\boldmath $\phi$})$ of 
(\ref{tdsigmasyphi}) are written as 
\begin{equation}
\gamma_{sy}(\mbox{\boldmath $\phi$})=\frac{e^2}{L_xL_y}\frac{1}{q}\sum_{\mu=1}^q 
\frac{\partial}{\partial\phi_y}
\left\langle{\hat \Phi}_{0,\mu}^{(N)}(\mbox{\boldmath $\phi$}),
v_{{\rm tot},s}^{(0)}(\mbox{\boldmath $\phi$})
{\hat \Phi}_{0,\mu}^{(N)}(\mbox{\boldmath $\phi$})\right\rangle\quad \mbox{for}\ s=x,y.
\label{anosigmayytphi}
\end{equation}

Next let us rewrite the Hall conductance $\sigma_{xy}(\mbox{\boldmath $\phi$})$ of 
(\ref{sigmaxyphi}) in terms of the vectors 
${\hat \Phi}_{0,\mu}^{(N)}(\mbox{\boldmath $\phi$})$. Note that 
\begin{eqnarray}
\frac{\partial}{\partial\phi_i}
{\hat \Phi}_{0,\mu}^{(N)}(\mbox{\boldmath $\phi$})&=&
\frac{\partial}{\partial\phi_i}
Q(E_0^{(N)}(\mbox{\boldmath $\phi$})){\hat \Phi}_{0,\mu}^{(N)}(\mbox{\boldmath $\phi$})\ret
&=&
Q_i(E_0^{(N)}(\mbox{\boldmath $\phi$})){\hat \Phi}_{0,\mu}^{(N)}(\mbox{\boldmath $\phi$})
+Q(E_0^{(N)}(\mbox{\boldmath $\phi$}))\frac{\partial}{\partial\phi_i}
{\hat \Phi}_{0,\mu}^{(N)}(\mbox{\boldmath $\phi$}).
\label{derivePhi0}
\end{eqnarray}
Therefore one has 
\begin{equation}
\left[1-Q(E_0^{(N)}(\mbox{\boldmath $\phi$}))\right]\frac{\partial}{\partial\phi_i}
{\hat \Phi}_{0,\mu}^{(N)}(\mbox{\boldmath $\phi$})=
Q_i(E_0^{(N)}(\mbox{\boldmath $\phi$})){\hat \Phi}_{0,\mu}^{(N)}(\mbox{\boldmath $\phi$}). 
\label{Phidiff}
\end{equation}
Using this identity (\ref{Phidiff}), the Hall 
conductance $\sigma_{xy}(\mbox{\boldmath $\phi$})$ 
of (\ref{sigmaxyphi}) can be written as 
\begin{eqnarray}
\sigma_{xy}(\mbox{\boldmath $\phi$})
&=&-\frac{i\hbar e^2}{L_xL_y}\frac{1}{q}\sum_{\mu=1}^q\left[
\left\langle\frac{\partial}{\partial\phi_x}{\hat\Phi}_{0,\mu}^{(N)}(\mbox{\boldmath $\phi$}),
\left[1-Q(E_0^{(N)}(\mbox{\boldmath $\phi$}))\right]
\frac{\partial}{\partial\phi_y}{\hat\Phi}_{0,\mu}^{(N)}(\mbox{\boldmath $\phi$})
\right\rangle-(x\leftrightarrow y)\right]\ret
&=&-\frac{i\hbar e^2}{L_xL_y}\frac{1}{q}\sum_{\mu=1}^q\left[
\left\langle\frac{\partial}{\partial\phi_x}{\hat\Phi}_{0,\mu}^{(N)}(\mbox{\boldmath $\phi$}),
\frac{\partial}{\partial\phi_y}{\hat\Phi}_{0,\mu}^{(N)}(\mbox{\boldmath $\phi$})
\right\rangle-(x\leftrightarrow y)\right]\ret
&=&-\frac{i\hbar e^2}{L_xL_y}\frac{1}{q}\sum_{\mu=1}^q
\left[\frac{\partial}{\partial\phi_x}
\left\langle{\hat\Phi}_{0,\mu}^{(N)}(\mbox{\boldmath $\phi$}),
\frac{\partial}{\partial\phi_y}{\hat\Phi}_{0,\mu}^{(N)}(\mbox{\boldmath $\phi$})
\right\rangle-(x\leftrightarrow y)\right],
\label{sigmaxyphi2}
\end{eqnarray}
where we have used the identity 
\begin{eqnarray}
& &\left\langle\frac{\partial}{\partial\phi_y}
{\hat\Phi}_{0,\mu}^{(N)}(\mbox{\boldmath $\phi$}),
{\hat \Phi}_{0,\mu'}^{(N)}(\mbox{\boldmath $\phi$})\right\rangle
+\left\langle{\hat\Phi}_{0,\mu}^{(N)}(\mbox{\boldmath $\phi$}),
\frac{\partial}{\partial\phi_y}
{\hat\Phi}_{0,\mu'}^{(N)}(\mbox{\boldmath $\phi$})\right\rangle\ret
&=&\frac{\partial}{\partial\phi_y}
\left\langle{\hat\Phi}_{0,\mu}^{(N)}(\mbox{\boldmath $\phi$}),
{\hat\Phi}_{0,\mu'}^{(N)}(\mbox{\boldmath $\phi$})\right\rangle
=\frac{\partial}{\partial\phi_y}\delta_{\mu,\mu'}=0
\end{eqnarray}
for getting the second equality. The Hall conductance expressed in terms 
of the derivative of wavefunctions was first introduced in ref.~\cite{TKNN}. 

Following Kunz \cite{Kunz}, we shall introduce a gauge transformation as 
\begin{equation}
{\tilde \Phi}_{0,\mu}^{(N)}(\mbox{\boldmath $\phi$})
=G^{(N)}(\mbox{\boldmath $\phi$}){\hat\Phi}_{0,\mu}^{(N)}(\mbox{\boldmath $\phi$})
\label{tildePhi}
\end{equation}
with 
\begin{equation}
G^{(N)}(\mbox{\boldmath $\phi$})=\prod_{j=1}^N 
\exp\left[\frac{i}{\hbar}(x_j\phi_x+y_j\phi_y)\right].
\label{Gaugetr}
\end{equation}
Then the  Hamiltonian is transformed as 
\begin{eqnarray}
{\tilde H}_0^{(N)}(\mbox{\boldmath $\phi$})&:=&G^{(N)}(\mbox{\boldmath $\phi$})
H_0^{(N)}(\mbox{\boldmath $\phi$})
\left[G^{(N)}(\mbox{\boldmath $\phi$})\right]^{-1}\ret
&=&\sum_{j=1}^N\left\{\frac{1}{m_e}\left[{\bf p}_j+e{\bf A}({\bf r}_j)\right]^2
+W({\bf r}_j)\right\}+\sum_{1\le i<j\le N}W^{(2)}({\bf r}_i-{\bf r}_j). 
\end{eqnarray}
The expression of the right-hand side does not include the gauge parameters 
$\mbox{\boldmath $\phi$}$ explicitly but the Hamiltonian indeed depends on 
$\mbox{\boldmath $\phi$}$ through 
the boundary conditions. Namely the boundary conditions are twisted 
with the angles $\mbox{\boldmath $\phi$}$. Here we stress that the boundary condition 
in the $s$ direction becomes periodic for the special values 
$\phi_s=m\Delta\phi_s$ of the gauge parameter with an integer $m$ for $s=x,y$, 
where $\Delta\phi_s={2\pi\hbar}/{L_s}$ are given in (\ref{defTg}). 
Clearly the sector of the ground state has this periodicity 
from Assumption~\ref{gapassumption} on the ground state. 
Therefore the ground state vectors ${\tilde \Phi}_{0,\mu}^{(N)}(\mbox{\boldmath $\phi$})$ 
must satisfy the relations, 
\begin{equation}
{\tilde \Phi}_{0,\mu}^{(N)}(\Delta\phi_x,\phi_y)
=\sum_{\mu'=1}^q C^{(x)}_{\mu,\mu'}(\phi_y)
{\tilde \Phi}_{0,\mu'}^{(N)}(0,\phi_y)
\label{relationtildePhi1}
\end{equation}
and
\begin{equation}
{\tilde \Phi}_{0,\mu}^{(N)}(\phi_x,\Delta\phi_y)
=\sum_{\mu'=1}^q C^{(y)}_{\mu,\mu'}(\phi_x)
{\tilde \Phi}_{0,\mu'}^{(N)}(\phi_x,0),
\label{relationtildePhi2}
\end{equation}
where $C^{(x)}(\phi_y)$ and $C^{(y)}(\phi_x)$ are 
a $q\times q$ unitary matrix as a function of 
$\phi_y$ and $\phi_x$, respectively. 

In terms of the vectors ${\tilde \Phi}_{0,\mu}^{(N)}(\mbox{\boldmath $\phi$})$, 
the acceleration coefficients $\gamma_{sy}(\mbox{\boldmath $\phi$})$ 
of (\ref{anosigmayytphi}) can be written as 
\begin{equation}
\gamma_{sy}(\mbox{\boldmath $\phi$})=\frac{e^2}{L_xL_y}\frac{1}{q}\sum_{\mu=1}^q
\frac{\partial}{\partial\phi_y}
\left\langle{\tilde \Phi}_{0,\mu}^{(N)}(\mbox{\boldmath $\phi$}),v_{{\rm tot},s}^{(0)}(0)
{\tilde \Phi}_{0,\mu}^{(N)}(\mbox{\boldmath $\phi$})\right\rangle\qquad\mbox{for}\ s=x,y,
\end{equation}
where $v_{{\rm tot},s}^{(0)}(0)=\sum_{j=1}^N[p_{s,j}+eA_s({\bf r}_j)]/m_e$. 
Combining this with the relation (\ref{relationtildePhi2}), 
the averaged acceleration coefficients 
$\overline{\gamma_{sy}(\mbox{\boldmath $\phi$})}$ of (\ref{avgammasy}) vanish as 
\begin{eqnarray}
{\overline{\gamma_{sy}(\mbox{\boldmath $\phi$})}}
&=&\frac{1}{\Delta\phi_x\Delta\phi_y}\frac{e^2}{L_xL_y}
\frac{1}{q}\sum_{\mu=1}^q\int_0^{\Delta\phi_x}d\phi_x
\left[\left\langle{\tilde \Phi}_{0,\mu}^{(N)}(\phi_x,\Delta\phi_y),
v_{{\rm tot},s}^{(0)}(0)
{\tilde \Phi}_{0,\mu}^{(N)}(\phi_x,\Delta\phi_y)\right\rangle\right.\ret
& &\qquad\qquad\qquad\qquad\qquad
-\left.\left\langle{\tilde \Phi}_{0,\mu}^{(N)}(\phi_x,0),
v_{{\rm tot},s}^{(0)}(0)
{\tilde \Phi}_{0,\mu}^{(N)}(\phi_x,0)\right\rangle\right]=0
\end{eqnarray}
for $s=x,y$. Here we have used the unitarity of $C^{(y)}(\phi_x)$. 

Similarly the Hall conductance $\sigma_{xy}(\mbox{\boldmath $\phi$})$ 
of (\ref{sigmaxyphi2}) becomes 
\begin{eqnarray}
\sigma_{xy}(\mbox{\boldmath $\phi$})
&=&-\frac{i\hbar e^2}{L_xL_y}\frac{1}{q}\sum_{\mu=1}^q
\left[\frac{\partial}{\partial\phi_x}
\left\langle{\tilde \Phi}_{0,\mu}^{(N)}(\mbox{\boldmath $\phi$}),
\frac{\partial}{\partial\phi_y}{\tilde \Phi}_{0,\mu}^{(N)}(\mbox{\boldmath $\phi$})
\right\rangle
-\frac{\partial}{\partial\phi_y}
\left\langle{\tilde \Phi}_{0,\mu}^{(N)}(\mbox{\boldmath $\phi$}),
\frac{\partial}{\partial\phi_x}{\tilde \Phi}_{0,\mu}^{(N)}(\mbox{\boldmath $\phi$})
\right\rangle\right.\ret
&-&\left.\frac{i}{\hbar}\frac{\partial}{\partial\phi_x}
\left\langle{\tilde \Phi}_{0,\mu}^{(N)}(\mbox{\boldmath $\phi$}),
\sum_{i=1}^Ny_i{\tilde \Phi}_{0,\mu}^{(N)}(\mbox{\boldmath $\phi$})\right\rangle
+\frac{i}{\hbar}\frac{\partial}{\partial\phi_y}
\left\langle{\tilde \Phi}_{0,\mu}^{(N)}(\mbox{\boldmath $\phi$}),
\sum_{j=1}^Nx_j{\tilde \Phi}_{0,\mu}^{(N)}(\mbox{\boldmath $\phi$})\right\rangle\right]. 
\end{eqnarray}
By averaging over the gauge parameters $\mbox{\boldmath $\phi$}$, 
${\overline {\sigma_{xy}(\mbox{\boldmath $\phi$})}}$ of (\ref{avsigmaxy}) 
is written as  
\begin{equation}
{\overline {\sigma_{xy}(\mbox{\boldmath $\phi$})}} 
=\frac{e^2}{h}\frac{1}{2\pi i}
\frac{1}{q}\sum_{\mu=1}^q
\int_{{\cal T}_{\rm g}}d\phi_xd\phi_y
\left[\frac{\partial}{\partial\phi_x}
\left\langle{\tilde \Phi}_{0,\mu}^{(N)}(\mbox{\boldmath $\phi$}),
\frac{\partial}{\partial\phi_y}{\tilde \Phi}_{0,\mu}^{(N)}(\mbox{\boldmath $\phi$})
\right\rangle-(x\leftrightarrow y)\right],
\label{geoinvsigmaxy}
\end{equation}
where we have used the relations (\ref{relationtildePhi1}) and 
(\ref{relationtildePhi2}). Thus one get 
the geometrically invariant form of the Hall conductance. 
To see this more explicitly, we write 
\begin{equation}
{\cal A}_{\mu,\mu',s}(\mbox{\boldmath $\phi$})=
\left\langle{\tilde \Phi}_{0,\mu}^{(N)}(\mbox{\boldmath $\phi$}),
\frac{\partial}{\partial\phi_s}{\tilde \Phi}_{0,\mu'}^{(N)}(\mbox{\boldmath $\phi$})
\right\rangle\quad\mbox{for}\ s=x,y,
\end{equation}
and
\begin{equation}
{\cal F}(\mbox{\boldmath $\phi$})=
\frac{\partial}{\partial \phi_x}{\cal A}_y(\mbox{\boldmath $\phi$})
-\frac{\partial}{\partial \phi_y}{\cal A}_x(\mbox{\boldmath $\phi$})
+[{\cal A}_x(\mbox{\boldmath $\phi$}),{\cal A}_y(\mbox{\boldmath $\phi$})], 
\end{equation}
which are, respectively, the connections and the curvature 
in the language of differential geometry \cite{Atiyah}. 
In the language of the corresponding non-Abelian gauge theory 
\cite{Weinberg}, these corresponds to the gauge fields and the 
field strength tensor. The connections ${\cal A}_s(\mbox{\boldmath $\phi$})$ are 
the $q\times q$ matrix with the matrix elements 
${\cal A}_{\mu,\mu',s}(\mbox{\boldmath $\phi$})$. 
Then the geometric invariant on the gauge torus\footnote{It is necessary to 
impose the periodic boundary conditions on the rectangular region 
${\cal T}_{\rm g}$ for identifying it as a torus.}
${\cal T}_{\rm g}$ is given by 
\begin{equation}
{\cal I}=\frac{1}{2\pi i}\int_{{\cal T}_{\rm g}}d\phi_xd\phi_y\ 
{\rm tr}\ {\cal F}(\mbox{\boldmath $\phi$}),
\label{1stCnumber} 
\end{equation}
where ${\rm tr}$ is the trace of $q\times q$ matrix. 
Since ${\rm tr}[{\cal A}_x(\mbox{\boldmath $\phi$}),{\cal A}_y(\mbox{\boldmath $\phi$})]=0$, 
one has 
\begin{equation}
\overline{\sigma_{xy}(\mbox{\boldmath $\phi$})}=\frac{e^2}{h}\frac{{\cal I}}{q}.
\end{equation}
In passing, we remark that the connection ${\cal A}_s(\mbox{\boldmath $\phi$})$ is 
not necessarily periodic at the boundaries of the rectangular region 
${\cal T}_{\rm g}$. 
If the connection ${\cal A}_s(\mbox{\boldmath $\phi$})$ satisfies the periodic boundary 
conditions, then the quantity ${\cal I}$ vanishes, i.e., 
the averaged Hall conductance $\overline{\sigma_{xy}(\mbox{\boldmath $\phi$})}$ becomes 
zero. 

Following the way of computing a geometric invariant 
in refs.~\cite{Kohmoto,Kunz}, 
we shall show that the geometric invariant ${\cal I}$ takes an integer value, 
i.e., the averaged Hall conductance ${\overline {\sigma_{xy}(\mbox{\boldmath $\phi$})}}$ 
is quantized to a rational number $p/q$. For this purpose, 
we rewrite the averaged Hall conductance 
${\overline {\sigma_{xy}(\mbox{\boldmath $\phi$})}}$ 
of (\ref{geoinvsigmaxy}) as 
\begin{eqnarray}
{\overline {\sigma_{xy}(\mbox{\boldmath $\phi$})}}
&=&\frac{e^2}{h}\frac{1}{2\pi i}\frac{1}{q}\sum_{\mu=1}^q
\int_0^{\Delta\phi_y}d\phi_y\left[\left\langle
{\tilde \Phi}_{0,\mu}^{(N)}(\Delta\phi_x,\phi_y),
\frac{\partial}{\partial\phi_y}
{\tilde \Phi}_{0,\mu}^{(N)}(\Delta\phi_x,\phi_y)\right\rangle\right.
\ret& &\quad\qquad\qquad\qquad\qquad 
-\left.
\left\langle{\tilde \Phi}_{0,\mu}^{(N)}(0,\phi_y),
\frac{\partial}{\partial\phi_y}
{\tilde \Phi}_{0,\mu}^{(N)}(0,\phi_y)\right\rangle\right]\ret
&-&\frac{e^2}{h}\frac{1}{2\pi i}
\frac{1}{q}\sum_{\mu=1}^q
\int_0^{\Delta\phi_x}d\phi_x\left[\left\langle
{\tilde \Phi}_{0,\mu}^{(N)}(\phi_x,\Delta\phi_y),
\frac{\partial}{\partial\phi_x}
{\tilde \Phi}_{0,\mu}^{(N)}(\phi_x,\Delta\phi_y)\right\rangle
\right.
\ret& &\quad\qquad\qquad\qquad\qquad
-\left.\left\langle
{\tilde \Phi}_{0,\mu}^{(N)}(\phi_x,0),\frac{\partial}{\partial\phi_x}
{\tilde \Phi}_{0,\mu}^{(N)}(\phi_x,0)\right\rangle\right].
\end{eqnarray}
In order to compute this right-hand side, we introduce 
$q\times q$ matrices $\theta^{(x)}(\phi_y)$ and $\theta^{(y)}(\phi_x)$ as 
\begin{equation}
C^{(x)}(\phi_y)=\exp\left[i\theta^{(x)}(\phi_y)\right]
\quad
\mbox{and} \quad
C^{(y)}(\phi_x)=\exp\left[i\theta^{(y)}(\phi_x)\right]. 
\label{thetas}
\end{equation}
Using these and the relations (\ref{relationtildePhi1}) and 
(\ref{relationtildePhi2}) 
again, one has 
\begin{eqnarray}
{\overline {\sigma_{xy}(\mbox{\boldmath $\phi$})}}&=&\frac{e^2}{h}\frac{1}{2\pi i}
\frac{1}{q}\sum_{\mu=1}^q\left[\int_0^{\Delta\phi_y}d\phi_y
\frac{\partial}{\partial\phi_y}
i\theta_{\mu,\mu}^{(x)}(\phi_y)-\int_0^{\Delta\phi_x}d\phi_x
\frac{\partial}{\partial\phi_x}i\theta_{\mu,\mu}^{(y)}(\phi_x)\right]\ret
&=&\frac{e^2}{2\pi h}\frac{1}{q}\sum_{\mu=1}^q
\left[\theta_{\mu,\mu}^{(x)}(\Delta\phi_y)-
\theta_{\mu,\mu}^{(x)}(0)-\theta_{\mu,\mu}^{(y)}(\Delta\phi_x)
+\theta_{\mu,\mu}^{(y)}(0)\right]\ret
&=&\frac{e^2}{2\pi h}\frac{1}{q}{\rm tr}\left[\theta^{(x)}(\Delta\phi_y)-
\theta^{(x)}(0)-\theta^{(y)}(\Delta\phi_x)
+\theta^{(y)}(0)\right],
\label{avsigmaxytheta}
\end{eqnarray}
where  we have used 
\begin{equation}
\left\langle{\tilde \Phi}_{0,\mu}^{(N)}(\mbox{\boldmath $\phi$}),
{\tilde \Phi}_{0,\mu'}^{(N)}(\mbox{\boldmath $\phi$})\right\rangle=\delta_{\mu,\mu'}. 
\label{othonormal}
\end{equation}

On the other hand, {from} the relations (\ref{relationtildePhi1}) and 
(\ref{relationtildePhi2}) and (\ref{thetas}), one has 
\begin{equation}
{\tilde \Phi}_0^{(N)}(\Delta\phi_x,\Delta\phi_y)
=\exp\left[i\theta_x(\Delta\phi_y)\right]
{\tilde \Phi}_0^{(N)}(0,\Delta\phi_y),
\end{equation}
\begin{equation}
{\tilde \Phi}_0^{(N)}(\Delta\phi_x,0)
=\exp\left[i\theta_x(0)\right]
{\tilde \Phi}_0^{(N)}(0,0),
\end{equation}
\begin{equation}
{\tilde \Phi}_0^{(N)}(\Delta\phi_x,\Delta\phi_y)
=\exp\left[i\theta_y(\Delta\phi_x)\right]
{\tilde \Phi}_0^{(N)}(\Delta\phi_x,0),
\end{equation}
and 
\begin{equation}
{\tilde \Phi}_0^{(N)}(0,\Delta\phi_y)
=\exp\left[i\theta_y(0)\right]
{\tilde \Phi}_0^{(N)}(0,0),
\end{equation}
where ${\tilde \Phi}_0^{(N)}(\phi_x,\phi_y)$ is the $q$ component 
vector whose $\mu$-th component is 
${\tilde \Phi}_{0,\mu}^{(N)}(\phi_x,\phi_y)$. 
These four equations yield 
\begin{equation}
\exp\left[i\theta^{(x)}(\Delta\phi_y)\right]
\exp\left[-i\theta^{(x)}(0)\right]
\exp\left[-i\theta^{(y)}(\Delta\phi_x)\right]
\exp\left[i\theta^{(y)}(0)\right]=1,
\end{equation}
where we have used the relation (\ref{othonormal}). 
Taking the determinant of both sides of this equation and using 
${\rm det}\exp[i\theta^{(s)}(\cdots)]=\exp\left[i{\rm tr}\ \theta^{(s)}(\cdots)
\right]$ for $s=x,y$, one has 
\begin{equation}
{\rm tr}\left[\theta^{(x)}(\Delta\phi_y)-
\theta^{(x)}(0)-\theta^{(y)}(\Delta\phi_x)
+\theta^{(y)}(0)\right]=-2\pi p \quad \mbox{with an integer}\ p. 
\end{equation}
Owing to this relation, the averaged Hall conductance 
${\overline {\sigma_{xy}(\mbox{\boldmath $\phi$})}}$ of (\ref{avsigmaxytheta}) 
must satisfy 
\begin{equation}
{\overline{\sigma_{xy}(\mbox{\boldmath $\phi$})}}
=-\frac{e^2}{h}\frac{p}{q}\quad\mbox{with\ \ an integer}\ p.
\label{fqsigmaxy} 
\end{equation}

%%%%%%%%%%%%%%%%%%%%%%%%%%%%%%%%%%%%%%%%%%%%%%%%%%%%%%%%%%%%%%
\subsection{Fractional quantization and Atiyah-Singer index theorem}
\label{AStheorem}

In this subsection, we will show that the integer $-p$ of the fractional 
quantization (\ref{fqsigmaxy}) of the averaged Hall conductance  
${\overline{\sigma_{xy}(\mbox{\boldmath $\phi$})}}$ is equal to 
an index of a Pauli-Dirac operator ${\cal D}$ with the gauge field 
${\cal A}(\mbox{\boldmath $\phi$})=({\cal A}_x(\mbox{\boldmath $\phi$}),
{\cal A}_y(\mbox{\boldmath $\phi$}))$ on the gauge torus ${\cal T}_{\rm g}$. 
This is nothing but a special case of Atiyah-Singer index theorem. 
Namely the first Chern number ${\cal I}$ of (\ref{1stCnumber}) is equal to 
the index, ind~${\cal D}$, of the Pauli-Dirac operator as we will see 
in Theorem~\ref{ASindextheorem} below. 
The corresponding system described by the Hamiltonian $H=-{\cal D}^2$ 
is equivalent to 
a single electron system with spin-$1/2$ and with $q$ flavors in the gauge field 
${\cal A}(\mbox{\boldmath $\phi$})$ on the torus ${\cal T}_{\rm g}$, 
and the index, $-p={\rm ind}~{\cal D}$, is equal to the difference of 
the degeneracy between  
the up-spin and the down-spin ground states. Thus the integer $p$ of 
the fractionally quantized Hall conductance of the interacting 
electrons is closely related to the ground state property of the single electron 
coupled to the gauge field ${\cal A}(\mbox{\boldmath $\phi$})$ 
determined by the ground state of the original interacting electron system.  

Throughout the present subsection, for simplicity, we will write 
${\tilde \Phi}_\mu(\mbox{\boldmath $\phi$})=
{\tilde \Phi}_{0,\mu}^{(N)}(\mbox{\boldmath $\phi$})$, 
$\mu=1,2,\ldots,q$, for the ground state wavefunctions, by dropping 
the subscript $0$ and the superscript $(N)$. 
Then the matrix elements of the gauge field 
${\cal A}(\mbox{\boldmath $\phi$})=({\cal A}_x(\mbox{\boldmath $\phi$}),
{\cal A}_y(\mbox{\boldmath $\phi$}))$ 
are written in terms of ${\tilde\Phi}_\mu(\mbox{\boldmath $\phi$})$ as 
\begin{equation}
{\cal A}_{\mu,\nu,s}(\mbox{\boldmath $\phi$})=
\left\langle{\tilde \Phi}_\mu(\mbox{\boldmath $\phi$}),\partial_s
{\tilde \Phi}_\nu(\mbox{\boldmath $\phi$})\right\rangle \quad \mbox{for}\ \ 
s=x,y, 
\label{defA}
\end{equation}
where we have written 
\begin{equation}
\partial_s=\frac{\partial}{\partial\phi_s}\quad \mbox{for } s=x,y. 
\end{equation}
We define gauge transformations as 
\begin{equation}
{\cal G}_{\mu,\nu}^{(x)}(\phi_x,\phi_y):=
\left\langle{\tilde \Phi}_\mu(\phi_x+\Delta\phi_x,\phi_y),
{\tilde \Phi}_\nu(\phi_x,\phi_y)\right\rangle \quad \mbox{for small}\ \ \phi_x
\label{defGx}
\end{equation}
and 
\begin{equation}
{\cal G}_{\mu,\nu}^{(y)}(\phi_x,\phi_y):=
\left\langle{\tilde \Phi}_\mu(\phi_x,\phi_y+\Delta\phi_y),
{\tilde \Phi}_\nu(\phi_x,\phi_y)\right\rangle 
\quad \mbox{for small}\ \ \phi_y. 
\label{defGy}
\end{equation}
These are $q\times q$ unitary matrices. Actually, 
\begin{eqnarray}
\sum_{\alpha=1}^q {\cal G}_{\mu,\alpha}^{(x)}(\mbox{\boldmath $\phi$})
{{\cal G}_{\nu,\alpha}^{(x)}(\mbox{\boldmath $\phi$})}^\ast&=&
\sum_{\alpha=1}^q \left\langle{\tilde \Phi}_\mu(\phi_x+\Delta\phi_x,\phi_y),
{\tilde \Phi}_\alpha(\mbox{\boldmath $\phi$})\right\rangle
\left\langle{\tilde \Phi}_\alpha(\mbox{\boldmath $\phi$}),
{\tilde \Phi}_\nu(\phi_x+\Delta\phi_x,\phi_y)\right\rangle\ret
&=&\left\langle{\tilde \Phi}_\mu(\phi_x+\Delta\phi_x,\phi_y),
{\tilde \Phi}_\nu(\phi_x+\Delta\phi_x,\phi_y)\right\rangle
=\delta_{\mu,\nu}. 
\end{eqnarray}
Similarly, one has 
\begin{equation}
\sum_{\alpha=1}^q {\cal G}_{\mu,\alpha}^{(y)}(\mbox{\boldmath $\phi$})
{{\cal G}_{\nu,\alpha}^{(y)}(\mbox{\boldmath $\phi$})}^\ast=\delta_{\mu,\nu}.
\end{equation}

\begin{lemma}
\label{transA}
The gauge field ${\cal A}(\mbox{\boldmath $\phi$})=
({\cal A}_x(\mbox{\boldmath $\phi$}),{\cal A}_y(\mbox{\boldmath $\phi$}))$ 
satisfies 
\begin{equation}
{\cal A}_s(\phi_x+\Delta\phi_x,\phi_y)={\cal G}^{(x)}(\mbox{\boldmath $\phi$})
{\cal A}_s(\mbox{\boldmath $\phi$})
{{\cal G}^{(x)}(\mbox{\boldmath $\phi$})}^\ast
+{\cal G}^{(x)}(\mbox{\boldmath $\phi$})\partial_s
{{\cal G}^{(x)}(\mbox{\boldmath $\phi$})}^\ast \quad 
\mbox{for small} \ \ \phi_x
\end{equation}
and 
\begin{equation}
{\cal A}_s(\phi_x,\phi_y+\Delta\phi_y)={\cal G}^{(y)}(\mbox{\boldmath $\phi$})
{\cal A}_s(\mbox{\boldmath $\phi$}){{\cal G}^{(y)}
(\mbox{\boldmath $\phi$})}^\ast
+{\cal G}^{(y)}(\mbox{\boldmath $\phi$})\partial_s
{{\cal G}^{(y)}(\mbox{\boldmath $\phi$})}^\ast \quad 
\mbox{for small} \ \ \phi_y. 
\label{Atransy}
\end{equation}
\end{lemma}

\begin{proof}{Proof} From the definitions (\ref{defA}), (\ref{defGx}) and 
(\ref{defGy}) for the gauge field ${\cal A}(\mbox{\boldmath $\phi$})$ 
and the gauge transformations ${\cal G}^{(s)}(\mbox{\boldmath $\phi$})$ 
for $s=x,y$, we have 
\begin{eqnarray}
& &{\cal A}_{\mu,\nu,s}(\phi_x+\Delta\phi_x,\phi_y)\ret
&=&\left\langle{\tilde \Phi}_\mu(\phi_x+\Delta\phi_x,\phi_y),
\partial_s{\tilde \Phi}_\nu(\phi_x+\Delta\phi_x,\phi_y)
\right\rangle\ret
&=&\sum_{\alpha,\beta=1}^q 
\left\langle{\tilde \Phi}_\mu(\phi_x+\Delta\phi_x,\phi_y),
{\tilde \Phi}_\alpha(\mbox{\boldmath $\phi$})\right\rangle
\left\langle{\tilde \Phi}_\alpha(\mbox{\boldmath $\phi$}),
\partial_s{\tilde \Phi}_\beta(\mbox{\boldmath $\phi$})\right\rangle
\left\langle{\tilde \Phi}_\beta(\mbox{\boldmath $\phi$}),
{\tilde \Phi}_\nu(\phi_x+\Delta\phi_x,\phi_y)\right\rangle\ret
&=&\sum_{\alpha,\beta}\left\{{\cal G}_{\mu,\alpha}^{(x)}
(\mbox{\boldmath $\phi$})
{\cal A}_{\alpha,\beta,s}(\mbox{\boldmath $\phi$})
{{\cal G}_{\nu,\beta}^{(x)}(\mbox{\boldmath $\phi$})}^\ast
+{\cal G}_{\mu,\alpha}^{(x)}(\mbox{\boldmath $\phi$})\delta_{\alpha,\beta}
\partial_s{{\cal G}_{\nu,\beta}^{(x)}(\mbox{\boldmath $\phi$})}^\ast
\right\}\ret 
&=&\left({\cal G}^{(x)}(\mbox{\boldmath $\phi$}){\cal A}_s
(\mbox{\boldmath $\phi$})
{{\cal G}^{(x)}(\mbox{\boldmath $\phi$})}^\ast\right)_{\mu,\nu}+
\left({\cal G}^{(x)}(\mbox{\boldmath $\phi$})\partial_s
{{\cal G}^{(x)}(\mbox{\boldmath $\phi$})}^\ast\right)_{\mu,\nu}. 
\end{eqnarray}
In the same way, the other relation (\ref{Atransy}) can be obtained. 
\end{proof}

We define covariant derivatives $\nabla_s$ as 
\begin{equation}
\nabla_s:=\partial_s+{\cal A}_s(\mbox{\boldmath $\phi$})
\quad \mbox{for} \quad s=x,y. 
\end{equation}
These covariant derivatives $\nabla_s$ act on a vector 
field $f(\mbox{\boldmath $\phi$})=(f_1(\mbox{\boldmath $\phi$}),
f_2(\mbox{\boldmath $\phi$}),\cdots, f_q(\mbox{\boldmath $\phi$}))$ on 
the torus ${\cal T}_{\rm g}$.  
We require that such a vector field $f(\mbox{\boldmath $\phi$})$ is transformed as 
\begin{equation}
f_\mu(\phi_x+\Delta\phi_x,\phi_y)=\sum_{\alpha=1}^q 
{\cal G}_{\mu,\alpha}^{(x)}(\mbox{\boldmath $\phi$})
f_\alpha(\mbox{\boldmath $\phi$}) \quad \mbox{for small} \ \ 
\phi_x
\label{ftransx}
\end{equation}
and 
\begin{equation}
f_\mu(\phi_x,\phi_y+\Delta\phi_y)=\sum_{\alpha=1}^q 
{\cal G}_{\mu,\alpha}^{(y)}(\mbox{\boldmath $\phi$})
f_\alpha(\mbox{\boldmath $\phi$}) \quad \mbox{for small} \ \ 
\phi_y,
\label{ftransy}
\end{equation}
by the gauge transformation ${\cal G}^{(s)}(\mbox{\boldmath $\phi$})$.
Then one has 

\begin{lemma}
Assume the vector field $f(\mbox{\boldmath $\phi$})$ is continuously 
differentiable with respect 
to the gauge parameters $\mbox{\boldmath $\phi$}=(\phi_x,\phi_y)$. 
Then $f(\mbox{\boldmath $\phi$})$ satisfies 
\begin{equation}
\left(\nabla_sf\right)(\phi_x+\Delta\phi_x,\phi_y)=
{\cal G}^{(x)}(\mbox{\boldmath $\phi$})\left(\nabla_sf\right)(\phi_x,\phi_y)
\quad \mbox{for small}\ \ \phi_x, 
\label{codevfx}
\end{equation}
and 
\begin{equation}
\left(\nabla_sf\right)(\phi_x,\phi_y+\Delta\phi_y)=
{\cal G}^{(y)}(\mbox{\boldmath $\phi$})\left(\nabla_sf\right)(\phi_x,\phi_y)
\quad \mbox{for small} \ \ \phi_y. 
\end{equation}
\end{lemma}

\begin{proof}{Proof}
{From} the transformation (\ref{ftransx}) for the vector 
field $f(\mbox{\boldmath $\phi$})$, one has 
\begin{equation}
\partial_sf_\mu(\phi_x+\Delta\phi_x,\phi_y)
=\sum_{\alpha=1}^q \left[\partial_s
{\cal G}_{\mu,\alpha}^{(x)}(\mbox{\boldmath $\phi$})
\cdot f_\alpha(\mbox{\boldmath $\phi$})
+{\cal G}_{\mu,\alpha}^{(x)}(\mbox{\boldmath $\phi$})
\partial_sf_\alpha(\mbox{\boldmath $\phi$})
\right]
\label{partialftrans}
\end{equation}
for small $\phi_x$. 
{From} the transformation (\ref{ftransx}) and Lemma~\ref{transA}, one has 
\begin{eqnarray}
& &\sum_{\alpha=1}^q {\cal A}_{\mu,\alpha,s}(\phi_x+\Delta\phi_x,\phi_y)
f_\alpha(\phi_x+\Delta\phi_x,\phi_y)\ret
&=&\left({\cal G}^{(x)}(\mbox{\boldmath $\phi$})
{\cal A}_s(\mbox{\boldmath $\phi$})f(\mbox{\boldmath $\phi$})\right)_\mu+
\sum_{\alpha=1}^q\left({\cal G}^{(x)}(\mbox{\boldmath $\phi$})\partial_s
{{\cal G}^{(x)}(\mbox{\boldmath $\phi$})}^\ast\right)_{\mu,\alpha}
\left({\cal G}^{(x)}(\mbox{\boldmath $\phi$})
f(\mbox{\boldmath $\phi$})\right)_\alpha\ret
&=&\left({\cal G}^{(x)}(\mbox{\boldmath $\phi$})
{\cal A}_s(\mbox{\boldmath $\phi$})f(\mbox{\boldmath $\phi$})\right)_\mu
-\left(\partial_s{\cal G}^{(x)}(\mbox{\boldmath $\phi$})\cdot 
f(\mbox{\boldmath $\phi$})\right)_\mu,
\label{Aftrans}
\end{eqnarray}
where we have used the identity, 
\begin{equation}
{\cal G}^{(s)}(\mbox{\boldmath $\phi$})\partial_t{{\cal G}^{(s)}
(\mbox{\boldmath $\phi$})}^\ast+
\partial_t{\cal G}^{(s)}(\mbox{\boldmath $\phi$})\cdot
{{\cal G}^{(s)}(\mbox{\boldmath $\phi$})}^\ast=0
\quad \mbox{for}\ \ s,t=x,y,
\end{equation}
for getting the second equality. Combining (\ref{partialftrans}) and 
(\ref{Aftrans}), the relation (\ref{codevfx}) is derived. The other relation 
is also derived in the same way. 
\end{proof}

We introduce the Pauli-Dirac operator ${\cal D}$ as 
\begin{equation}
{\cal D}=\sigma_x\nabla_x+\sigma_y\nabla_y, 
\end{equation}
where $\sigma_x$ and $\sigma_y$ are the Pauli matrices given by 
\begin{equation}
\sigma_x=\left(\matrix{0 & 1 \cr 1 & 0 \cr}\right)\quad\mbox{and}\quad 
\sigma_y=\left(\matrix{0 & -i\ \cr i & \ 0 \cr}\right).
\end{equation}
Clearly the Pauli-Dirac operator ${\cal D}$ is written as 
\begin{equation}
{\cal D}=\left(\matrix{0 & {\cal D}^- \cr {\cal D}^+ & 0 \cr}\right), 
\end{equation}
where 
\begin{equation}
{\cal D}^+=\nabla_x+i\nabla_y \quad \mbox{and} \quad 
{\cal D}^-=\nabla_x-i\nabla_y. 
\end{equation}
The Hamiltonian of the corresponding system is given by 
\begin{equation}
H:=-{\cal D}^2=\left(\matrix{-{\cal D}^-{\cal D}^+ & 0 \cr 
0 & -{\cal D}^+{\cal D}^-\cr}\right). 
\end{equation}
Note that 
\begin{eqnarray}
{\cal D}^-{\cal D}^+&=&\left(\nabla_x-i\nabla_y\right)
\left(\nabla_x+i\nabla_y\right)\ret
&=&\nabla_x^2+\nabla_y^2+i\nabla_x\nabla_y-i\nabla_y\nabla_x\ret
&=&\nabla_x^2+\nabla_y^2+i\left[\nabla_x,\nabla_y\right].
\end{eqnarray}
Since the commutator in the last line is equal to the curvature ${\cal F}$ as 
\begin{equation}
\left[\nabla_x,\nabla_y\right]=
\partial_x{\cal A}_y -\partial_y{\cal A}_x+\left[{\cal A}_x,{\cal A}_y\right]
=:{\cal F}, 
\end{equation}
one has 
\begin{equation}
{\cal D}^-{\cal D}^+=\triangle +i{\cal F} 
\label{D-D+}
\end{equation}
with the Laplacian, 
\begin{equation}
\triangle=\nabla_x^2+\nabla_y^2=(\partial_x+{\cal A}_x)^2
+(\partial_y+{\cal A}_y)^2. 
\end{equation}
In the same way, one has 
\begin{equation}
{\cal D}^+{\cal D}^-=\triangle -i{\cal F}. 
\label{D+D-}
\end{equation}
Therefore the Hamiltonian $H$ becomes 
\begin{equation}
H=-\triangle -i{\cal F}\sigma_z
\quad\mbox{with}\quad
\sigma_z:=\left(\matrix{1 & 0\cr 0 & -1 \cr}\right). 
\end{equation}
Namely the system is equivalent to the single electron with spin-1/2 and 
$q$ flavors in the ``magnetic field" ${\cal F}$ in two dimensions. 
Here we stress that the number $q$ of the flavors is equal to the degeneracy 
of the ground state of the original interacting electron system. 

We define the index of the Pauli-Dirac operator ${\cal D}$ as 
\begin{equation}
{\rm ind}\ {\cal D}:={\rm dim}\ {\rm ker}\ {\cal D}^+
-{\rm dim}\ {\rm ker}\ {\cal D}^-.
\label{defindex}
\end{equation} 
In the standard way, one has 
\begin{lemma}
\label{lemma:indDTr}
For any positive constant $\beta$, the following equality is valid: 
\begin{equation}
{\rm ind}\ {\cal D}={\rm Tr}\ \sigma_z e^{-\beta H}. 
\end{equation}
\end{lemma}

\begin{proof}{Proof}
Let $\psi$ be an eigenvector of the operator ${\cal D}^-{\cal D}^+$ with 
the eigenvalue $\lambda\ne 0$, i.e., 
\begin{equation}
{\cal D}^-{\cal D}^+\psi=\lambda\psi. 
\end{equation}
Then the vector ${\cal D}^+\psi$ is the eigenvector of the operator 
${\cal D}^+{\cal D}^-$ with the same eigenvalue $\lambda$. Actually, one has 
\begin{equation}
\left\Vert {\cal D}^+\psi\right\Vert^2=
\left\langle {\cal D}^+\psi,{\cal D}^+\psi\right\rangle
=-\left\langle \psi, {\cal D}^-{\cal D}^+\psi\right\rangle
=-\lambda\left\Vert \psi\right\Vert^2\ne 0
\end{equation}
and 
\begin{equation}
{\cal D}^+{\cal D}^-({\cal D}^+\psi)
={\cal D}^+({\cal D}^-{\cal D}^+\psi)=\lambda {\cal D}^+\psi. 
\end{equation}
Let $\psi_1,\psi_2$ be eigenvectors of ${\cal D}^-{\cal D}^+$ with 
the eigenvalue $\lambda$, and assume the two vectors are orthogonal to 
each other. Then 
\begin{equation}
\left\langle {\cal D}^+\psi_1,{\cal D}^+\psi_2\right\rangle 
=-\left\langle\psi_1,{\cal D}^-{\cal D}^+\psi_2\right\rangle
=-\lambda\left\langle \psi_1,\psi_2\right\rangle=0. 
\end{equation}
We denote by $d^+(\lambda)$ the dimension of the eigenspace of 
the operator ${\cal D}^-{\cal D}^+$ with the eigenvalue $\lambda$, and 
by $d^-(\lambda)$ the dimension of the eigenspace of ${\cal D}^+{\cal D}^-$ 
with $\lambda$. From the above observations, one has 
$d^+(\lambda)\le d^-(\lambda)$ for $\lambda\ne 0$. Further 
$d^-(\lambda)\le d^+(\lambda)$ for $\lambda\ne 0$ in the same way. 
Combining these two inequalities, one has 
$d^+(\lambda)=d^-(\lambda)$ for $\lambda\ne 0$. 
{From} this fact, one gets 
\begin{equation}
{\rm Tr}\ \sigma_ze^{-\beta H}={\rm Tr}\ e^{\beta {\cal D}^-{\cal D}^+}
-{\rm Tr}\ e^{\beta {\cal D}^+{\cal D}^-}=
{\rm dim}\ {\rm ker}\ {\cal D}^-{\cal D}^+ 
-{\rm dim}\ {\rm ker}\ {\cal D}^+{\cal D}^-.
\label{sTrdimker}
\end{equation}
This right-hand side is equal to the index of the operator ${\cal D}$ 
by the definition (\ref{defindex}).  
\end{proof}

The following theorem is a special case of the Atiyah-Singer index 
theorem \cite{CGKS,Gilkey}: 
\begin{theorem}
\label{ASindextheorem}
\begin{equation}
{\rm ind}\ {\cal D}=\frac{1}{2\pi i}\int_{{\cal T}_{\rm g}}d\phi_xd\phi_y\ 
{\rm tr}\ {\cal F}
\end{equation}
\end{theorem}
\begin{proof}{Proof}
We shall prove this statement along the same line as in ref.~\cite{Fujikawa}. 
First let us introduce the complete system of the orthonormal functions as 
\begin{equation}
\psi(k_x,k_y):=\frac{1}{\sqrt{\Delta\phi_x\Delta\phi_y}}
\exp\left[i(k_x\phi_x+k_y\phi_y)\right]\quad \mbox{with}\ 
k_x=\frac{2\pi n_x}{\Delta\phi_x}, k_y=\frac{2\pi n_y}{\Delta\phi_y}
\end{equation}
with $n_x,n_y\in{\bf Z}$. Using this system of the functions, one has 
\begin{eqnarray}
{\rm Tr}\ e^{\beta(\triangle \pm i{\cal F})}&=&\sum_{k_x,k_y}
{\rm tr}\ \left\langle\psi(k_x,k_y),e^{\beta(\triangle \pm i{\cal F})}
\psi(k_x,k_y)\right\rangle\ret
&=&\frac{1}{\Delta\phi_x\Delta\phi_y}\sum_{k_x,k_y}
{\rm tr}\ \int_{{\cal T}_{\rm g}} d\phi_x d\phi_y\ 
\exp\left[\beta\left(\triangle({\bf k})\pm i{\cal F}\right)\right], 
\end{eqnarray}
where we have written 
\begin{equation}
\triangle({\bf k})=(\nabla_x+ik_x)^2+(\nabla_y+ik_y)^2
\end{equation}
and have used 
\begin{equation}
e^{-ik_s\phi_s}\partial_se^{ik_s\phi_s}=\partial_s+ik_s \quad \mbox{for}\ 
s=x,y.
\end{equation}
Moreover, by using DuHamel's formula 
\begin{equation}
e^{-t(X+Y)}=e^{-tX}+\int_0^tds\ e^{-(t-s)X}Ye^{-s(X+Y)},
\end{equation}
one has 
\begin{eqnarray}
& &{\rm Tr}\ e^{\beta (\triangle+i{\cal F})}-
{\rm Tr}\ e^{\beta (\triangle-i{\cal F})}\ret
&=&\frac{\beta}{i\Delta\phi_x\Delta\phi_y}\sum_{k_x,k_y}{\rm tr}
\int_{{\cal T}_{\rm g}}d\phi_xd\phi_y\ \frac{1}{\beta}\int_0^\beta 
d\beta'\ e^{(\beta-\beta')\triangle({\bf k})}{\cal F}
\left[e^{\beta'(\triangle({\bf k})+i{\cal F})}+
e^{\beta'(\triangle({\bf k})-i{\cal F})}\right]\ret
&=&\frac{\beta}{i\Delta\phi_x\Delta\phi_y}\sum_{k_x,k_y}{\rm tr}
\int_{{\cal T}_{\rm g}}d\phi_xd\phi_y\ \int_0^1 ds\ 
e^{(1-s)\beta\triangle({\bf k})}{\cal F}
\left[e^{s\beta(\triangle({\bf k})+i{\cal F})}+
e^{s\beta(\triangle({\bf k})-i{\cal F})}\right]\ret
&=&\frac{\beta}{i\Delta\phi_x\Delta\phi_y}\sum_{{\tilde k}_x,{\tilde k}_y}
{\rm tr}
\int_{{\cal T}_{\rm g}}d\phi_xd\phi_y\ \int_0^1 ds\ 
e^{(1-s){\tilde \triangle}({\tilde {\bf k}})}{\cal F}
\left[e^{s({\tilde \triangle}({\tilde {\bf k}})+i\beta{\cal F})}+
e^{s({\tilde \triangle}({\tilde {\bf k}})-i\beta{\cal F})}\right], 
\label{diffTr}
\end{eqnarray}
where we have introduced the variable $s$ as $\beta'=\beta s$ with 
$0\le s\le 1$, and we have written 
\begin{equation}
{\tilde k}_x=\sqrt{\beta}k_x \quad \mbox{and}\quad 
{\tilde k}_y=\sqrt{\beta}k_y, 
\end{equation}
and 
\begin{equation}
{\tilde \triangle}({\tilde {\bf k}})=
\left(\sqrt{\beta}\nabla_x+i{\tilde k}_x\right)^2+
\left(\sqrt{\beta}\nabla_y+i{\tilde k}_y\right)^2. 
\end{equation}
Here, since the sum of the operators 
\begin{equation}
\frac{\beta}{\Delta\phi_x\Delta\phi_y}\sum_{{\tilde k}_x,{\tilde k}_y}
\exp\left[{\tilde \triangle}({\tilde {\bf k}})\right]
\end{equation}
is uniformly bounded in $\beta$, the right-hand side of (\ref{diffTr}) 
converges to 
\begin{equation}
\frac{1}{(2\pi)^2i} \int d{\tilde k}_xd{\tilde k}_y\ 
{\rm tr} \int_{{\cal T}_{\rm g}}d\phi_xd\phi_y\ 
\exp\left[-\left({\tilde k}_x^2+{\tilde k}_y^2\right)\right] 
\times 2{\cal F}=\frac{1}{2\pi i} 
\int_{{\cal T}_{\rm g}}d\phi_xd\phi_y\ {\rm tr}\ {\cal F}
\end{equation}
in the limit $\beta\downarrow 0$. Combining this observation, (\ref{D-D+}), 
(\ref{D+D-}), Lemma~\ref{lemma:indDTr} and (\ref{sTrdimker}), one obtains 
the desired relation, 
\begin{equation}
{\rm ind}\ {\cal D}=
{\rm Tr}\ e^{\beta (\triangle+i{\cal F})}-
{\rm Tr}\ e^{\beta (\triangle-i{\cal F})}
=\frac{1}{2\pi i} 
\int_{{\cal T}_{\rm g}}d\phi_xd\phi_y\ {\rm tr}\ {\cal F}. 
\end{equation}
\end{proof}

%%%%%%%%%%%%%%%%%%%%%%%%%%%%%%%%%%%%%%%%%%%%%%%%%%
\Section{The non-interacting case}
\label{noninteracting}

As a demonstration, we first treat the simplest model of the quantum 
Hall system, and determine the explicit forms of the function 
$\theta^{(s)}(\mbox{\boldmath $\phi$})$, $s=x,y$, introduced in Section~\ref{ProofQHC}. 
As a result, the well-known integral quantization of the Hall conductance 
is obtained. In the next subsection, we will treat the general non-interacting case, 
and give the proof of Theorem~\ref{theorem:noninteracting}. 

%%%%%%%%%%%%%%%%%%%%%%%%%%%%%%%%%%%%%%%%%%%%%%%%%%%%%%%%%
\subsection{The single electron Landau Hamiltonian}
\label{singleLandau}

The single electron Hamiltonian in two dimensions with the uniform magnetic 
field only is given by 
\begin{equation}
{\cal H}_0(\mbox{\boldmath $\phi$})=\frac{1}{2m_e}
\left[(p_x-eBy+\phi_x)^2+(p_y+\phi_y)^2\right]
\label{hamsingleLandau}
\end{equation}
with the gauge parameters $\mbox{\boldmath $\phi$}=(\phi_x,\phi_y)\in {\bf R}^2$. 
The eigenvectors on ${\bf R}^2$ are given by 
\begin{equation}
\varphi_{n,k}(x,y;\mbox{\boldmath $\phi$})=e^{ikx}
\exp\left[-\frac{i}{\hbar}\phi_yy\right]v_n(y-y(k,\phi_x)),
\end{equation}
where $k$ is the real wavenumber in the $x$ direction, and 
\begin{equation}
v_{n,k}(y;\phi_x):=v_n(y-y(k,\phi_x)):=
N_n\exp\left[-(y-y(k,\phi_x))^2/(2\ell_B^2)\right]
\wp_n\left[(y-y(k,\phi_x))/\ell_B\right]
\end{equation}
with 
\begin{equation}
y(k,\phi_x):=\frac{1}{eB}(\hbar k+\phi_x).
\label{defykphi} 
\end{equation}
Here $\wp_n$ is the $n$-th Hermite polynomial, and the normalization 
constant $N_n$ is taken to satisfy 
\begin{equation}
\int_{-\infty}^{+\infty}dy|v_{n,k}(y;\phi_x)|^2=1.
\end{equation}
 
Now consider the $L_x\times L_y$ rectangular box 
${\cal T}=[-L_x/2,L_x/2]\times[-L_y/2,L_y/2]$ satisfying 
$L_xL_y=2\pi M\ell_B^2$ with a sufficiently large positive and 
even integer $M$. We impose the periodic boundary conditions 
\begin{equation}
\varphi(x,y;\mbox{\boldmath $\phi$})=t^{(x)}(L_x)\varphi(x,y;\mbox{\boldmath $\phi$}), 
\quad 
\varphi(x,y;\mbox{\boldmath $\phi$})=t^{(y)}(L_y)\varphi(x,y;\mbox{\boldmath $\phi$})
\label{PBCvarphi}
\end{equation}
for the wavefunctions $\varphi$. Then the complete system 
of the eigenvectors \cite{Koma1,Koma2} of 
the Hamiltonian satisfying the periodic boundary conditions is given by 
\begin{equation}
\varphi_{n,k}^{\rm P}(x,y;\mbox{\boldmath $\phi$})=L_x^{-1/2}
\sum_{\ell=-\infty}^{+\infty}e^{i(k+\ell K)x}
\exp\left[-\frac{i}{\hbar}\phi_y(y-\ell L_y)\right]
v_n(y-y(k,\phi_x)-\ell L_y)
\label{varphiP}
\end{equation}
for $k=2\pi m/L_x$ with $m=-M/2+1,\ldots,M/2-1,M/2$, and 
with $K=L_y/\ell_B^2$. The energy eigenvalue is given by $(n+1/2)\hbar\omega_c$ 
for $n=0,1,2\ldots$. By the gauge transformation 
\begin{equation}
G(x,y;\mbox{\boldmath $\phi$})=\exp\left[\frac{i}{\hbar}(x\phi_x+y\phi_y)\right] 
\end{equation}
corresponding to the transformation (\ref{Gaugetr}) of $N$ body, 
the eigenvectors are transformed as 
\begin{eqnarray}
{\tilde \varphi}_{n,k}^{\rm P}(x,y;\mbox{\boldmath $\phi$})&=&
G(x,y;\mbox{\boldmath $\phi$})\varphi_{n,k}^{\rm P}(x,y;\mbox{\boldmath $\phi$})\ret
&=&L_x^{-1/2}
\sum_{\ell=-\infty}^{+\infty}e^{i(k+\ell K)x}
\exp\left[\frac{i}{\hbar}\left(x\phi_x+\ell L_y\phi_y\right)\right]
v_n(y-y(k,\phi_x)-\ell L_y).\ret
\label{tildevarphi}
\end{eqnarray}

In order to determine the explicit forms of the phases 
$\theta^{(x)}$ and $\theta^{(y)}$ in (\ref{thetas}), we first show that 
\begin{equation}
{\tilde \varphi}_{n,k}^{\rm P}(x,y;\Delta\phi_x,\phi_y)
={\tilde \varphi}_{n,k+2\pi/L_x}^{\rm P}(x,y;0,\phi_y)\quad 
\mbox{for}\ k\neq k_{\rm max}, 
\label{tildevarphirelation1}
\end{equation}
\begin{equation}
{\tilde \varphi}_{n,k_{\rm max}}^{\rm P}(x,y;\Delta\phi_x,\phi_y)
=\exp\left[-\frac{i}{\hbar}L_y\phi_y\right]
{\tilde \varphi}_{n,k_{\rm min}}^{\rm P}(x,y;0,\phi_y),
\label{tildevarphirelation2}
\end{equation}
and 
\begin{equation}
{\tilde \varphi}_{n,k}^{\rm P}(x,y;\phi_x,\Delta\phi_y)
={\tilde \varphi}_{n,k}^{\rm P}(x,y;\phi_x,0)\quad\mbox{for any }k,
\label{tildevarphirelation3}
\end{equation}
where $k_{\rm max}:={\pi M}/{L_x}$ and $k_{\rm min}:=2\pi(-M/2+1)/L_x$.
This last relation (\ref{tildevarphirelation3}) is immediately obtained from 
the expression (\ref{tildevarphi}) of 
${\tilde \varphi}_{n,k}^{\rm P}$. By the definition (\ref{defykphi}), one has  
\begin{equation}
y(k,\Delta\phi_x)=y(k+2\pi/L_x,0)\quad\mbox{for }
k\neq k_{\rm max}. 
\end{equation}
Combining this with the expression (\ref{tildevarphi}), 
one obtains the first relation (\ref{tildevarphirelation1}). 
For the largest wavenumber $k_{\rm max}=\pi M/L_x$, one has  
\begin{equation}
k_{\rm max}+\frac{2\pi}{L_x}=k_{\rm min}+K,
\end{equation}
where we have used the relations $K=L_y/\ell_B^2$ and $L_xL_y=2\pi M\ell_B^2$. 
Substituting this into (\ref{defykphi}), one gets 
\begin{equation}
y(k_{\rm max},\Delta\phi_x)=y(k_{\rm min},0)+L_y. 
\end{equation}
Combining these with the expression (\ref{tildevarphi}),  
the second relation (\ref{tildevarphirelation2}) is obtained.  

Let us consider the ground state ${\tilde \Phi}_0^{(N)}(\mbox{\boldmath $\phi$})$ 
with $N=\ell M$ electrons, 
where $\ell$ is a positive integer and $M$ is the number of the states in 
a single Landau level. Namely all the states in the lowest $\ell$ Landau 
levels are occupied with the $\ell M$ electrons and the rest of the higher 
Landau levels are all empty. Clearly the ground state is unique, i.e., $q=1$, 
and the energy gap $\hbar\omega_c$ appears above the ground state. 
Combining (\ref{relationtildePhi1}), (\ref{thetas}) and the above results, 
(\ref{tildevarphirelation1}) and (\ref{tildevarphirelation2}), one gets 
\begin{equation}
\theta^{(x)}(\phi_y)=-\frac{L_y}{\hbar}\phi_y \ell +\delta^{(x)}
\label{thetaxresult}
\end{equation}
for the $N=\ell M$ electrons ground state ${\tilde \Phi}_0^{(N)}(\mbox{\boldmath $\phi$})$. 
Here $\delta^{(x)}$ is a real constant which is independent of $\mbox{\boldmath $\phi$}$. 
Combining (\ref{relationtildePhi2}) and (\ref{thetas}) and (\ref{tildevarphirelation3}), 
one obtains 
\begin{equation}
\theta^{(y)}(\phi_x)=0\quad \mbox{for all}\ \ \phi_x. 
\label{thetayresult}
\end{equation}
{From} (\ref{avsigmaxytheta}), (\ref{thetaxresult}) and (\ref{thetayresult}), 
the averaged Hall conductance is given by 
\begin{equation}
\overline{\sigma_{xy}(\mbox{\boldmath $\phi$})}=-\frac{e^2}{h}\ell. 
\end{equation}
Since the present system is translationally invariant, the non-averaged 
Hall conductance is also quantized to the same integer 
as we already obtained the more general result 
(\ref{trinvsxy}). 

%%%%%%%%%%%%%%%%%%%%%%%%%%%%%%%%%%%%%%%%%%%%%%%%
\subsection{The general electron gases}

Consider the general non-interacting case, i.e., $W^{(2)}=0$ 
in the Hamiltonian (\ref{Ham0phi}). 
The corresponding single electron Hamiltonian is given by 
\begin{equation}
{\cal H}(\mbox{\boldmath $\phi$})=\frac{1}{2m_e}\left[{\bf p}+e{\bf A}({\bf r})+
\mbox{\boldmath $\phi$}\right]^2+W({\bf r})
\label{hamphiAW}
\end{equation}
with the vector potential ${\bf A}={\bf A}_0+{\bf A}_{\rm P}$ and 
with the periodic boundary conditions (\ref{PBCvarphi}). 
Consider the ground state $\Phi_0^{(N)}(\mbox{\boldmath $\phi$})$ 
with $N=\ell M$ electrons, 
where $\ell$ is a positive integer and $M$ is the number of the states in 
a single Landau level. This is the same situation as in the preceding 
subsection except for the potentials. The aim of this subsection is to prove 
all the statements of 
Theorem~\ref{theorem:noninteracting}. 

First of all, we show that the gap condition (\ref{gapcondition}) 
is valid if the vector potential ${\bf A}_{\rm P}$ and 
the electrostatic potential $W$ satisfy the condition (\ref{nonintgapcondition}) in 
Theorem~\ref{theorem:noninteracting}.
For this purpose, we first rewrite the single electron Hamiltonian (\ref{hamphiAW}) as 
\begin{eqnarray}
{\cal H}(\mbox{\boldmath $\phi$})
&=&{\cal H}_0(\mbox{\boldmath $\phi$})+\frac{e}{2m_e}{\bf A}_{\rm P}({\bf r})\cdot
\left[{\bf p}+e{\bf A}_0({\bf r})+\mbox{\boldmath $\phi$}\right]
+\frac{e}{2m_e}\left[{\bf p}+e{\bf A}_0({\bf r})+\mbox{\boldmath $\phi$}\right]\cdot
{\bf A}_{\rm P}({\bf r})\ret
&+&\frac{e^2}{2m_e}\left|{\bf A}_{\rm P}({\bf r})\right|^2+W({\bf r}),
\label{Hphiexpand}
\end{eqnarray}
where ${\cal H}_0(\mbox{\boldmath $\phi$})$ is given by (\ref{hamsingleLandau}).   
Using the Schwarz inequality, one has   
\begin{eqnarray}
\left|\left(\psi,\frac{e}{2m_e}{\bf A}_{\rm P}\cdot({\bf p}+e{\bf A}_0)\psi\right)\right|
&\le&\frac{e}{2m_e}\sqrt{\left(\psi,\left|{\bf A}_{\rm P}\right|^2\psi\right)}
\sqrt{\left(\psi,({\bf p}+e{\bf A}_0)^2\psi\right)}\ret
&\le&\frac{e}{2m_e}\left\Vert|{\bf A}_{\rm P}|\right\Vert_\infty
\sqrt{\left(\psi,({\bf p}+e{\bf A}_0)^2\psi\right)}
\end{eqnarray}
for the vector $\psi$ in the domain of the Hamiltonian. From this inequality, 
the energy expectation can be evaluated as  
\begin{eqnarray}
\left(\psi,{\cal H}(\mbox{\boldmath $\phi$})\psi\right)
&\le&\left(\psi,{\cal H}_0(\mbox{\boldmath $\phi$})\psi\right)
+\frac{\sqrt{2}e}{\sqrt{m_e}}\left\Vert|{\bf A}_{\rm P}|\right\Vert_\infty
\sqrt{\left(\psi,{\cal H}_0(\mbox{\boldmath $\phi$})\psi\right)}+\frac{e^2}{2m_e}
\left\Vert|{\bf A}_{\rm P}|\right\Vert_\infty^2+\left\Vert W^+\right\Vert_\infty\ret
\label{calHphibound}
\end{eqnarray}
and 
\begin{eqnarray}
\left(\psi,{\cal H}(\mbox{\boldmath $\phi$})\psi\right)
&\ge&\left(\psi,{\cal H}_0(\mbox{\boldmath $\phi$})\psi\right)
-\frac{\sqrt{2}e}{\sqrt{m_e}}\left\Vert|{\bf A}_{\rm P}|\right\Vert_\infty
\sqrt{\left(\psi,{\cal H}_0(\mbox{\boldmath $\phi$})\psi\right)}
-\left\Vert W^-\right\Vert_\infty.
\label{lowerboundexpectation}
\end{eqnarray}
Let us denote by ${\cal E}_{n,+}^{\rm edge}$ and ${\cal E}_{n,-}^{\rm edge}$ 
the upper and lower edges of the Landau band with the index $n$, respectively. 
{From} the standard argument about the min-max principle,\footnote{See, for 
example, Section~XIII.1 of the book \cite{RSIV} by M.~Reed and B.~Simon.}
one has 
\begin{equation}
{\cal E}_{n,+}^{\rm edge}\le (n+1/2)\hbar\omega_c
+\frac{\sqrt{2}e}{\sqrt{m_e}}\left\Vert|{\bf A}_{\rm P}|\right\Vert_\infty
\sqrt{(n+1/2)\hbar\omega_c}+\frac{e^2}{2m_e}
\left\Vert|{\bf A}_{\rm P}|\right\Vert_\infty^2+\left\Vert W^+\right\Vert_\infty
\label{Eedge+}
\end{equation}
for $n=0,1,2,\ldots$. For the lower edge, we assume 
\begin{equation}
\frac{e}{\sqrt{2m_e}}\left\Vert|{\bf A}_{\rm P}|\right\Vert_\infty
\le \sqrt{\frac{1}{2}\hbar\omega_c}.
\label{monotone}
\end{equation}
Then the right-hand side of the bound (\ref{lowerboundexpectation}) 
is a strictly monotone increasing function of the expectation
$(\psi,{\cal H}_0(\mbox{\boldmath $\phi$})\psi)$. Therefore, the same argument yields 
\begin{equation}
{\cal E}_{n,-}^{\rm edge}\ge (n+1/2)\hbar\omega_c
-\frac{\sqrt{2}e}{\sqrt{m_e}}\left\Vert|{\bf A}_{\rm P}|\right\Vert_\infty
\sqrt{(n+1/2)\hbar\omega_c}-\left\Vert W^-\right\Vert_\infty 
\end{equation}
for $n=0,1,2,\ldots$. If this right-hand side with the index $n+1$ is strictly 
larger than the right-hand side of (\ref{Eedge+}) with the index $n$, 
then there exists a spectral gap above the Landau band with the index $n$, i.e., 
${\cal E}_{n+1,-}^{\rm edge}>{\cal E}_{n,+}^{\rm edge}$. This gap condition 
can be written as the desired form (\ref{nonintgapcondition}) with $\ell=n+1$. 
Clearly the condition (\ref{nonintgapcondition}) is stronger than the above 
(\ref{monotone}) for the vector potential ${\bf A}_{\rm P}$. 
Therefore we have no need to take 
into account the condition (\ref{monotone}).

Next we check that the corrections 
$\delta\sigma_{sy}(t;\mbox{\boldmath $\phi$},\eta,T)$ 
in (\ref{sigmatotxy0}) and (\ref{sigmatotyy0}) to the dominant parts 
of the linear response coefficients are negligibly small 
for the slow switching process. Because of the excitation energy gap   
above the ground state, one can easily prove 
the bound,
\begin{equation}
|\delta\sigma_{sy}(t;\mbox{\boldmath $\phi$},\eta,T)|
\le \frac{e^2}{h}\nu\left[({\cal C}_1+{\cal C}_2\omega_cT)e^{-\eta T}+
({\cal C}_3+{\cal C}_4\eta/\omega_c)\eta/\omega_c\right], 
\label{deltasigmasyboundnonint}
\end{equation}
by using 
the expressions (\ref{deltasigmasy0})--(\ref{M4}). 
Here all the positive constants ${\cal C}_j$, $j=1,2,3,4$, are independent 
of the system sizes $L_x,L_y$ and of the number $N$ of the electrons for 
a fixed filling factor $\nu$ of the electrons.

In the rest of this section, we derive two bounds, 
(\ref{fscsigmaxy}) and (\ref{fscgammasy}), in Theorem~\ref{theorem:noninteracting}.
To begin with, we rewrite the Hall conductance 
$\sigma_{xy}(\mbox{\boldmath $\phi$})$ of 
(\ref{sigmaxy}) as 
\begin{equation}
\sigma_{xy}(\mbox{\boldmath $\phi$})=-\frac{i\hbar e^2}{L_xL_y}\ 
\sum_{m,n:{\cal E}_m\le {\cal E}_{\rm F}<{\cal E}_n}\ 
\left[\frac{\left\langle\psi_m,v_x^{(0)}\psi_n\right\rangle
\left\langle\psi_n,v_y^{(0)}\psi_m
\right\rangle}{({\cal E}_m-{\cal E}_n)^2}-(x\leftrightarrow y)\right]
\label{ssigmaxyphi}
\end{equation}
in terms of the eigenvector $\psi_n$ of 
the single electron Hamiltonian ${\cal H}(\mbox{\boldmath $\phi$})$ of (\ref{hamphiAW}) 
with the energy eigenvalue ${\cal E}_n$. 
Here ${\cal E}_{\rm F}$ is the Fermi energy, and 
\begin{equation}
{\bf v}^{(0)}=\frac{1}{m_e}\left[{\bf p}+e{\bf A}({\bf r})\right].
\end{equation}
We have dropped $\mbox{\boldmath $\phi$}$ in ${\bf v}^{(0)}$ by relying on 
the orthogonality between $\psi_m$ and $\psi_n$ with ${\cal E}_m\ne {\cal E}_n$. 
Further we introduce some operators as follows: 
\begin{equation}
Q({\cal E}\le{\cal E}_{\rm F};\mbox{\boldmath $\phi$})
=\frac{1}{2\pi i}\int_\gamma R(z;\mbox{\boldmath $\phi$})dz 
\quad
\mbox{with the resolvent} \quad 
R(z;\mbox{\boldmath $\phi$})=\frac{1}{z-{\cal H}(\mbox{\boldmath $\phi$})}
\end{equation}
and 
\begin{equation}
Q_i({\cal E}\le{\cal E}_{\rm F};\mbox{\boldmath $\phi$})
=\frac{1}{2\pi i}\int_\gamma R(z;\mbox{\boldmath $\phi$})
v_i^{(0)}R(z;\mbox{\boldmath $\phi$})dz,
\end{equation}
where we have chosen the closed path $\gamma$ so that 
the operator $Q({\cal E}\le{\cal E}_{\rm F};\mbox{\boldmath $\phi$})$ is 
the projection onto the subspace spanned by all the levels below the Fermi energy  
${\cal E}_{\rm F}$. Here the path $\gamma$ is taken to be independent of 
the gauge parameters $\mbox{\boldmath $\phi$}$ by the gap condition (\ref{gapcondition}). 
Note that 
\begin{eqnarray}
& &{\rm Tr}\left[Q({\cal E}\le{\cal E}_{\rm F};\mbox{\boldmath $\phi$})
Q_x({\cal E}\le{\cal E}_{\rm F};\mbox{\boldmath $\phi$})
Q_y({\cal E}\le{\cal E}_{\rm F};\mbox{\boldmath $\phi$})\right]\ret
&=&\sum_{n:{\cal E}_n\le{\cal E}_{\rm F}}
\left\langle\psi_n,Q_x({\cal E}\le{\cal E}_{\rm F};\mbox{\boldmath $\phi$})
Q_y({\cal E}\le{\cal E}_{\rm F};\mbox{\boldmath $\phi$})\psi_n\right\rangle\ret
&=&\sum_{n:{\cal E}_n\le{\cal E}_{\rm F}}
\left\langle\psi_n,
\frac{1}{2\pi i}\int_\gamma \frac{1}{z_1-{\cal E}_n}v_x^{(0)}
\frac{1}{z_1-{\cal H}(\mbox{\boldmath $\phi$})}dz_1
\frac{1}{2\pi i}\int_\gamma \frac{1}{z_2-{\cal H}(\mbox{\boldmath $\phi$})}v_y^{(0)}
\frac{1}{z_2-{\cal E}_n}dz_2\psi_n\right\rangle\ret
&=&\sum_{n:{\cal E}_n\le{\cal E}_{\rm F}}\sum_m
\left\langle\psi_n,v_x^{(0)}\psi_m\right\rangle
\left\langle\psi_m,v_y^{(0)}\psi_n\right\rangle
\left(\frac{1}{2\pi i}\int_\gamma \frac{1}{z-{\cal E}_n}
\frac{1}{z-{\cal E}_m}dz\right)^2\ret
&=&\sum_{n:{\cal E}_n\le{\cal E}_{\rm F}}
\sum_{m:{\cal E}_m>{\cal E}_{\rm F}}
\left\langle\psi_n,v_x^{(0)}\psi_m\right\rangle
\left\langle\psi_m,v_y^{(0)}\psi_n\right\rangle
\frac{1}{\left({\cal E}_n-{\cal E}_m\right)^2}.
\end{eqnarray}
Combining this with the expression (\ref{ssigmaxyphi}), 
the Hall conductance $\sigma_{xy}(\mbox{\boldmath $\phi$})$ can be rewritten as 
\begin{equation}
\sigma_{xy}(\mbox{\boldmath $\phi$})=-\frac{i\hbar e^2}{L_xL_y}
{\rm Tr}\left[Q({\cal E}\le{\cal E}_{\rm F};\mbox{\boldmath $\phi$})
\left\{Q_x({\cal E}\le{\cal E}_{\rm F};\mbox{\boldmath $\phi$})
Q_y({\cal E}\le{\cal E}_{\rm F};\mbox{\boldmath $\phi$})-(x\leftrightarrow y)
\right\}\right].
\label{ssigmaxyphi2}
\end{equation}

\begin{theorem}
\label{diffsigmaxy}
The following inequality is valid: 
\begin{equation}
\left|\sigma_{xy}(\mbox{\boldmath $\phi$}+\delta\mbox{\boldmath $\phi$})
-\sigma_{xy}(\mbox{\boldmath $\phi$})\right|
\le{\cal C}\max_{i=x,y}|\delta\phi_i|,
\end{equation}
where ${\cal C}$ is a positive constant which is independent of 
the system sizes $L_x,L_y$ and of the number $N$ of the electrons. 
\end{theorem}
This Theorem follows from the next Proposition~\ref{pro:Trdiff} with 
the expression (\ref{ssigmaxyphi2}) of $\sigma_{xy}(\mbox{\boldmath $\phi$})$. 

\begin{pro}
\label{pro:Trdiff}
The following bound is valid: 
\begin{eqnarray}
& &\left|{\rm Tr}\left[Q({\cal E}\le{\cal E}_{\rm F};\mbox{\boldmath $\phi$}
+\delta\mbox{\boldmath $\phi$})
Q_i({\cal E}\le{\cal E}_{\rm F};\mbox{\boldmath $\phi$}+\delta\mbox{\boldmath $\phi$})
Q_j({\cal E}\le{\cal E}_{\rm F};\mbox{\boldmath $\phi$}
+\delta\mbox{\boldmath $\phi$})\right]\right.\ret
& &-\left.{\rm Tr}\left[Q({\cal E}\le{\cal E}_{\rm F};\mbox{\boldmath $\phi$})
Q_i({\cal E}\le{\cal E}_{\rm F};\mbox{\boldmath $\phi$})
Q_j({\cal E}\le{\cal E}_{\rm F};\mbox{\boldmath $\phi$})\right]\right|
\le{\cal C}N\max_{\ell=x,y}|\delta\phi_\ell|
\label{TrdiffQ}
\end{eqnarray}
for $i,j=x,y$. Here ${\cal C}$ is a positive constant which is independent of 
the system sizes $L_x,L_y$ and of the number $N$ of the electrons. 
\end{pro}
The proof is given in Appendix~\ref{appendix:pro:Trdiff}. 

Fix $\mbox{\boldmath $\phi$}_0\in [0,\Delta\phi_x]\times[0,\Delta\phi_y]$. Then we have 
\begin{equation}
\sigma_{xy}(\mbox{\boldmath $\phi$}_0)-\overline{\sigma_{xy}(\mbox{\boldmath $\phi$})}
=\frac{1}{\Delta\phi_x\Delta\phi_y}\int_0^{\Delta\phi_x}d\phi_x
\int_0^{\Delta\phi_y}d\phi_y\left[\sigma_{xy}(\mbox{\boldmath $\phi$}_0)
-\sigma_{xy}(\mbox{\boldmath $\phi$})\right]. 
\end{equation}
Using Theorem~\ref{diffsigmaxy}, the difference can be evaluated as 
\begin{equation}
\left|\sigma_{xy}(\mbox{\boldmath $\phi$}_0)
-\overline{\sigma_{xy}(\mbox{\boldmath $\phi$})}\right|
\le {\cal C}\max\left\{L_x^{-1},L_y^{-1}\right\}. 
\end{equation}
As shown in Section~\ref{Sec:QAHC}, the averaged Hall conductance is quantized 
as $\overline{\sigma_{xy}(\mbox{\boldmath $\phi$})}=-e^2p/h$ 
because the present ground state 
is non-degenerate, i.e., $q=1$. Hence we obtain 
\begin{equation}
\left|\sigma_{xy}(\mbox{\boldmath $\phi$}_0)+\frac{e^2}{h}p\right|
\le {\cal C}\max\left\{L_x^{-1},L_y^{-1}\right\} 
\end{equation}
with an integer $p$. 

In order to determine the integer $p$, consider the Hamiltonian 
\begin{equation}
{\cal H}(\mbox{\boldmath $\phi$},\mbox{\boldmath $\lambda$})=\frac{1}{2m_e}
\left[{\bf p}+e{\bf A}_0({\bf r})+e\lambda_A{\bf A}_{\rm P}({\bf r})
+\mbox{\boldmath $\phi$}\right]^2+\lambda_WW({\bf r})
\end{equation}
with the parameters $\mbox{\boldmath $\lambda$}=(\lambda_A,\lambda_W)\in[0,1]^2$. 
This Hamiltonian ${\cal H}(\mbox{\boldmath $\phi$},\mbox{\boldmath $\lambda$})$ 
continuously connects ${\cal H}(\mbox{\boldmath $\phi$})$ 
with ${\cal H}_0(\mbox{\boldmath $\phi$})$ 
of (\ref{hamsingleLandau}) by varying $\mbox{\boldmath $\lambda$}$ from $(1,1)$ to 
$(0,0)$ continuously. 
Then, in the same way as in the proof of Proposition~\ref{pro:Trdiff}, 
one can prove that the averaged Hall 
conductance $\overline{\sigma_{xy}(\mbox{\boldmath $\phi$})}$ for 
the Hamiltonian ${\cal H}(\mbox{\boldmath $\phi$},\mbox{\boldmath $\lambda$})$ 
is a continuous function 
of the parameters $\mbox{\boldmath $\lambda$}$. As shown in Section~\ref{singleLandau}, 
$\overline{\sigma_{xy}(\mbox{\boldmath $\phi$})}=-e^2\ell/h$ for $\lambda=(0,0)$. 
Therefore the integer $p$ must be equal to the filling factor $\ell$ 
of the Landau levels. Consequently we obtain the bound (\ref{fscsigmaxy}). 

Next consider the acceleration coefficients $\gamma_{sy}(\mbox{\boldmath $\phi$})$. 
In terms of the projection operator $Q({\cal E}\le{\cal E}_F;\mbox{\boldmath $\phi$})$, 
these can be expressed as 
\begin{eqnarray}
\gamma_{sy}(\mbox{\boldmath $\phi$})&=&\frac{e^2}{L_xL_y}
\frac{\partial}{\partial\phi_y}{\rm Tr}\ v_s^{(0)}(\mbox{\boldmath $\phi$})
Q({\cal E}\le{\cal E}_F;\mbox{\boldmath $\phi$})\ret
&=&\frac{e^2}{L_xL_y}\left[{\rm Tr}\ \frac{1}{m_e}\delta_{s,y}
Q({\cal E}\le{\cal E}_F;\mbox{\boldmath $\phi$})
+{\rm Tr}\ v_s^{(0)}(\mbox{\boldmath $\phi$})
Q_y({\cal E}\le{\cal E}_F;\mbox{\boldmath $\phi$})\right]\ret
&=&\frac{e^2}{L_xL_y}\left[\frac{N}{m_e}\delta_{s,y}
+{\rm Tr}\ v_s^{(0)}(\mbox{\boldmath $\phi$})
Q_y({\cal E}\le{\cal E}_F;\mbox{\boldmath $\phi$})\right]. 
\label{ano2sigmasytphi}
\end{eqnarray}

\begin{theorem}
\label{tildesigmasydiff}
The following inequality is valid: 
\begin{equation}
\left|\gamma_{sy}(\mbox{\boldmath $\phi$}+\delta\mbox{\boldmath $\phi$})
-\gamma_{sy}(\mbox{\boldmath $\phi$})\right|
\le {\cal C}\max_{i=x,y}|\delta\phi_i|,
\end{equation}
where the positive constant ${\cal C}$ is independent of the time $t$, 
the number $N$ of the electrons, and of the system sizes $L_x,L_y$. 
\end{theorem}
Owing to the expression (\ref{ano2sigmasytphi}), this theorem follows 
from the proposition: 

\begin{pro}
\label{proTrvQydiff}
The following inequality is valid: 
\begin{equation}
\left|{\rm Tr}\ v_s^{(0)}(\mbox{\boldmath $\phi$}+\delta\mbox{\boldmath $\phi$})
Q_y({\cal E}\le{\cal E}_F;\mbox{\boldmath $\phi$}+\delta\mbox{\boldmath $\phi$})
-{\rm Tr}\ v_s^{(0)}(\mbox{\boldmath $\phi$})Q_y({\cal E}
\le{\cal E}_F;\mbox{\boldmath $\phi$})\right|\le {\cal C}N\max_{i=x,y}|\delta\phi_i|,
\end{equation}
where the positive constant ${\cal C}$ is independent of the time $t$, 
the number $N$ of the electrons, and of the system sizes $L_x,L_y$. 
\end{pro}
The proof is given in Appendix~\ref{pproTrvQydiff}. 

Fix $\mbox{\boldmath $\phi$}_0\in[0,\Delta\phi_x]\times[0,\Delta\phi_y]$. Then 
\begin{equation}
\gamma_{sy}(\mbox{\boldmath $\phi$}_0)-\overline{\gamma_{sy}(\mbox{\boldmath $\phi$})}
=\frac{1}{\Delta\phi_x\Delta\phi_y}\int_0^{\Delta\phi_x}d\phi_x
\int_0^{\Delta\phi_y}d\phi_y\left[\gamma_{sy}(\mbox{\boldmath $\phi$}_0)
-\gamma_{sy}(\mbox{\boldmath $\phi$})\right]. 
\end{equation}
Using Theorem~\ref{tildesigmasydiff} and the fact 
$\overline{\gamma_{sy}(\mbox{\boldmath $\phi$})}=0$ which was shown 
in Section~\ref{Sec:QAHC}, we have the bound (\ref{fscgammasy}). 

%%%%%%%%%%%%%%%%%%%%%%%%%%%%%%%%%%%%%%%%%%%%%%%%%%%%%%%%%%%%%%%%%%%%
\Section{The interacting case}
\label{IntCase}

In this section we study the interacting case in detail. 
In Sec.~\ref{Sec:QAHC}, we obtained that the averaged Hall conductance 
is quantized as 
\begin{equation}
\overline{\sigma_{xy}(\mbox{\boldmath $\phi$})}=-\frac{e^2}{h}\frac{p}{q}
\label{sigmaxyfraction1}
\end{equation}
with integers $p,q$. Here the integer $q$ is the dimension 
of the sector of the ground state. 
Since the Hall conductance $\sigma_{xy}(\mbox{\boldmath $\phi$})$ is a continuous 
function of the gauge parameters $\mbox{\boldmath $\phi$}$, 
we can find special gauge parameters 
$\mbox{\boldmath $\phi$}_0\in{\cal T}_{\rm g}$ satisfying 
\begin{equation}
\sigma_{xy}(\mbox{\boldmath $\phi$}_0)=-\frac{e^2}{h}\frac{p}{q}, 
\label{sigmaxyfraction2}
\end{equation}
by using the mean value theorem about integration. Namely the Hall 
conductance is exactly quantized for the special value 
$\mbox{\boldmath $\phi$}=\mbox{\boldmath $\phi$}_0$ 
of the gauge parameters. But we cannot necessarily expect 
the same exact quantization for general fixed values of the gauge parameters 
in ${\cal T}_{\rm g}$. 
Besides, the value $\mbox{\boldmath $\phi$}_0$ may strongly depend on the potentials, 
the system sizes $L_x,L_y$ and on the number $N$ of the electrons 
although $\mbox{\boldmath $\phi$}_0$ tends to zero 
as $L_x,L_y\rightarrow\infty$.\footnote{
The space ${\cal T}_{\rm g}$ itself contracts into the single point $(0,0)$ 
in the infinite volume limit from the definition (\ref{defTg}) 
of the gauge torus ${\cal T}_{\rm g}$.}

As we treated the non-interacting case 
in the preceding section, we want to resolve the following two issues 
for the interacting case: 
\begin{itemize}
\item Estimating the finite size correction 
for $\sigma_{xy}(\mbox{\boldmath $\phi$})$ for 
fixed gauge parameters $\mbox{\boldmath $\phi$}\ne\mbox{\boldmath $\phi$}_0$.
\item Elucidating the relation between the integer $p$ and the filling factor $\nu$ 
of the Landau levels. 
\end{itemize}
Unfortunately we cannot estimate the finite size correction because of 
certain technical difficulty. Therefore we focus on the second issue only 
in this paper. 

%%%%%%%%%%%%%%%%%%%%%%%%%%%%%%%%%%%%%%%%%%%%%%%%%%%%%%%%%%%%%%%%%%%%%%
\subsection{Boundedness of the Hall conductance $\sigma_{xy}(\mbox{\boldmath $\phi$})$}

Firstly we prove that the Hall conductance $\sigma_{xy}(\mbox{\boldmath $\phi$})$ is 
uniformly bounded in the number $N$ of the electrons and the system sizes 
$L_x,L_y$ for any fixed filling factor $\nu$ under the assumptions below. 
If the dimension $q$ of the sector of the ground state 
is also uniformly bounded in $L_x,L_y,N$ in addition to the boundedness 
of $\sigma_{xy}(\mbox{\boldmath $\phi$})$, then 
there exist a sequence $\{(L_x^{(i)},L_y^{(i)})\}_i$ 
of the system sizes going to infinity 
and two integers $p^{(\infty)},q^{(\infty)}$ such that 
\begin{equation}
\sigma_{xy}^{(\infty)}:=\lim_{L_x^{(i)},L_y^{(i)}\rightarrow\infty}
\sigma_{xy}(\mbox{\boldmath $\phi$}_0)=-\frac{e^2}{h}\frac{p^{(\infty)}}{q^{(\infty)}}. 
\end{equation}
Namely the quantization of the Hall conductance occurs 
in the infinite volume limit although the number $p^{(\infty)}/q^{(\infty)}$ 
may be equal to an integer. Unfortunately we cannot determine the explicit 
values of the integers $p^{(\infty)}$ and $q^{(\infty)}$ 
for the infinite volume ground state for a given filling factor $\nu$. 

In order to prove the boundedness of 
the Hall conductance $\sigma_{xy}(\mbox{\boldmath $\phi$})$, we need some technical 
assumptions: 

\begin{assumption}
\label{assumption:dif1}
The the electrostatic potential $W$ and 
the electron-electron interaction $W^{(2)}$ of the present model satisfy 
higher differentiability as 
$W\in C^2({\bf R}^2)$ and $W^{(2)}\in C^1({\bf R}^2)$. 
The norms,
\begin{equation}
\left\Vert \frac{\partial^2 W}{\partial x^2}\right\Vert_\infty\quad 
\mbox{and}\quad 
\left\Vert \frac{\partial^2 W}{\partial y^2}\right\Vert_\infty,
\end{equation}
are bounded uniformly in the sizes $L_x,L_y$ of the system. 
\end{assumption}

\begin{theorem}
\label{theorem:bounded1}
Suppose ${\bf A}_{\rm P}=0$, and require Assumption~\ref{assumption:dif1} 
in addition to the assumptions (including Assumption~\ref{gapassumption}) 
in Sec.~\ref{modelresults}. Then the 
the Hall conductance 
$\sigma_{xy}(\mbox{\boldmath $\phi$})$ of (\ref{sigmaxy}) is uniformly bounded 
in the number $N$ of the electrons 
and in the system sizes $L_x,L_y$ for any fixed filling factor $\nu$. 
\end{theorem} 
The proof is given in Appendix~\ref{Boundsigmaxy}. In the same way, we can prove 
the boundedness of the acceleration coefficients 
$\gamma_{sy}(\mbox{\boldmath $\phi$})$ of (\ref{gammaxy}) 
and (\ref{gammayy}), and can get the bound,  
\begin{equation}
|\delta\sigma_{sy}(t;\mbox{\boldmath $\phi$},\eta,T)|
\le \frac{e^2}{h}\nu\left[({\cal C}_1+{\cal C}_2\omega_cT)e^{-\eta T}+
({\cal C}_3+{\cal C}_4\eta/\omega_c)\eta/\omega_c\right],
\label{deltasigmasybound2}
\end{equation}
for the corrections $\delta\sigma_{sy}(t;\mbox{\boldmath $\phi$},\eta,T)$ 
to the dominant parts of the linear response coefficients 
(\ref{sigmatotxy0}) and (\ref{sigmatotyy0}) 
under the same assumptions. 
Here all the positive constants ${\cal C}_j$, $j=1,2,3,4$, are independent 
of the system sizes $L_x,L_y$ and of the number $N$ of the electrons for 
a fixed filling factor $\nu$ of the electrons.   

Next consider the case with ${\bf A}_{\rm P}\ne 0$. 
We write the $z$ component of the magnetic field for the 
vector potential ${\bf A}_{\rm P}$ as 
\begin{equation}
B_{{\rm P},z}=\frac{\partial A_{{\rm P},y}}{\partial x}
-\frac{\partial A_{{\rm P},x}}{\partial y}.
\end{equation}

\begin{assumption}
\label{assumption:difB1}
The magnetic field $B_{{\rm P},z}$ for the vector potential ${\bf A}_{\rm P}$ 
satisfies $B_{{\rm P},z}\in C^2({\bf R}^2)$, and the norms, 
\begin{equation}
\left\Vert B_{{\rm P},z}\right\Vert_\infty, 
\left\Vert \frac{\partial B_{{\rm P},z}}{\partial x}\right\Vert_\infty 
\quad \mbox{and}\quad 
\left\Vert \frac{\partial B_{{\rm P},z}}{\partial y}\right\Vert_\infty, 
\end{equation}
are bounded uniformly in the sizes $L_x,L_y$ of the system. 
\end{assumption}

\begin{assumption}
\label{assumption:decayW2}
The electron-electron interaction $W^{(2)}$ 
is non-negative (repulsive) and satisfies the decay condition, 
\begin{equation}
W^{(2)}(x,y)\le W^{(2)}_0\left\{1+[{\rm dist}(x,y)/r_0]^2\right\}^{-\gamma/2}
\ \ \mbox{with the constants}\ W^{(2)}_0>0,\gamma>2,r_0>0,
\label{decayconditionW2}
\end{equation}
where the distance is given by 
\begin{equation}
{\rm dist}(x,y):=\sqrt{\min_{m\in{\bf Z}}|x-mL_x|^2
+\min_{n\in{\bf Z}}|y-nL_y|^2}.
\end{equation}
\end{assumption}

For the case with ${\bf A}_{\rm P}\ne 0$, we have the following theorem: 

\begin{theorem}
\label{theorem:bounded2}
Require Assumptions~\ref{assumption:dif1}, 
\ref{assumption:difB1} and \ref{assumption:decayW2} in 
addition to the assumptions (including Assumption~\ref{gapassumption}) 
in Sec.~\ref{modelresults}. Then the Hall conductance 
$\sigma_{xy}(\mbox{\boldmath $\phi$})$ of (\ref{sigmaxy}) is uniformly bounded 
in the number $N$ of the electrons 
and in the system sizes $L_x,L_y$ for any fixed filling factor $\nu$ of 
the electrons. 
\end{theorem} 
The proof is given in Appendix~\ref{Boundsigmaxy}. Under the same assumptions, 
we can also prove the boundedness of the acceleration coefficients 
$\gamma_{sy}(\mbox{\boldmath $\phi$})$, and can get 
the same bound for $\delta\sigma_{sy}(t,\mbox{\boldmath $\phi$};\eta,T)$ 
as (\ref{deltasigmasybound2}) with different constants ${\cal C}_j$. 

%%%%%%%%%%%%%%%%%%%%%%%%%%%%%%%%%%%%%%%%%%%%%%%%%%%
\subsection{Fractional quantization of the Hall conductance 
$\sigma_{xy}(\mbox{\boldmath $\phi$})$}
\label{FQHE}

The rational number $p/q$ in the right-hand side of 
(\ref{sigmaxyfraction1}) or (\ref{sigmaxyfraction2}) 
may be equal to an integer, i.e., the number $p$ may equal 
a multiple of $q$. But, from the result (\ref{trinvsxy}), we can expect 
that the Hall conductance exhibits purely fractional quantization as 
\begin{equation}
\sigma_{xy}(\mbox{\boldmath $\phi$}_0)=-\frac{e^2}{h}\nu 
\end{equation}
for a fractional filling factor $\nu=p/q\notin{\bf N}$ when a spectral 
gap exists above the sector of the ground state with weak disorder. 
Next we shall show that this expectation holds under certain assumptions. 

When the system is translationally invariant in one direction \cite{AY}, 
we can obtain the desired result as follows: 

\begin{theorem}
\label{1dtranslationinv}
In addition to the assumptions (including Assumption~\ref{gapassumption}) 
in Sec.~\ref{modelresults}, 
assume that the electrostatic potential $W$ is a function of the 
single variable $x$ only or $y$ only as $W(x)$ or $W(y)$, and assume 
${\bf A}_{\rm P}=0$, $W\in C^{1}({\bf R})$ and $W^{(2)}\in C^1({\bf R}^2)$. 
Then there exist gauge parameters $\mbox{\boldmath $\phi$}_0\in{\cal T}_{\rm g}$ 
such that 
\begin{equation}
\sigma_{xy}(\mbox{\boldmath $\phi$}_0)=-\frac{e^2}{h}\frac{p}{q}
\end{equation}
for the fractional filling factor $\nu=p/q$ of the Landau levels. 
\end{theorem}
The proof is given in Appendix~\ref{fractionalproofs}. 

In order to proceed to more generic situations, we write 
\begin{equation}
D^{(m,n)}:=\frac{\partial^{m+n}}{\partial x^m\partial y^n} 
\end{equation}
for non-negative integers $m,n$. 

\begin{assumption}
\label{assumption:difW2}
The electrostatic potential $W$ and the electron-electron interaction 
$W^{(2)}$ satisfy higher differentiability as 
$W\in C^3({\bf R}^2)$ and $W^{(2)}\in C^1({\bf R}^2)$. 
\end{assumption}

\begin{theorem}
\label{fractionTheorem1}
Suppose ${\bf A}_{\rm P}=0$, i.e., $B_{{\rm P},z}=0$, and require 
Assumption~\ref{assumption:difW2} in addition to 
the assumptions (including Assumption~\ref{gapassumption}) 
in Sec.~\ref{modelresults}. Then the Hall conductance 
$\overline{\sigma_{xy}(\mbox{\boldmath $\phi$})}$ averaged over 
the gauge parameters $\mbox{\boldmath $\phi$}$ 
satisfies the bound, 
\begin{equation}
-\frac{e^2}{h}\nu(1+\delta)\le \overline{\sigma_{xy}(\mbox{\boldmath $\phi$})}
\le -\frac{e^2}{h}\nu(1-\delta),
\label{fractionalbound2} 
\end{equation}
with 
\begin{equation}
\delta=2\ell_B^4\frac{\hbar\omega_c}{(\Delta E)^3}
\max_{m+n=2}
\left\Vert D^{(m,n)}W\right\Vert_\infty^2.
\label{defdelta} 
\end{equation}
Here $\nu=N/M$ is the fixed filling factor of the electrons, and 
the definition of the energy gap $\Delta E$ is given 
in Assumption~\ref{gapassumption}. 
\end{theorem}
The proof is given in Appendix~\ref{fractionalproofs}. 
We can obtain a similar bound for the non-averaged Hall 
conductance $\sigma_{xy}(\mbox{\boldmath $\phi$})$ 
to (\ref{fractionalbound2}). See Appendix~\ref{fractionalproofs}. 

\begin{coro}
\label{fractioncoro}
Under the same assumptions as in the above theorem, there exist 
gauge parameters $\mbox{\boldmath $\phi$}_0\in{\cal T}_{\rm g}$ such that 
\begin{equation}
\sigma_{xy}(\mbox{\boldmath $\phi$}_0)=-\frac{e^2}{h}\frac{p}{q}
\end{equation}
with the integers $p,q$ satisfying 
\begin{equation}
\nu(1-\delta)\le \frac{p}{q} \le \nu(1+\delta) 
\label{fractionalbound}
\end{equation}
with the same $\delta$ as in the theorem. 
\end{coro}
Consider again the situation that the interval $[\nu(1-\delta),\nu(1+\delta)]$ 
does not include any integer. Then the number $p/q$ 
must be equal to purely a fractional number, i.e., a non-integer. 
In addition to this condition, if the number $\delta$ and 
the dimension $q$ of the sector 
of the ground state are uniformly bounded in the sizes $L_x,L_y$ 
of the system, then there exist a sequence $\{(L_x^{(i)},L_y^{(i)})\}_i$ 
of the system sizes going to infinity 
and two integers $p^{(\infty)},q^{(\infty)}$ such that 
\begin{equation}
\sigma_{xy}^{(\infty)}:=\lim_{L_x^{(i)},L_y^{(i)}\rightarrow\infty}
\sigma_{xy}(\mbox{\boldmath $\phi$}_0)
=-\frac{e^2}{h}\frac{p^{(\infty)}}{q^{(\infty)}}, 
\end{equation}
and that $\sigma_{xy}^{(\infty)}$ satisfies the bound derived from 
(\ref{fractionalbound}). 
Therefore the number $p^{(\infty)}/q^{(\infty)}$ equals a purely fractional 
number also in the infinite volume limit. 

In order to get a similar bound for the Hall conductance 
in the case with the vector potential ${\bf A}_{\rm P}\ne 0$, 
we need stronger assumptions as follows: 

\begin{assumption}
\label{assumption:dif3}
The magnetic field $B_{{\rm P},z}$ for the vector potential 
${\bf A}_{\rm P}$ and the electrostatic potential $W$ 
of the present model satisfy higher differentiability as 
$B_{{\rm P},z}\in C^4({\bf R}^2)$ and $W \in C^4({\bf R}^2)$. 
\end{assumption}

\begin{assumption}
\label{assumption:W23}
The electron-electron interaction 
$W^{(2)}$ of the present model satisfies 
$W^{(2)}\in C^2({\bf R}^2)$, and the following two conditions: 
\begin{equation}
\ell_B\left[\left|\left(\frac{\partial W^{(2)}}{\partial x}
\right)({\bf r})\right|+
\left|\left(\frac{\partial W^{(2)}}{\partial y}
\right)({\bf r})\right|\right]
\le \alpha_{\rm int}
W^{(2)}({\bf r})
\label{assumW21}
\end{equation}
and 
\begin{equation}
\ell_B^2\left[\left|\left(\frac{\partial^2 W^{(2)}}{\partial x^2}
\right)({\bf r})\right|+
\left|\left(\frac{\partial^2 W^{(2)}}{\partial y^2}
\right)({\bf r})\right|\right]
\le \alpha_{\rm int}
W^{(2)}({\bf r})
\label{assumW22}
\end{equation}
for any ${\bf r}$, with a positive constant $\alpha_{\rm int}$ 
which is independent of the sizes $L_x,L_y$ of the system. 
\end{assumption}

\begin{theorem}
\label{fractionTheorem2}
In addition to the assumptions (including 
Assumption~\ref{gapassumption}) in Sec.~\ref{modelresults}, 
we require Assumptions~\ref{assumption:decayW2}, \ref{assumption:dif3} 
and \ref{assumption:W23}. 
Then there exists a positive number $\delta$ such that 
$\delta$ is a continuous function 
of the norms, $\Vert D^{(k,\ell)}B_{{\rm P},z}\Vert_\infty$ for 
$k+\ell\le 4$ and $\Vert D^{(m,n)}W\Vert_\infty$ for $m+n\le 3$, 
and satisfies $\delta=0$ for the special point with ${\bf A}_{\rm P}=0$ and 
$W=0$, and that the Hall conductance 
$\overline{\sigma_{xy}(\mbox{\boldmath $\phi$})}$ satisfies the bound, 
\begin{equation}
-\frac{e^2}{h}\nu(1+\delta)\le \overline{\sigma_{xy}(\mbox{\boldmath $\phi$})}
\le \frac{e^2}{h}\nu(1-\delta),
\end{equation}
where $\nu=N/M$ is the fixed filling factor of the electrons. 
\end{theorem} 
The proof is given in Appendix~\ref{fractionalproofs}. 
Clearly we get the following corollary similar to 
Corollary~\ref{fractioncoro}: 

\begin{coro}
Under the same assumptions as in the above theorem, there exists 
gauge parameters $\mbox{\boldmath $\phi$}_0\in{\cal T}_{\rm g}$ such that 
\begin{equation}
\sigma_{xy}(\mbox{\boldmath $\phi$}_0)=-\frac{e^2}{h}\frac{p}{q}
\end{equation}
with integers $p,q$ satisfying 
\begin{equation}
\nu(1-\delta)\le \frac{p}{q}\le \nu(1+\delta) 
\end{equation}
with the same $\delta$ as in the above theorem. 
\end{coro}
For this case, we can also make the same remarks as those after 
Corollary~\ref{fractioncoro}. But we omit to make those remarks again.

%%%%%%%%%%%%%%%%%%%%%%%%%%%%%%%%%%%%%%%%%%%%%%%%%%%%%%%%%%%%%%%%%%%
\appendix
%%%%%%%%%%%%%%%%%%%%%%%%%%%%%%%%%%%%%%%%%%%%%%%%%%%%%%%%%%%%%%%%%%
\section{Differentiability of the ground-state wavefunctions}
\label{appendix:prodiffPhi}
\setcounter{equation}{0}
\setcounter{theorem}{0}

Following Kato,\footnote{See Sec.~4.2 of Chap.~II of 
his book \cite{Kato}.} we give the proof 
of Proposition~\ref{prodiffPhi} in this appendix. 
For this purpose, it is enough to construct an operator-valued 
function $U_{\rm g}(\mbox{\boldmath $\phi$})$ of the gauge 
parameters $\mbox{\boldmath $\phi$}\in{\cal T}_{\rm g}$ 
with the following two conditions: 
\begin{itemize}
\item The inverse $U_{\rm g}^{-1}(\mbox{\boldmath $\phi$})$ exists and both of 
$U_{\rm g}(\mbox{\boldmath $\phi$})$ and 
$U_{\rm g}^{-1}(\mbox{\boldmath $\phi$})$ are 
infinitely differentiable with respect to $\mbox{\boldmath $\phi$}$;
\item $U_{\rm g}(\mbox{\boldmath $\phi$})Q(E_0^{(N)}(0))
U_{\rm g}^{-1}(\mbox{\boldmath $\phi$})
=Q(E_0^{(N)}(\mbox{\boldmath $\phi$}))$.
\end{itemize}
{From} this second property, one obtain that, if the vectors 
${\hat \Phi}_{0,\mu}^{(N)}(0)$, $\mu=1,2,\ldots,q$, span 
the sector of the degenerate ground state for $\mbox{\boldmath $\phi$}=0$, then 
the vectors 
${\hat \Phi}_{0,\mu}^{(N)}(\mbox{\boldmath $\phi$})=
U_{\rm g}(\mbox{\boldmath $\phi$}){\hat \Phi}_{0,\mu}^{(N)}(0)$, 
$\mu=1,2,\ldots,q$, also span the sector of the degenerate ground state 
for any given $\mbox{\boldmath $\phi$}\in{\cal T}_{\rm g}$. 

We begin with the following lemma:

\begin{lemma}
\label{lemma:pRNnormbound}
Let $z\notin\sigma(H_0^{(N)}(\mbox{\boldmath $\phi$}))$, i.e., 
the complex number $z$ is not 
in the spectrum $\sigma(H_0^{(N)}(\mbox{\boldmath $\phi$}))$ of the Hamiltonian 
$H_0^{(N)}(\mbox{\boldmath $\phi$})$. Then 
\begin{eqnarray}
& &\left\Vert
\frac{1}{m_e}\left[p_{s,j}+eA_s({\bf r}_j)+\phi_s\right]
\frac{1}{z-H_0^{(N)}(\mbox{\boldmath $\phi$})}
\right\Vert\ret
&\le& \sqrt{\frac{2}{m_e{\rm dist}(z,\sigma(H_0^{(N)}(\mbox{\boldmath $\phi$})))}
\left[1+
\frac{|z|+N\Vert W\Vert_\infty+N(N-1)\Vert W^{(2)}\Vert_\infty/2}
{{\rm dist}(z,\sigma(H_0^{(N)}(\mbox{\boldmath $\phi$})))}
\right]}
\label{pRNnormbound}
\end{eqnarray}
for $s=x,y$ and $j=1,2,\ldots,N$. 
\end{lemma}
\begin{proof}{Proof}
Let $\Phi$ be an $N$ electrons vector with norm one. Then one has 
\begin{eqnarray}
& &\left\langle\Phi,\frac{1}{z^\ast-H_0^{(N)}(\mbox{\boldmath $\phi$})}
\left\{\frac{1}{m_e}[p_{s,j}+eA_s({\bf r}_j)+\phi_s]\right\}^2
\frac{1}{z-H_0^{(N)}(\mbox{\boldmath $\phi$})}\Phi\right\rangle\ret
&\le&\left\langle\Phi,\frac{1}{z^\ast-H_0^{(N)}(\mbox{\boldmath $\phi$})}
\frac{2}{m_e}\left[H_0^{(N)}(\mbox{\boldmath $\phi$})+N\Vert W\Vert_\infty
+\frac{N(N-1)}{2}\Vert W^{(2)}\Vert_\infty\right]
\frac{1}{z-H_0^{(N)}(\mbox{\boldmath $\phi$})}\Phi\right\rangle\ret
&=&\frac{2}{m_e}\sum_n 
\left|\left\langle\Phi,\Phi_n^{(N)}(\mbox{\boldmath $\phi$})\right\rangle\right|^2
\frac{1}{|z-E_n^{(N)}(\mbox{\boldmath $\phi$})|^2}
\left[E_n^{(N)}(\mbox{\boldmath $\phi$})+N\Vert W\Vert_\infty
+\frac{N(N-1)}{2}\Vert W^{(2)}\Vert_\infty\right],\ret
\label{vRNbound}
\end{eqnarray}
where $\Phi_n^{(N)}(\mbox{\boldmath $\phi$})$ are the eigenvectors 
of $H_0^{(N)}(\mbox{\boldmath $\phi$})$ with 
the eigenvalue $E_n^{(N)}(\mbox{\boldmath $\phi$})$ counting degenerate eigenvalues 
a number of times equal to their multiplicity. Note that 
\begin{eqnarray}
& &\frac{1}{|z-E_n^{(N)}(\mbox{\boldmath $\phi$})|^2}
\left[E_n^{(N)}(\mbox{\boldmath $\phi$})+N\Vert W\Vert_\infty
+\frac{N(N-1)}{2}\Vert W^{(2)}\Vert_\infty\right]\ret
&=&\frac{1}{|z-E_n^{(N)}(\mbox{\boldmath $\phi$})|^2}
\left[E_n^{(N)}(\mbox{\boldmath $\phi$})-z+z+N\Vert W\Vert_\infty
+\frac{N(N-1)}{2}\Vert W^{(2)}\Vert_\infty\right]\ret
&=&-\frac{1}{z^\ast-E_n^{(N)}(\mbox{\boldmath $\phi$})}+
\frac{1}{|z-E_n^{(N)}(\mbox{\boldmath $\phi$})|^2}\left[z+N\Vert W\Vert_\infty
+\frac{N(N-1)}{2}\Vert W^{(2)}\Vert_\infty\right]\ret
&\le&\frac{1}{{\rm dist}(z,\sigma(H_0^{(N)}(\mbox{\boldmath $\phi$})))}
+\frac{|z|+N\Vert W\Vert_\infty+N(N-1)\Vert W^{(2)}\Vert_\infty/2}
{{\rm dist}(z,\sigma(H_0^{(N)}(\mbox{\boldmath $\phi$})))^2}.
\end{eqnarray}
Substituting this into (\ref{vRNbound}), the desired bound 
(\ref{pRNnormbound}) is obtained. 
\end{proof}

For any $z\notin\sigma(H_0^{(N)}(\mbox{\boldmath $\phi$}))$, one has 
\begin{equation}
\frac{\partial}{\partial\phi_s}R^{(N)}(z;\mbox{\boldmath $\phi$})
=\sum_{j=1}^NR^{(N)}(z;\mbox{\boldmath $\phi$})
\frac{1}{m_e}[p_{s,j}+eA_s({\bf r}_j)+\phi_s]R^{(N)}(z;\mbox{\boldmath $\phi$}),
\end{equation}
where we have written 
\begin{equation}
R^{(N)}(z;\mbox{\boldmath $\phi$})=\frac{1}{z-H_0^{(N)}(\mbox{\boldmath $\phi$})}.
\end{equation}
Here, since the product 
$[p_{s,j}+eA_s({\bf r}_j)+\phi_s]R^{(N)}(z;\mbox{\boldmath $\phi$})$ of the two operators 
is bounded owing to the above Lemma~\ref{lemma:pRNnormbound}, the resolvent 
$R^{(N)}(z;\mbox{\boldmath $\phi$})$ is infinitely differentiable with respect 
to $\mbox{\boldmath $\phi$}$. 
Therefore, by the integral representation (\ref{intrepQphi}), 
the projection $Q(E_0^{(N)}(\mbox{\boldmath $\phi$}))$ is also infinitely differentiable 
with respect to $\mbox{\boldmath $\phi$}$. 

We introduce abbreviations as 
\begin{equation}
P(\mbox{\boldmath $\phi$})=Q(E_0^{(N)}(\mbox{\boldmath $\phi$})), 
\end{equation}
and 
\begin{equation}
P_s(\mbox{\boldmath $\phi$})=\frac{\partial}{\partial\phi_s}
Q(E_0^{(N)}(\mbox{\boldmath $\phi$})) 
\quad\mbox{for}\ s=x,y, 
\end{equation}
and write the commutators for them as 
\begin{equation}
F^{(s)}(\mbox{\boldmath $\phi$})=[P_s(\mbox{\boldmath $\phi$}),P(\mbox{\boldmath $\phi$})]. 
\label{defFsphi}
\end{equation}
Note that, for $s=x,y$, 
\begin{equation}
P_s(\mbox{\boldmath $\phi$})P(\mbox{\boldmath $\phi$})+
P(\mbox{\boldmath $\phi$})P_s(\mbox{\boldmath $\phi$})=P_s(\mbox{\boldmath $\phi$}),
\label{PsPPPs}
\end{equation}
and
\begin{equation}
P(\mbox{\boldmath $\phi$})P_s(\mbox{\boldmath $\phi$})P(\mbox{\boldmath $\phi$})=0. 
\label{productPPsP}
\end{equation}
These are equivalent to the equations (\ref{idQdiff}) 
and (\ref{QQsQ}), respectively. 
Combining the latter with (\ref{defFsphi}), one has 
\begin{equation}
P(\mbox{\boldmath $\phi$})F^{(s)}(\mbox{\boldmath $\phi$})=-P(\mbox{\boldmath $\phi$})
P_s(\mbox{\boldmath $\phi$}), \quad 
F^{(s)}(\mbox{\boldmath $\phi$})P(\mbox{\boldmath $\phi$})=
P_s(\mbox{\boldmath $\phi$})P(\mbox{\boldmath $\phi$}).
\end{equation}
Further, from these and (\ref{PsPPPs}), one obtains 
\begin{equation}
P_s(\mbox{\boldmath $\phi$})=[F^{(s)}(\mbox{\boldmath $\phi$}),P(\mbox{\boldmath $\phi$})].
\label{Pscommu}
\end{equation}

Now introduce the ordinary differential equation, 
\begin{equation}
\frac{d}{d\phi_x}X_+=F^{(x)}(\phi_x,0)X_+,
\label{ODE1+onGT}
\end{equation}
for the unknown operator-valued function $X_+=X_+(\phi_x)$ of $\phi_x$. 
Here we have fixed the argument $\phi_y$ to zero in $F^{(x)}$. 
Let $X_+=U_{\rm g}^{(x)}(\phi_x)$ be the unique solution satisfying 
the initial condition $U_{\rm g}^{(x)}(0)=1$. 
Since $F^{(x)}(\phi_x,0)$ is infinitely differentiable with respect to 
$\phi_x$, the solution $U_{\rm g}^{(x)}(\phi_x)$ is also infinitely 
differentiable. 
The existence and the infinite differentiability of the unique solution 
follow from the standard theory for differential equations. 
Further introduce the ordinary differential equation, 
\begin{equation}
\frac{d}{d\phi_x}X_-=-X_-F^{(x)}(\phi_x,0),
\label{ODE1-onGT}
\end{equation}
for the unknown operator-valued function $X_-=X_-(\phi_x)$. 
Let $X_-=V_{\rm g}^{(x)}(\phi_x)$ be the unique solution satisfying 
the initial condition $V_{\rm g}^{(x)}(0)=1$. 

Note that 
\begin{eqnarray}
& &\frac{d}{d\phi_x}
\left(V_{\rm g}^{(x)}(\phi_x)U_{\rm g}^{(x)}(\phi_x)\right)\ret
&=&\frac{d V_{\rm g}^{(x)}}{d\phi_x}(\phi_x)U_{\rm g}^{(x)}(\phi_x)
+V_{\rm g}^{(x)}(\phi_x)\frac{d U_{\rm g}^{(x)}}{d\phi_x}(\phi_x)\ret
&=&-V_{\rm g}^{(x)}(\phi_x)F^{(x)}(\phi_x,0)U_{\rm g}^{(x)}(\phi_x)
+V_{\rm g}^{(x)}(\phi_x)F^{(x)}(\phi_x,0)U_{\rm g}^{(x)}(\phi_x)=0. 
\end{eqnarray}
Hence $V_{\rm g}^{(x)}U_{\rm g}^{(x)}$ is a constant and 
\begin{equation}
V_{\rm g}^{(x)}(\phi_x)U_{\rm g}^{(x)}(\phi_x)
=V_{\rm g}^{(x)}(0)U_{\rm g}^{(x)}(0)=1
\quad\mbox{for all}\ \phi_x. 
\label{VU=1}
\end{equation}
Similarly one has 
\begin{eqnarray}
& &\frac{d}{d\phi_x}
\left(U_{\rm g}^{(x)}(\phi_x)V_{\rm g}^{(x)}(\phi_x)\right)\ret
&=&\frac{d U_{\rm g}^{(x)}}{d\phi_x}(\phi_x)V_{\rm g}^{(x)}(\phi_x)
+U_{\rm g}^{(x)}(\phi_x)\frac{d V_{\rm g}^{(x)}}{d\phi_x}(\phi_x)\ret
&=&F^{(x)}(\phi_x,0)U_{\rm g}^{(x)}(\phi_x)V_{\rm g}^{(x)}(\phi_x)
+U_{\rm g}^{(x)}(\phi_x)\left[-V_{\rm g}^{(x)}(\phi_x)
F^{(x)}(\phi_x,0)\right]\ret
&=&\left[F^{(x)}(\phi_x,0),U_{\rm g}^{(x)}(\phi_x)V_{\rm g}^{(x)}(\phi_x)
\right]. 
\end{eqnarray}
This is also an ordinary differential equation 
for $Z(\phi_x)=U_{\rm g}^{(x)}(\phi_x)V_{\rm g}^{(x)}(\phi_x)$ 
with the initial condition $Z(0)=U_{\rm g}^{(x)}(0)V_{\rm g}^{(x)}(0)=1$. 
Clearly $Z(\phi_x)=1$ satisfies the equation and the initial condition. 
Combining this with the uniqueness of the solution, the following relation 
must hold:
\begin{equation}
U_{\rm g}^{(x)}(\phi_x)V_{\rm g}^{(x)}(\phi_x)=1
\quad\mbox{for all}\ \phi_x. 
\end{equation}
By combining this with (\ref{VU=1}), one gets 
$V_{\rm g}^{(x)}=\left(U_{\rm g}^{(x)}\right)^{-1}$. 

Consider the operator-valued 
function $P(\phi_x,0)U_{\rm g}^{(x)}(\phi_x)$, and by differentiation, one has 
\begin{eqnarray}
\frac{d}{d \phi_x}\left(P(\phi_x,0)U_{\rm g}^{(x)}(\phi_x)\right)&=&
P_x(\phi_x,0)U_{\rm g}^{(x)}(\phi_x)+
P(\phi_x,0)\frac{d U_{\rm g}^{(x)}}{d\phi_x}(\phi_x)\ret
&=&P_x(\phi_x,0)U_{\rm g}^{(x)}(\phi_x)+
P(\phi_x,0)F^{(x)}(\phi_x,0)U_{\rm g}^{(x)}(\phi_x)\ret
&=&\left[P_x(\phi_x,0)+P(\phi_x,0)F^{(x)}(\phi_x,0)\right]
U_{\rm g}^{(x)}(\phi_x)\ret
&=&F^{(x)}(\phi_x,0)P(\phi_x,0)U_{\rm g}^{(x)}(\phi_x),
\end{eqnarray}
where we have used the relation (\ref{Pscommu}). Thus 
the function $X_+=P(\phi_x,0)U_{\rm g}^{(x)}(\phi_x)$ is a solution 
of the differential equation (\ref{ODE1+onGT}) with the initial 
condition $P(0,0)U_{\rm g}^{(x)}(0)=P(0,0)$. By the uniqueness 
of the solution, $P(\phi_x,0)U_{\rm g}^{(x)}(\phi_x)$ must coincide 
with $U_{\rm g}^{(x)}(\phi_x)P(0,0)$ which has 
the same initial condition $X_+(0)=P(0,0)$. Namely one has 
\begin{equation}
P(\phi_x,0)U_{\rm g}^{(x)}(\phi_x)
=U_{\rm g}^{(x)}(\phi_x)P(0,0). 
\label{xPmap}
\end{equation}

Next consider the ordinary differential equation, 
\begin{equation}
\frac{d}{d\phi_y}Y_+=F^{(y)}(\phi_x,\phi_y)Y_+, 
\label{ODEY+}
\end{equation}
for the unknown operator-valued function $Y_+=Y_+(\phi_y;\phi_x)$ of 
$\phi_y$ and with the parameter $\phi_x$. 
Let $Y_+=U_{\rm g}^{(y)}(\phi_x,\phi_y)$ 
be the unique solution satisfying the initial condition 
$U_{\rm g}^{(y)}(\phi_x,0)=1$. The solution 
$U_{\rm g}^{(y)}(\phi_x,\phi_y)$ is infinitely differentiable with 
respect to both $\phi_x$ and $\phi_y$ because of the infinitely 
differentiability of $F^{(y)}$. Further introduce the ordinary 
differential equation, 
\begin{equation}
\frac{d}{d\phi_y}Y_-= -Y_-F^{(y)}(\phi_x,\phi_y), 
\label{ODEY-}
\end{equation}
for the unknown operator-valued function $Y_-=Y_-(\phi_y;\phi_x)$ of 
$\phi_y$ and with the parameter $\phi_x$. 
Let $Y_-=V_{\rm g}^{(y)}(\phi_x,\phi_y)$ be the unique solution satisfying 
the initial condition $V_{\rm g}^{(y)}(\phi_x,0)=1$. Then one has 
$V_{\rm g}^{(y)}=(U_{\rm g}^{(y)})^{-1}$ in the same way as in the above. 
Consider the function $P(\phi_x,\phi_y)U_{\rm g}^{(y)}(\phi_x,\phi_y)$. 
By differentiation, one has 
\begin{eqnarray}
\frac{\partial}{\partial\phi_y}\left(
P(\phi_x,\phi_y)U_{\rm g}^{(y)}(\phi_x,\phi_y)\right)&=&
P_y(\phi_x,\phi_y)U_{\rm g}^{(y)}(\phi_x,\phi_y)
+P(\phi_x,\phi_y)\frac{\partial U_{\rm g}^{(y)}}{\partial\phi_y}
(\phi_x,\phi_y)\ret
&=&\left[P_y(\phi_x,\phi_y)+P(\phi_x,\phi_y)F^{(y)}(\phi_x,\phi_y)
\right]U_{\rm g}^{(y)}(\phi_x,\phi_y)\ret
&=&F^{(y)}(\phi_x,\phi_y)P(\phi_x,\phi_y)U_{\rm g}^{(y)}(\phi_x,\phi_y),
\end{eqnarray}
where we have used the relation (\ref{Pscommu}). This equation may be 
treated as the ordinary differential equation (\ref{ODEY+}) for the function 
$Y_+(\phi_y;\phi_x)=P(\phi_x,\phi_y)U_{\rm g}^{(y)}(\phi_x,\phi_y)$ 
of $\phi_y$ and with the parameter $\phi_x$. 
The initial condition is $P(\phi_x,0)U_{\rm g}^{(y)}(\phi_x,0)=P(\phi_x,0)$. 
By the uniqueness of the solution, $P(\phi_x,\phi_y)U_{\rm g}^{(y)}
(\phi_x,\phi_y)$ must coincide with $U_{\rm g}^{(y)}
(\phi_x,\phi_y)P(\phi_x,0)$ which has the same initial condition 
$U_{\rm g}^{(y)}(\phi_x,0)P(\phi_x,0)=P(\phi_x,0)$. Namely one has 
\begin{equation}
P(\phi_x,\phi_y)U_{\rm g}^{(y)}(\phi_x,\phi_y)=
U_{\rm g}^{(y)}(\phi_x,\phi_y)P(\phi_x,0).
\end{equation}
Combining this with (\ref{xPmap}), one gets 
\begin{eqnarray}
P(\phi_x,\phi_y)U_{\rm g}^{(y)}(\phi_x,\phi_y)U_{\rm g}^{(x)}(\phi_x)&=&
U_{\rm g}^{(y)}(\phi_x,\phi_y)P(\phi_x,0)U_{\rm g}^{(x)}(\phi_x)\ret
&=&U_{\rm g}^{(y)}(\phi_x,\phi_y)U_{\rm g}^{(x)}(\phi_x)P(0,0).
\end{eqnarray}
Let $U_{\rm g}(\mbox{\boldmath $\phi$})=
U_{\rm g}^{(y)}(\mbox{\boldmath $\phi$})U_{\rm g}^{(x)}(\phi_x)$. 
Then the operator $U_{\rm g}(\mbox{\boldmath $\phi$})$ is invertible 
and both the operator 
$U_{\rm g}(\mbox{\boldmath $\phi$})$ and its inverse are infinitely differentiable 
with respect to $\mbox{\boldmath $\phi$}$. Besides, the above result is 
rewritten in the desired form as 
\begin{equation}
P(\mbox{\boldmath $\phi$})=U_{\rm g}(\mbox{\boldmath $\phi$})
P(0,0)U_{\rm g}^{-1}(\mbox{\boldmath $\phi$}).
\end{equation}

%%%%%%%%%%%%%%%%%%%%%%%%%%%%%%%%%%%%%%%%%%%%%%%%%%%%%%%%%%%%%%%%%%
\section[Proof of Proposition 6.2]{Proof of Proposition~\ref{pro:Trdiff}}
\label{appendix:pro:Trdiff}
\setcounter{equation}{0}
\setcounter{theorem}{0}

In order to prove Proposition~\ref{pro:Trdiff}, we need some lemmas. 

\begin{lemma}
\label{vRbound}
Let $z\notin\sigma({\cal H}(\mbox{\boldmath $\phi$}))$. Then 
\begin{equation}
\left\Vert v_i^{(0)}(\mbox{\boldmath $\phi$})
R(z;\mbox{\boldmath $\phi$})\right\Vert\le \sqrt{\frac{2}{m_e}
\left[\frac{1}{{\rm dist}(z,\sigma({\cal H}(\mbox{\boldmath $\phi$})))}
+\frac{|z|+\Vert W\Vert_\infty}{{\rm dist}(z,\sigma({\cal H}(\mbox{\boldmath $\phi$})))^2}
\right]}
\end{equation}
with the velocity operator $v_i^{(0)}(\mbox{\boldmath $\phi$})
=[p_i+eA_i({\bf r})+\phi_i]/m_e$ for $i=x,y$. 
\end{lemma}
\begin{proof}{Proof}
For any vector $\psi$, one has 
\begin{eqnarray}
\left\Vert v_i^{(0)}(\mbox{\boldmath $\phi$})
\frac{1}{z-{\cal H}(\mbox{\boldmath $\phi$})}\psi\right\Vert^2
&=&\left\langle\psi,\frac{1}{z^\ast-{\cal H}(\mbox{\boldmath $\phi$})}
[v_i^{(0)}(\mbox{\boldmath $\phi$})]^2\frac{1}{z-{\cal H}(\mbox{\boldmath $\phi$})}
\psi\right\rangle\ret
&\le&\frac{2}{m_e}\left\langle\psi,\frac{1}{z^\ast-{\cal H}(\mbox{\boldmath $\phi$})}
\left[{\cal H}(\mbox{\boldmath $\phi$})+\left\Vert W\right\Vert_\infty\right]
\frac{1}{z-{\cal H}(\mbox{\boldmath $\phi$})}\psi\right\rangle\ret
&=&\frac{2}{m_e}\sum_n\left|\left\langle\psi,\psi_n\right\rangle\right|^2
\frac{1}{z^\ast-{\cal E}_n}\left({\cal E}_n+\Vert W\Vert_\infty\right)
\frac{1}{z-{\cal E}_n}.\ret
\end{eqnarray}
Here, since 
\begin{eqnarray}
\left|\frac{1}{z^\ast-{\cal E}_n}\left({\cal E}_n+\Vert W\Vert_\infty\right)
\frac{1}{z-{\cal E}_n}\right|&=&
\left|\frac{1}{z^\ast-{\cal E}_n}\left({\cal E}_n-z+z
+\Vert W\Vert_\infty\right)\frac{1}{z-{\cal E}_n}\right|\ret
&=&\left|-\frac{1}{z^\ast-{\cal E}_n}
+\frac{z+\Vert W\Vert_\infty}{|z-{\cal E}_n|^2}\right|\ret
&\le&\frac{1}{{\rm dist}(z,\sigma({\cal H}(\mbox{\boldmath $\phi$})))}
+\frac{|z|+\Vert W\Vert_\infty}{{\rm dist}(z,\sigma({\cal H}(\mbox{\boldmath $\phi$})))^2},
\end{eqnarray}
the desired bound is obtained. 
\end{proof}

\begin{lemma}
\label{lemma:diffRbound}
Let $z\notin \sigma({\cal H}(\mbox{\boldmath $\phi$}+\delta\mbox{\boldmath $\phi$}))\cup
\sigma({\cal H}(\mbox{\boldmath $\phi$}))$. 
Then 
\begin{equation}
\left\Vert R(z;\mbox{\boldmath $\phi$}+\delta\mbox{\boldmath $\phi$})-
R(z;\mbox{\boldmath $\phi$})\right\Vert
\le {\cal C}\max_{i=x,y}|\delta\phi_i|,
\label{diffRbound}
\end{equation}
where the positive constant ${\cal C}$ depends on 
${\rm dist}(z,\sigma({\cal H}(\mbox{\boldmath $\phi$}+\delta\mbox{\boldmath $\phi$}))
\cup\sigma({\cal H}(\mbox{\boldmath $\phi$})))$. 
\end{lemma}
\begin{proof}{Proof} From 
\begin{equation}
{\cal H}(\mbox{\boldmath $\phi$}+\delta\mbox{\boldmath $\phi$})-
{\cal H}(\mbox{\boldmath $\phi$})=
\frac{1}{m_e}\delta\mbox{\boldmath $\phi$}\cdot\left[{\bf p}+e{\bf A}({\bf r})+
\mbox{\boldmath $\phi$}\right]
+\frac{1}{2m_e}(\delta\mbox{\boldmath $\phi$})^2,
\end{equation}
one has 
\begin{eqnarray}
R(z;\mbox{\boldmath $\phi$}+\delta\mbox{\boldmath $\phi$})-R(z;\mbox{\boldmath $\phi$})&=&
\sum_{i=x,y}R(z;\mbox{\boldmath $\phi$}+\delta\mbox{\boldmath $\phi$})
v_i^{(0)}(\mbox{\boldmath $\phi$})
R(z;\mbox{\boldmath $\phi$})\delta\phi_i\ret
&+&\frac{1}{2m_e}R(z;\mbox{\boldmath $\phi$}+\delta\mbox{\boldmath $\phi$})
R(z;\mbox{\boldmath $\phi$})(\delta\phi_x^2+\delta\phi_y^2).
\label{diffReq}
\end{eqnarray}
The norm is evaluated as 
\begin{eqnarray}
\left\Vert R(z;\mbox{\boldmath $\phi$}+\delta\mbox{\boldmath $\phi$})
-R(z;\mbox{\boldmath $\phi$})\right\Vert&\le&
\sum_{i=x,y}\left\Vert R(z;\mbox{\boldmath $\phi$}+\delta\mbox{\boldmath $\phi$})\right\Vert
\left\Vert v_i^{(0)}(\mbox{\boldmath $\phi$})R(z;\mbox{\boldmath $\phi$})
\right\Vert|\delta\phi_i|\ret
&+&\frac{1}{2m_e}
\left\Vert R(z;\mbox{\boldmath $\phi$}+\delta\mbox{\boldmath $\phi$})\right\Vert 
\left\Vert R(z;\mbox{\boldmath $\phi$})\right\Vert(\delta\phi_x^2+\delta\phi_y^2).
\end{eqnarray}
Consequently, from Lemma~\ref{vRbound}, the bound (\ref{diffRbound}) 
is obtained. 
\end{proof}

\begin{lemma}
\label{lemma:diffQbound}
The following bound is valid: 
\begin{equation}
\left\Vert Q({\cal E}\le{\cal E}_{\rm F};\mbox{\boldmath $\phi$}
+\delta\mbox{\boldmath $\phi$})
-Q({\cal E}\le{\cal E}_{\rm F};\mbox{\boldmath $\phi$})\right\Vert\le {\cal C}
\max_{i=x,y}|\delta\phi_i|
\label{diffQbound}
\end{equation}
with a positive constant ${\cal C}$. 
\end{lemma}
\begin{proof}{Proof}
By definitions, 
\begin{eqnarray}
\left\Vert Q({\cal E}\le{\cal E}_{\rm F};\mbox{\boldmath $\phi$}
+\delta\mbox{\boldmath $\phi$})
-Q({\cal E}\le{\cal E}_{\rm F};\mbox{\boldmath $\phi$})\right\Vert 
&=&\left\Vert\frac{1}{2\pi i}
\int_\gamma \left[R(z;\mbox{\boldmath $\phi$}+\delta\mbox{\boldmath $\phi$})
-R(z;\mbox{\boldmath $\phi$})\right]dz\right\Vert\ret
&\le&\frac{|\gamma|}{2\pi}\max_{z\in\gamma}
\left\Vert R(z;\mbox{\boldmath $\phi$}+\delta\mbox{\boldmath $\phi$})
-R(z;\mbox{\boldmath $\phi$})\right\Vert,
\end{eqnarray}
where $|\gamma|$ denotes the total length of the path $\gamma$. 
Therefore the inequality (\ref{diffQbound}) follows from 
Lemma~\ref{lemma:diffRbound}.
\end{proof}

\begin{lemma}
\label{lemma:diffQibound}
The following bound is valid: 
\begin{equation}
\left\Vert Q_i({\cal E}\le{\cal E}_{\rm F};\mbox{\boldmath $\phi$}+
\delta\mbox{\boldmath $\phi$})
-Q_i({\cal E}\le{\cal E}_{\rm F};\mbox{\boldmath $\phi$})\right\Vert\le {\cal C}
\max_{j=x,y}|\delta\phi_j|
\label{diffQibound}
\end{equation}
with a positive constant ${\cal C}$. 
\end{lemma}
\begin{proof}{Proof}
Since 
\begin{equation}
\int_\gamma [R(z;\mbox{\boldmath $\phi$})]^2dz=0,
\end{equation}
one has another expression 
\begin{equation}
Q_i({\cal E}\le{\cal E}_{\rm F};\mbox{\boldmath $\phi$})
=\frac{1}{2\pi i}\int_\gamma R(z;\mbox{\boldmath $\phi$})
v_i^{(0)}(\mbox{\boldmath $\phi$}')R(z;\mbox{\boldmath $\phi$})dz 
\quad\mbox{with}\quad 
v_i^{(0)}(\mbox{\boldmath $\phi$}')=\frac{1}{m_e}[p_i+eA_i({\bf r})+\phi_i'] 
\label{expQi}
\end{equation}
for $i=x,y$ and for any $\mbox{\boldmath $\phi$}'$. Using this expression 
(\ref{expQi}), one has 
\begin{eqnarray}
& &Q_i({\cal E}\le{\cal E}_{\rm F};\mbox{\boldmath $\phi$}+\delta\mbox{\boldmath $\phi$})
-Q_i({\cal E}\le{\cal E}_{\rm F};\mbox{\boldmath $\phi$})\ret
&=&\frac{1}{2\pi i}\int_\gamma 
\left[R(z;\mbox{\boldmath $\phi$}+\delta\mbox{\boldmath $\phi$})
-R(z;\mbox{\boldmath $\phi$})\right]
v_i^{(0)}(\mbox{\boldmath $\phi$}+\delta\mbox{\boldmath $\phi$})
R(z;\mbox{\boldmath $\phi$}+\delta\mbox{\boldmath $\phi$})dz \ret
&+&\frac{1}{2\pi i}\int_\gamma 
R(z;\mbox{\boldmath $\phi$})v_i^{(0)}(\mbox{\boldmath $\phi$}
+\delta\mbox{\boldmath $\phi$})
\left[R(z;\mbox{\boldmath $\phi$}+\delta\mbox{\boldmath $\phi$})
-R(z;\mbox{\boldmath $\phi$})\right]dz.
\label{diffQieq}
\end{eqnarray}
The norm can be evaluated as 
\begin{eqnarray}
& &\left\Vert Q_i({\cal E}\le{\cal E}_{\rm F};\mbox{\boldmath $\phi$}+
\delta\mbox{\boldmath $\phi$})
-Q_i({\cal E}\le{\cal E}_{\rm F};\mbox{\boldmath $\phi$})\right\Vert\ret
&\le&\frac{|\gamma|}{2\pi}
\max_{z\in\gamma}\left\Vert R(z;\mbox{\boldmath $\phi$}
+\delta\mbox{\boldmath $\phi$})-R(z;\mbox{\boldmath $\phi$})\right\Vert
\left\Vert v_i^{(0)}(\mbox{\boldmath $\phi$}+\delta\mbox{\boldmath $\phi$})
R(z;\mbox{\boldmath $\phi$}+\delta\mbox{\boldmath $\phi$})\right\Vert\ret
&+&\frac{|\gamma|}{2\pi}\max_{z\in\gamma}\left\Vert R(z;\mbox{\boldmath $\phi$})\right\Vert
\left\Vert v_i^{(0)}(\mbox{\boldmath $\phi$}+\delta\mbox{\boldmath $\phi$})
\left[R(z;\mbox{\boldmath $\phi$}+\delta\mbox{\boldmath $\phi$})
-R(z;\mbox{\boldmath $\phi$})\right]\right\Vert. 
\label{normdiffQibound}
\end{eqnarray}
Using Lemmas~\ref{vRbound} and \ref{lemma:diffRbound}, the first term 
in the right-hand side can be evaluated, and one can get the desired bound 
for the first term. Thus it is sufficient to evaluated the second term. 
Using the identity (\ref{diffReq}), one has 
\begin{eqnarray}
& &\left\Vert v_i^{(0)}(\mbox{\boldmath $\phi$}+\delta\mbox{\boldmath $\phi$})
\left[R(z;\mbox{\boldmath $\phi$}+\delta\mbox{\boldmath $\phi$})
-R(z;\mbox{\boldmath $\phi$})\right]\right\Vert\ret
&\le& \sum_{j=x,y}
\left\Vert v_i^{(0)}(\mbox{\boldmath $\phi$}+\delta\mbox{\boldmath $\phi$})
R(z;\mbox{\boldmath $\phi$}+\delta\mbox{\boldmath $\phi$})\right\Vert
\left\Vert v_j^{(0)}(\mbox{\boldmath $\phi$})
R(z;\mbox{\boldmath $\phi$})\right\Vert|\delta\phi_j|\ret
&+&\frac{1}{2m_e}\left\Vert v_i^{(0)}(\mbox{\boldmath $\phi$}
+\delta\mbox{\boldmath $\phi$})
R(z;\mbox{\boldmath $\phi$}+\delta\mbox{\boldmath $\phi$})
\right\Vert \left\Vert R(z;\mbox{\boldmath $\phi$})\right\Vert(\delta\phi_x^2
+\delta\phi_y^2).
\label{vdiffRbound}
\end{eqnarray}
Combining this with Lemma~\ref{vRbound}, one can get the desired bound also 
for the second term in the right-hand side of (\ref{normdiffQibound}). 
\end{proof}

Now we shall give the proof of Proposition~\ref{pro:Trdiff}. 
Since all the cases can be treated in the same way, we treat the case with 
$i=x,j=y$ only. Note that 
\begin{eqnarray}
& &{\rm Tr}\left[Q({\cal E}\le{\cal E}_{\rm F};\mbox{\boldmath $\phi$}+
\delta\mbox{\boldmath $\phi$})
Q_x({\cal E}\le{\cal E}_{\rm F};\mbox{\boldmath $\phi$}+\delta\mbox{\boldmath $\phi$})
Q_y({\cal E}\le{\cal E}_{\rm F};\mbox{\boldmath $\phi$}+\delta\mbox{\boldmath $\phi$})
\right]\ret
& &-{\rm Tr}\left[Q({\cal E}\le{\cal E}_{\rm F};\mbox{\boldmath $\phi$})
Q_x({\cal E}\le{\cal E}_{\rm F};\mbox{\boldmath $\phi$})
Q_y({\cal E}\le{\cal E}_{\rm F};\mbox{\boldmath $\phi$})\right]\ret
&=&{\rm Tr}\left[\left\{Q({\cal E}\le{\cal E}_{\rm F};\mbox{\boldmath $\phi$}
+\delta\mbox{\boldmath $\phi$})
-Q({\cal E}\le{\cal E}_{\rm F};\mbox{\boldmath $\phi$})\right\}
Q_x({\cal E}\le{\cal E}_{\rm F};\mbox{\boldmath $\phi$}+\delta\mbox{\boldmath $\phi$})
Q_y({\cal E}\le{\cal E}_{\rm F};\mbox{\boldmath $\phi$}+\delta\mbox{\boldmath $\phi$})
\right]\ret
&+&{\rm Tr}\left[Q({\cal E}\le{\cal E}_{\rm F};\mbox{\boldmath $\phi$})
\left\{Q_x({\cal E}\le{\cal E}_{\rm F};\mbox{\boldmath $\phi$}
+\delta\mbox{\boldmath $\phi$})
-Q_x({\cal E}\le{\cal E}_{\rm F};\mbox{\boldmath $\phi$})\right\}
Q_y({\cal E}\le{\cal E}_{\rm F};\mbox{\boldmath $\phi$}+\delta\mbox{\boldmath $\phi$})
\right]\ret
&+&{\rm Tr}\left[Q({\cal E}\le{\cal E}_{\rm F};\mbox{\boldmath $\phi$})
Q_x({\cal E}\le{\cal E}_{\rm F};\mbox{\boldmath $\phi$})\left\{
Q_y({\cal E}\le{\cal E}_{\rm F};\mbox{\boldmath $\phi$}+\delta\mbox{\boldmath $\phi$})
-Q_y({\cal E}\le{\cal E}_{\rm F};\mbox{\boldmath $\phi$})\right\}\right]. 
\label{diffTrQdecom}
\end{eqnarray}
Consider first the first term in the right-hand side. Using the identity, 
\begin{equation}
Q_i({\cal E}\le{\cal E}_{\rm F};\mbox{\boldmath $\phi$})
=Q({\cal E}\le{\cal E}_{\rm F};\mbox{\boldmath $\phi$})
Q_i({\cal E}\le{\cal E}_{\rm F};\mbox{\boldmath $\phi$})
+Q_i({\cal E}\le{\cal E}_{\rm F};\mbox{\boldmath $\phi$})
Q({\cal E}\le{\cal E}_{\rm F};\mbox{\boldmath $\phi$}), 
\label{Qyidentity}
\end{equation}
which is the non-interacting version of (\ref{idQdiff}), one has 
\begin{eqnarray}
& &{\rm Tr}\left[\delta Q_{\le}
Q_x({\cal E}\le{\cal E}_{\rm F};\mbox{\boldmath $\phi$}+\delta\mbox{\boldmath $\phi$})
Q_y({\cal E}\le{\cal E}_{\rm F};\mbox{\boldmath $\phi$}+\delta\mbox{\boldmath $\phi$})
\right]\ret
&=&\sum_{m:{\cal E}_m\le{\cal E}_{\rm F}}
\left\langle {\tilde \psi}_m,Q_x({\cal E}\le{\cal E}_{\rm F};
\mbox{\boldmath $\phi$}+\delta\mbox{\boldmath $\phi$})
Q_y({\cal E}\le{\cal E}_{\rm F};\mbox{\boldmath $\phi$}+\delta\mbox{\boldmath $\phi$})
\delta Q_{\le}{\tilde \psi}_m\right\rangle\ret
&+&\sum_{m:{\cal E}_m\le{\cal E}_{\rm F}}
\left\langle {\tilde \psi}_m,Q_y({\cal E}\le{\cal E}_{\rm F};
\mbox{\boldmath $\phi$}+\delta\mbox{\boldmath $\phi$})
\delta Q_{\le}
Q_x({\cal E}\le{\cal E}_{\rm F};\mbox{\boldmath $\phi$}+\delta\mbox{\boldmath $\phi$})
{\tilde \psi}_m\right\rangle,
\end{eqnarray}
where the vectors ${\tilde \psi}_m$ are the energy eigenvectors of the 
single electron Hamiltonian with $\mbox{\boldmath $\phi$}+\delta\mbox{\boldmath $\phi$}$, 
and we have written $\delta Q_{\le}
=Q({\cal E}\le{\cal E}_{\rm F};\mbox{\boldmath $\phi$}+\delta\mbox{\boldmath $\phi$})
-Q({\cal E}\le{\cal E}_{\rm F};\mbox{\boldmath $\phi$})$. Immediately, 
\begin{eqnarray}
& &\left|{\rm Tr}\left[\delta Q_{\le}
Q_x({\cal E}\le{\cal E}_{\rm F};\mbox{\boldmath $\phi$}+\delta\mbox{\boldmath $\phi$})
Q_y({\cal E}\le{\cal E}_{\rm F};\mbox{\boldmath $\phi$}+\delta\mbox{\boldmath $\phi$})
\right]\right|\ret
&\le&2N\left\Vert Q_x({\cal E}\le{\cal E}_{\rm F};\mbox{\boldmath $\phi$}+
\delta\mbox{\boldmath $\phi$})\right\Vert
\left\Vert Q_y({\cal E}\le{\cal E}_{\rm F};\mbox{\boldmath $\phi$}
+\delta\mbox{\boldmath $\phi$})\right\Vert
\left\Vert \delta Q_{\le}\right\Vert, 
\end{eqnarray}
where $N$ is the number of the electrons. Here the operators 
$Q_i({\cal E}\le{\cal E}_{\rm F};\mbox{\boldmath $\phi$}
+\delta\mbox{\boldmath $\phi$})$ are 
bounded because of Lemma~\ref{vRbound} and (\ref{expQi}). 
Combining these observations with 
Lemma~\ref{lemma:diffQbound}, one gets 
\begin{eqnarray}
& &\left|{\rm Tr}\left[\left\{Q({\cal E}\le{\cal E}_{\rm F};\mbox{\boldmath $\phi$}
+\delta\mbox{\boldmath $\phi$})
-Q({\cal E}\le{\cal E}_{\rm F};\mbox{\boldmath $\phi$})\right\}
Q_x({\cal E}\le{\cal E}_{\rm F};\mbox{\boldmath $\phi$}+\delta\mbox{\boldmath $\phi$})
Q_y({\cal E}\le{\cal E}_{\rm F};\mbox{\boldmath $\phi$}
+\delta\mbox{\boldmath $\phi$})\right]\right|\ret
&\le& {\cal C}N\max_{i=x,y}|\delta\phi_i|
\end{eqnarray}
with a positive constant ${\cal C}$. In a similar way, one has 
\begin{eqnarray}
& &\left|{\rm Tr}\left[Q({\cal E}\le{\cal E}_{\rm F};\mbox{\boldmath $\phi$})
\left\{Q_x({\cal E}\le{\cal E}_{\rm F};\mbox{\boldmath $\phi$}
+\delta\mbox{\boldmath $\phi$})
-Q_x({\cal E}\le{\cal E}_{\rm F};\mbox{\boldmath $\phi$})\right\}
Q_y({\cal E}\le{\cal E}_{\rm F};\mbox{\boldmath $\phi$}+\delta\mbox{\boldmath $\phi$})
\right]\right|\ret
&\le&\sum_{m:{\cal E}_m\le{\cal E}_{\rm F}}
\left|\left\langle\psi_m,\left\{Q_x({\cal E}\le{\cal E}_{\rm F};
\mbox{\boldmath $\phi$}+\delta\mbox{\boldmath $\phi$})
-Q_x({\cal E}\le{\cal E}_{\rm F};\mbox{\boldmath $\phi$})\right\}
Q_y({\cal E}\le{\cal E}_{\rm F};\mbox{\boldmath $\phi$}+\delta\mbox{\boldmath $\phi$})
\psi_m\right\rangle\right|\ret
&\le&N\left\Vert Q_x({\cal E}\le{\cal E}_{\rm F};
\mbox{\boldmath $\phi$}+\delta\mbox{\boldmath $\phi$})
-Q_x({\cal E}\le{\cal E}_{\rm F};\mbox{\boldmath $\phi$})\right\Vert
\left\Vert Q_y({\cal E}\le{\cal E}_{\rm F};\mbox{\boldmath $\phi$}
+\delta\mbox{\boldmath $\phi$})\right\Vert\ret
&\le& {\cal C}N\max_{i=x,y}|\delta\phi_i|,
\end{eqnarray}
where the vectors $\psi_m$ are the energy eigenvectors of the 
single electron Hamiltonian with $\mbox{\boldmath $\phi$}$, 
and ${\cal C}$ is a positive constant, and we have used 
Lemma~\ref{lemma:diffQibound} for getting the last inequality. 
In the same way, 
\begin{eqnarray}
& &\left|{\rm Tr}\left[Q({\cal E}\le{\cal E}_{\rm F};\mbox{\boldmath $\phi$})
Q_x({\cal E}\le{\cal E}_{\rm F};\mbox{\boldmath $\phi$})\left\{
Q_y({\cal E}\le{\cal E}_{\rm F};\mbox{\boldmath $\phi$}+\delta\mbox{\boldmath $\phi$})
-Q_y({\cal E}\le{\cal E}_{\rm F};\mbox{\boldmath $\phi$})\right\}\right]\right|\ret
&\le&{\cal C}N\max_{i=x,y}|\delta\phi_i|
\end{eqnarray}
with a positive constant ${\cal C}$. Combining these three inequalities 
with (\ref{diffTrQdecom}), one can obtain the desired bound (\ref{TrdiffQ}) 
in the case with $i=x,j=y$. 

%%%%%%%%%%%%%%%%%%%%%%%%%%%%%%%%%%%%%%%%%%%%%%%%%%%
\section[Proof of Proposition~6.4]{Proof of Proposition~\ref{proTrvQydiff}}
\setcounter{equation}{0}
\setcounter{theorem}{0}
\label{pproTrvQydiff}
Since ${\rm Tr}\ Q_y({\cal E}\le{\cal E}_F;\mbox{\boldmath $\phi$})=0$, 
it is sufficient to show 
\begin{equation}
\left|{\rm Tr}\ v_s^{(0)}(\mbox{\boldmath $\phi$}+\delta\mbox{\boldmath $\phi$})
Q_y({\cal E}\le{\cal E}_F;\mbox{\boldmath $\phi$}+\delta\mbox{\boldmath $\phi$})
-{\rm Tr}\ v_s^{(0)}(\mbox{\boldmath $\phi$}+\delta\mbox{\boldmath $\phi$})Q_y({\cal E}
\le{\cal E}_F;\mbox{\boldmath $\phi$})\right|\le {\cal C}N\max_{i=x,y}|\delta\phi_i| 
\label{TrvQydiffbound}
\end{equation}
with some positive constant ${\cal C}$. Using the identity (\ref{Qyidentity}), 
one has 
\begin{eqnarray}
& &{\rm Tr}\ v_s^{(0)}(\mbox{\boldmath $\phi$}+\delta\mbox{\boldmath $\phi$})
Q_y({\cal E}\le{\cal E}_F;\mbox{\boldmath $\phi$}+\delta\mbox{\boldmath $\phi$})
-{\rm Tr}\ v_s^{(0)}(\mbox{\boldmath $\phi$}+\delta\mbox{\boldmath $\phi$})Q_y({\cal E}
\le{\cal E}_F;\mbox{\boldmath $\phi$})\ret
&=&{\rm Tr}\ v_s^{(0)}(\mbox{\boldmath $\phi$}+\delta\mbox{\boldmath $\phi$})
\left\{Q_y({\cal E}\le{\cal E}_F;\mbox{\boldmath $\phi$}+\delta\mbox{\boldmath $\phi$})
Q({\cal E}\le{\cal E}_F;\mbox{\boldmath $\phi$}+\delta\mbox{\boldmath $\phi$})\right.\ret
& &\qquad\qquad\left.-Q_y({\cal E}\le{\cal E}_F;\mbox{\boldmath $\phi$})
Q({\cal E}\le{\cal E}_F;\mbox{\boldmath $\phi$})\right\}\ret
&+&{\rm Tr}\ v_s^{(0)}(\mbox{\boldmath $\phi$}+\delta\mbox{\boldmath $\phi$})
\left\{Q({\cal E}\le{\cal E}_F;\mbox{\boldmath $\phi$}+\delta\mbox{\boldmath $\phi$})
Q_y({\cal E}\le{\cal E}_F;\mbox{\boldmath $\phi$}
+\delta\mbox{\boldmath $\phi$})\right.\ret
& &\qquad\qquad\left.-Q({\cal E}\le{\cal E}_F;\mbox{\boldmath $\phi$})
Q_y({\cal E}\le{\cal E}_F;\mbox{\boldmath $\phi$})\right\}. 
\label{TrvQydiff}
\end{eqnarray}

We begin with estimating the first term in the right-hand side. Note that 
\begin{eqnarray}
& &[\mbox{1st term in r.h.s. of (\ref{TrvQydiff})}]\ret
&=&{\rm Tr}\ v_s^{(0)}(\mbox{\boldmath $\phi$}+\delta\mbox{\boldmath $\phi$})
\left\{Q_y({\cal E}\le{\cal E}_F;\mbox{\boldmath $\phi$}+\delta\mbox{\boldmath $\phi$})
-Q_y({\cal E}\le{\cal E}_F;\mbox{\boldmath $\phi$})\right\}
Q({\cal E}\le{\cal E}_F;\mbox{\boldmath $\phi$}+\delta\mbox{\boldmath $\phi$})\ret 
&+&{\rm Tr}\ v_s^{(0)}(\mbox{\boldmath $\phi$}+\delta\mbox{\boldmath $\phi$})
Q_y({\cal E}\le{\cal E}_F;\mbox{\boldmath $\phi$})
Q({\cal E}\le{\cal E}_F;\mbox{\boldmath $\phi$})
\left\{Q({\cal E}\le{\cal E}_F;\mbox{\boldmath $\phi$}+\delta\mbox{\boldmath $\phi$})
-Q({\cal E}\le{\cal E}_F;\mbox{\boldmath $\phi$})\right\}\ret
&+&{\rm Tr}\ v_s^{(0)}(\mbox{\boldmath $\phi$}+\delta\mbox{\boldmath $\phi$})
Q({\cal E}\le{\cal E}_F;\mbox{\boldmath $\phi$})
Q_y({\cal E}\le{\cal E}_F;\mbox{\boldmath $\phi$})
\left\{Q({\cal E}\le{\cal E}_F;\mbox{\boldmath $\phi$}+\delta\mbox{\boldmath $\phi$})
-Q({\cal E}\le{\cal E}_F;\mbox{\boldmath $\phi$})\right\},\ret
\label{rewritevQy1}
\end{eqnarray}
where we have used the identity (\ref{Qyidentity}) again. 
This first term in the right-hand side is rewritten as 
\begin{eqnarray}
& &{\rm Tr}\ v_s^{(0)}(\mbox{\boldmath $\phi$}+\delta\mbox{\boldmath $\phi$})
\left\{Q_y({\cal E}\le{\cal E}_F;\mbox{\boldmath $\phi$}+\delta\mbox{\boldmath $\phi$})
-Q_y({\cal E}\le{\cal E}_F;\mbox{\boldmath $\phi$})\right\}
Q({\cal E}\le{\cal E}_F;\mbox{\boldmath $\phi$}+\delta\mbox{\boldmath $\phi$})\ret
&=&\sum_{m:{\cal E}_m\le{\cal E}_F}\left\langle
{\tilde \psi}_m,v_s^{(0)}(\mbox{\boldmath $\phi$}+\delta\mbox{\boldmath $\phi$})
\left\{Q_y({\cal E}\le{\cal E}_F;\mbox{\boldmath $\phi$}+\delta\mbox{\boldmath $\phi$})
-Q_y({\cal E}\le{\cal E}_F;\mbox{\boldmath $\phi$})\right\}
{\tilde \psi}_m\right\rangle, 
\label{TrvdiffQyQ}
\end{eqnarray}
where the vectors ${\tilde \psi}_m$ are the energy eigenvectors of the 
single electron Hamiltonian with the gauge 
parameters $\mbox{\boldmath $\phi$}+\delta\mbox{\boldmath $\phi$}$. 
{From} Lemma~\ref{vRbound}, (\ref{diffQieq}) and (\ref{vdiffRbound}), 
one has 
\begin{equation}
\left\Vert v_s^{(0)}(\mbox{\boldmath $\phi$}+\delta\mbox{\boldmath $\phi$})
\left\{Q_y({\cal E}\le{\cal E}_F;\mbox{\boldmath $\phi$}+\delta\mbox{\boldmath $\phi$})
-Q_y({\cal E}\le{\cal E}_F;\mbox{\boldmath $\phi$})\right\}\right\Vert
\le {\cal C}\max_{i=x,y}|\delta\phi_i|,
\end{equation}
where the positive constant ${\cal C}$ is independent of the number $N$ of 
the electrons and the system sizes $L_x,L_y$. Using this bound for 
(\ref{TrvdiffQyQ}), one obtains 
\begin{eqnarray}
& &\left|{\rm Tr}\ v_s^{(0)}(\mbox{\boldmath $\phi$}+\delta\mbox{\boldmath $\phi$})
\left\{Q_y({\cal E}\le{\cal E}_F;\mbox{\boldmath $\phi$}+\delta\mbox{\boldmath $\phi$})
-Q_y({\cal E}\le{\cal E}_F;\mbox{\boldmath $\phi$})\right\}
Q({\cal E}\le{\cal E}_F;\mbox{\boldmath $\phi$}+\delta\mbox{\boldmath $\phi$})
\right|\ret&\le& 
{\cal C}N\max_{i=x,y}|\delta\phi_i|.
\label{TrvdiffQyQbound}
\end{eqnarray}
The second term in the right-hand side of (\ref{rewritevQy1}) is rewritten as 
$$
{\rm Tr}\ v_s^{(0)}(\mbox{\boldmath $\phi$}+\delta\mbox{\boldmath $\phi$})
Q_y({\cal E}\le{\cal E}_F;\mbox{\boldmath $\phi$})
Q({\cal E}\le{\cal E}_F;\mbox{\boldmath $\phi$})
\left\{Q({\cal E}\le{\cal E}_F;\mbox{\boldmath $\phi$}+\delta\mbox{\boldmath $\phi$})
-Q({\cal E}\le{\cal E}_F;\mbox{\boldmath $\phi$})\right\}
$$
\begin{equation}
=\sum_{m:{\cal E}_m\le{\cal E}_F}
\left\langle\psi_m,\left\{Q({\cal E}\le{\cal E}_F;\mbox{\boldmath $\phi$}
+\delta\mbox{\boldmath $\phi$})
-Q({\cal E}\le{\cal E}_F;\mbox{\boldmath $\phi$})\right\}
v_s^{(0)}(\mbox{\boldmath $\phi$}+\delta\mbox{\boldmath $\phi$})
Q_y({\cal E}\le{\cal E}_F;\mbox{\boldmath $\phi$})\psi_m\right\rangle,
\end{equation}
where $\psi_m$ are the energy eigenvectors of the single electron Hamiltonian 
with the gauge parameters $\mbox{\boldmath $\phi$}$. 
{From} Lemma~\ref{vRbound}, the operator 
$v_s^{(0)}(\mbox{\boldmath $\phi$}+\delta\mbox{\boldmath $\phi$})
Q_y({\cal E}\le{\cal E}_F;\mbox{\boldmath $\phi$})$ is bounded, 
and the difference $Q({\cal E}\le{\cal E}_F;\mbox{\boldmath $\phi$}
+\delta\mbox{\boldmath $\phi$})
-Q({\cal E}\le{\cal E}_F;\mbox{\boldmath $\phi$})$ was already estimated 
in Lemma~\ref{lemma:diffQbound}. Therefore 
\begin{eqnarray}
& &\left|{\rm Tr}\ v_s^{(0)}(\mbox{\boldmath $\phi$}+\delta\mbox{\boldmath $\phi$})
Q_y({\cal E}\le{\cal E}_F;\mbox{\boldmath $\phi$})
Q({\cal E}\le{\cal E}_F;\mbox{\boldmath $\phi$})
\left\{Q({\cal E}\le{\cal E}_F;\mbox{\boldmath $\phi$}+\delta\mbox{\boldmath $\phi$})
-Q({\cal E}\le{\cal E}_F;\mbox{\boldmath $\phi$})\right\}\right|\ret
&\le&{\cal C}N\max_{i=x,y}|\delta\phi_i|,
\label{TrvQyQdiffQ}
\end{eqnarray}
where the positive constant ${\cal C}$ is independent of the number $N$ of 
the electrons and the system sizes $L_x,L_y$. 
Similarly the third term in the right-hand side of (\ref{rewritevQy1}) is evaluated as 
\begin{eqnarray}
& &\left|{\rm Tr}\ v_s^{(0)}(\mbox{\boldmath $\phi$}+\delta\mbox{\boldmath $\phi$})
Q({\cal E}\le{\cal E}_F;\mbox{\boldmath $\phi$})
Q_y({\cal E}\le{\cal E}_F;\mbox{\boldmath $\phi$})
\left\{Q({\cal E}\le{\cal E}_F;\mbox{\boldmath $\phi$}+\delta\mbox{\boldmath $\phi$})
-Q({\cal E}\le{\cal E}_F;\mbox{\boldmath $\phi$})\right\}\right|\ret
&\le&\sum_{m:\atop {\cal E}_m\le{\cal E}_F}
\left|\left\langle\psi_m,Q_y({\cal E}\le{\cal E}_F;\mbox{\boldmath $\phi$})
\left\{Q({\cal E}\le{\cal E}_F;\mbox{\boldmath $\phi$}+\delta\mbox{\boldmath $\phi$})
-Q({\cal E}\le{\cal E}_F;\mbox{\boldmath $\phi$})\right\}
v_s^{(0)}(\mbox{\boldmath $\phi$}+\delta\mbox{\boldmath $\phi$})\psi_m
\right\rangle\right|\ret
&\le&\left\Vert Q_y({\cal E}\le{\cal E}_F;\mbox{\boldmath $\phi$})\right\Vert
\left\Vert Q({\cal E}\le{\cal E}_F;\mbox{\boldmath $\phi$}+\delta\mbox{\boldmath $\phi$})
-Q({\cal E}\le{\cal E}_F;\mbox{\boldmath $\phi$})\right\Vert\ret
&\times&\left\{\frac{N}{m_e}|\delta\phi_s|+\sum_{m:{\cal E}_m\le{\cal E}_F}
\sqrt{\left\langle\psi_m,\left[v_s^{(0)}(\mbox{\boldmath $\phi$})\right]^2\psi_m
\right\rangle}\right\}\ret
&\le&{\cal C}\max_{i=x,y}|\delta\phi_i|
\left\{\frac{N}{m_e}|\delta\phi_s|+
N\sqrt{\frac{2}{m_e}\left({\cal E}_F+\Vert W\Vert_\infty\right)}\right\}. 
\label{TrvQQydiffQ}
\end{eqnarray}
Thus all of the terms in the right-hand side of (\ref{rewritevQy1}) have been evaluated. 

Next consider the second term in the right-hand side of (\ref{TrvQydiff}). 
It can be rewritten as 
\begin{eqnarray}
& &[\mbox{2nd term in r.h.s. of (\ref{TrvQydiff})}]\ret
&=& {\rm Tr}\ v_s^{(0)}(\mbox{\boldmath $\phi$}+\delta\mbox{\boldmath $\phi$})
\delta Q_{\le}
Q_y({\cal E}\le{\cal E}_F;\mbox{\boldmath $\phi$}+\delta\mbox{\boldmath $\phi$})
Q({\cal E}\le{\cal E}_F;\mbox{\boldmath $\phi$}+\delta\mbox{\boldmath $\phi$})\ret
&+&{\rm Tr}\ v_s^{(0)}(\mbox{\boldmath $\phi$}+\delta\mbox{\boldmath $\phi$})
\delta Q_{\le}
Q({\cal E}\le{\cal E}_F;\mbox{\boldmath $\phi$}+\delta\mbox{\boldmath $\phi$})
Q_y({\cal E}\le{\cal E}_F;\mbox{\boldmath $\phi$}+\delta\mbox{\boldmath $\phi$})\ret
&+&{\rm Tr}\ v_s^{(0)}(\mbox{\boldmath $\phi$}+\delta\mbox{\boldmath $\phi$})
Q({\cal E}\le{\cal E}_F;\mbox{\boldmath $\phi$})
\left\{Q_y({\cal E}\le{\cal E}_F;\mbox{\boldmath $\phi$}+\delta\mbox{\boldmath $\phi$})
-Q_y({\cal E}\le{\cal E}_F;\mbox{\boldmath $\phi$})\right\},
\label{TrvQydiff2rewrite}
\end{eqnarray}
where we have used the identity (\ref{Qyidentity}), and we have written 
\begin{equation}
\delta Q_{\le}=Q({\cal E}\le{\cal E}_F;\mbox{\boldmath $\phi$}
+\delta\mbox{\boldmath $\phi$})
-Q({\cal E}\le{\cal E}_F;\mbox{\boldmath $\phi$}). 
\end{equation}
{From} Lemma~\ref{vRbound} and (\ref{diffReq}), one has 
\begin{equation}
\left\Vert v_s^{(0)}(\mbox{\boldmath $\phi$}+\delta\mbox{\boldmath $\phi$})
\delta Q_{\le}\right\Vert
\le {\cal C}\max_{i=x,y}|\delta\phi_i|,
\end{equation}
where the positive constant ${\cal C}$ is independent of the number $N$ 
of the electrons and the system sizes $L_x,L_y$. Hence, in the same way 
as in the above, the first and the second terms in the right-hand side of 
(\ref{TrvQydiff2rewrite}) are evaluated as 
\begin{eqnarray}
& &\left|{\rm Tr}\ v_s^{(0)}(\mbox{\boldmath $\phi$}+\delta\mbox{\boldmath $\phi$})
\delta Q_{\le}
Q_y({\cal E}\le{\cal E}_F;\mbox{\boldmath $\phi$}+\delta\mbox{\boldmath $\phi$})
Q({\cal E}\le{\cal E}_F;\mbox{\boldmath $\phi$}+\delta\mbox{\boldmath $\phi$})\right|\ret
&+&\left|{\rm Tr}\ v_s^{(0)}(\mbox{\boldmath $\phi$}+\delta\mbox{\boldmath $\phi$})
\delta Q_{\le}
Q({\cal E}\le{\cal E}_F;\mbox{\boldmath $\phi$}+\delta\mbox{\boldmath $\phi$})
Q_y({\cal E}\le{\cal E}_F;\mbox{\boldmath $\phi$}+\delta\mbox{\boldmath $\phi$})\right|
\le {\cal C}N\max_{i=x,y}|\delta\phi_i|\ret
\label{TrvdiffQQyQ}
\end{eqnarray}
with some positive constant ${\cal C}$. 
By using Lemma~\ref{lemma:diffQibound}, the third term in 
the right-hand side of (\ref{TrvQydiff2rewrite}) is evaluated as 
\begin{equation}
\left|{\rm Tr}\ v_s^{(0)}(\mbox{\boldmath $\phi$}+\delta\mbox{\boldmath $\phi$})
Q({\cal E}\le{\cal E}_F;\mbox{\boldmath $\phi$})
\left\{Q_y({\cal E}\le{\cal E}_F;\mbox{\boldmath $\phi$}+\delta\mbox{\boldmath $\phi$})
-Q_y({\cal E}\le{\cal E}_F;\mbox{\boldmath $\phi$})\right\}\right|\ret
\le{\cal C}N\max_{i=x,y}|\delta\phi_i|
\label{TrvQdiffQy}
\end{equation}
in the same way as in (\ref{TrvQQydiffQ}). 

Combining (\ref{TrvQydiff}), (\ref{rewritevQy1}), (\ref{TrvdiffQyQbound}), 
(\ref{TrvQyQdiffQ}), (\ref{TrvQQydiffQ}), 
(\ref{TrvQydiff2rewrite}), (\ref{TrvdiffQQyQ}) and (\ref{TrvQdiffQy}), 
the desired bound (\ref{TrvQydiffbound}) is obtained. 

%%%%%%%%%%%%%%%%%%%%%%%%%%%%%%%%%%%%%%%%%%%%%%%%%%%%%%%%%
\section[Proofs of Theorems~7.2 and 7.5]{Proofs of 
Theorems \ref{theorem:bounded1} and \ref{theorem:bounded2}}
\setcounter{equation}{0}
\setcounter{theorem}{0}
\label{Boundsigmaxy}

In order to give the proofs of 
Theorems \ref{theorem:bounded1} and \ref{theorem:bounded2}, 
we first recall the Hall conductance $\sigma_{xy}(\mbox{\boldmath $\phi$})$ 
of (\ref{sigmaxy}) which can be expressed as 
\begin{equation}
\sigma_{xy}(\mbox{\boldmath $\phi$})=\frac{i\hbar e^2}{L_xL_y}\frac{1}{q}\sum_{\mu=1}^q
\left[\left\langle\Phi_{0,\mu}^{(N)}(\mbox{\boldmath $\phi$}),
v_{{\rm tot},y}^{(0)}(\mbox{\boldmath $\phi$})
\frac{1-Q(E_0^{(N)}(\mbox{\boldmath $\phi$}))}{[E_{0,\mu}^{(N)}(\mbox{\boldmath $\phi$})
-H_0^{(N)}(\mbox{\boldmath $\phi$})]^2}
v_{{\rm tot},x}^{(0)}(\mbox{\boldmath $\phi$})\Phi_{0,\mu}^{(N)}(\mbox{\boldmath $\phi$})
\right\rangle
-{\rm c.c.}\right]. 
\label{sigmaxyphinew}
\end{equation}
Using the Schwarz inequality, the matrix element is estimated as 
\begin{eqnarray}
& &\left|\sum_{\mu=1}^q
\left\langle\Phi_{0,\mu}^{(N)}(\mbox{\boldmath $\phi$}),
v_{{\rm tot},y}^{(0)}(\mbox{\boldmath $\phi$})
\frac{1-Q(E_0^{(N)}(\mbox{\boldmath $\phi$}))}{[E_{0,\mu}^{(N)}(\mbox{\boldmath $\phi$})
-H_0^{(N)}(\mbox{\boldmath $\phi$})]^2}
v_{{\rm tot},x}^{(0)}(\mbox{\boldmath $\phi$})\Phi_{0,\mu}^{(N)}(\mbox{\boldmath $\phi$})
\right\rangle
\right|\ret
&\le&\frac{1}{(\Delta E)^2}\prod_{s=x,y}\sqrt{\sum_{\mu=1}^q
\left\langle\Phi_{0,\mu}^{(N)}(\mbox{\boldmath $\phi$}),
v_{{\rm tot},s}^{(0)}(\mbox{\boldmath $\phi$})
[1-Q(E_0^{(N)}(\mbox{\boldmath $\phi$}))]
v_{{\rm tot},s}^{(0)}(\mbox{\boldmath $\phi$})
\Phi_{0,\mu}^{(N)}(\mbox{\boldmath $\phi$})\right\rangle}.
\label{vQvbound}
\end{eqnarray}
Because of the factor $1/(L_xL_y)$ in the right-hand side of 
(\ref{sigmaxyphinew}), it is sufficient to show that this right-hand side 
is of order of $N$. But a simple estimate yields order $N^2$ because of 
the two total velocity operators in the matrix elements. 
In order to reduce the order $N^2$ to $N$, we use the method introduced 
in ref.~\cite{HvL}. See also refs.~\cite{Koma1,KomaTasaki}. 
Let $A$ be a symmetric operator. In the same way as in (\ref{doublecommuEq}), 
one has formally 
\begin{eqnarray}
& &\sum_{\mu=1}^q\left\langle A\Phi_{0,\mu}^{(N)}(\mbox{\boldmath $\phi$}),
[1-Q(E_0^{(N)}(\mbox{\boldmath $\phi$})]A
\Phi_{0,\mu}^{(N)}(\mbox{\boldmath $\phi$})\right\rangle\ret
&\le&\frac{1}{\Delta E}
\sum_{\mu=1}^q\left\langle A\Phi_{0,\mu}^{(N)}(\mbox{\boldmath $\phi$}),
[H_0^{(N)}(\mbox{\boldmath $\phi$})-E_{0,\mu}^{(N)}(\mbox{\boldmath $\phi$})]
A\Phi_{0,\mu}^{(N)}(\mbox{\boldmath $\phi$})\right\rangle\ret
&=&\frac{1}{2\Delta E}\sum_{\mu=1}^q\left\langle\Phi_{0,\mu}^{(N)}(\mbox{\boldmath $\phi$}),
[A,[H_0^{(N)}(\mbox{\boldmath $\phi$}),A]]
\Phi_{0,\mu}^{(N)}(\mbox{\boldmath $\phi$})\right\rangle.
\label{AQAbound}
\end{eqnarray}
We stress that this formal inequality can be justified in the same way 
as in Section~\ref{Sec:translationinv}. 
{From} the two inequalities (\ref{vQvbound}) and 
(\ref{AQAbound}), it is sufficient to show that the expectation values 
of the double commutators $[v_{{\rm tot},s}^{(0)}(\mbox{\boldmath $\phi$}),
[H_0^{(N)}(\mbox{\boldmath $\phi$}),v_{{\rm tot},s}^{(0)}(\mbox{\boldmath $\phi$})]]$ 
for the ground state are of order of $N$. 

Let us calculate those double commutators. Note that 
\begin{equation}
\left[v_{{\rm tot},x}^{(0)}(\mbox{\boldmath $\phi$}),
H_0^{(N)}(\mbox{\boldmath $\phi$})\right]
=-\frac{i\hbar eB}{m_e}v_{{\rm tot},y}^{(0)}(\mbox{\boldmath $\phi$})
-\frac{i\hbar eB}{m_e}I_y(\mbox{\boldmath $\phi$}),
\label{vyvxH0commu}
\end{equation}
and 
\begin{equation}
\left[v_{{\rm tot},y}^{(0)}(\mbox{\boldmath $\phi$}),
H_0^{(N)}(\mbox{\boldmath $\phi$})\right]
=\frac{i\hbar eB}{m_e}v_{{\rm tot},x}^{(0)}(\mbox{\boldmath $\phi$})-
\frac{i\hbar eB}{m_e}I_x(\mbox{\boldmath $\phi$}),
\label{vxvyH0commu}
\end{equation}
where 
\begin{eqnarray}
I_x(\mbox{\boldmath $\phi$})&=&\frac{1}{eB}\sum_{j=1}^N\left(\frac{\partial W}{\partial y}
\right)({\bf r}_j)-\frac{1}{2B}\sum_{j=1}^N\left[
B_{{\rm P},z}({\bf r}_j)v_{x,j}^{(0)}(\mbox{\boldmath $\phi$})
+v_{x,j}^{(0)}(\mbox{\boldmath $\phi$})B_{{\rm P},z}({\bf r}_j)\right]
\label{defIx}
\end{eqnarray}
and 
\begin{eqnarray}
I_y(\mbox{\boldmath $\phi$})&=&\frac{1}{eB}\sum_{j=1}^N\left(\frac{\partial W}{\partial x}
\right)({\bf r}_j)+\frac{1}{2B}\sum_{j=1}^N\left[
B_{{\rm P},z}({\bf r}_j)v_{y,j}^{(0)}(\mbox{\boldmath $\phi$})+
v_{y,j}^{(0)}(\mbox{\boldmath $\phi$})
B_{{\rm P},z}({\bf r}_j)\right]
\label{defIy}
\end{eqnarray}
with the velocity operator, 
\begin{equation}
{\bf v}_j^{(0)}(\mbox{\boldmath $\phi$})=\frac{1}{m_e}[{\bf p}_j+
e{\bf A}({\bf r}_j)+\mbox{\boldmath $\phi$}].
\end{equation}
Using the commutation relation (\ref{vyvxH0commu}), we have 
\begin{equation}
\left[v_{{\rm tot},x}^{(0)}(\mbox{\boldmath $\phi$}),
[H_0^{(N)}(\mbox{\boldmath $\phi$}),v_{{\rm tot},x}^{(0)}(\mbox{\boldmath $\phi$})]
\right]=\frac{i\hbar eB}{m_e}\left[v_{{\rm tot},x}^{(0)}(\mbox{\boldmath $\phi$}),
v_{{\rm tot},y}^{(0)}(\mbox{\boldmath $\phi$})\right]+\frac{i\hbar eB}{m_e}
\left[v_{{\rm tot},x}^{(0)}(\phi),I_y(\mbox{\boldmath $\phi$})\right].
\label{doublecommuvxH0}
\end{equation}
The commutator of the first term in the right-hand side is calculated as 
\begin{equation}
\left[v_{{\rm tot},x}^{(0)}(\mbox{\boldmath $\phi$}),
v_{{\rm tot},y}^{(0)}(\mbox{\boldmath $\phi$})\right]=-\frac{i\hbar eB}{m_e^2}N
-\frac{i\hbar e}{m_e^2}\sum_{j=1}^N B_{{\rm P},z}({\bf r}_j), 
\label{vxvycommu}
\end{equation}
and the commutator of the second term becomes 
\begin{eqnarray}
\left[v_{{\rm tot},x}^{(0)}(\mbox{\boldmath $\phi$}),I_y(\mbox{\boldmath $\phi$})\right]
&=&-\frac{i\hbar}{m_e eB}\sum_{j=1}^N\left(\frac{\partial^2 W}{\partial x^2}
\right)({\bf r}_j)-\frac{i\hbar e}{m_e^2 B}\sum_{j=1}^N
[B_{{\rm P},z}({\bf r}_j)]^2-\frac{i\hbar e}{m_e^2}\sum_{j=1}^N 
B_{{\rm P},z}({\bf r}_j)\ret
&-&\frac{i\hbar}{2m_e B}\sum_{j=1}^N 
\left[v_{y,j}^{(0)}(\mbox{\boldmath $\phi$})
\left(\frac{\partial B_{{\rm P},z}}{\partial x}\right)({\bf r}_j)
+\left(\frac{\partial B_{{\rm P},z}}{\partial x}\right)({\bf r}_j)
v_{y,j}^{(0)}(\mbox{\boldmath $\phi$})\right]. 
\end{eqnarray}
Substituting these into the right-hand side of (\ref{doublecommuvxH0}), 
we have 
\begin{eqnarray}
& &\left[v_{{\rm tot},x}^{(0)}(\mbox{\boldmath $\phi$}),
[H_0^{(N)}(\mbox{\boldmath $\phi$}),v_{{\rm tot},x}^{(0)}
(\mbox{\boldmath $\phi$})]\right]\ret
&=&\frac{\hbar^2e^2B^2}{m_e^3}N+\frac{\hbar^2}{m_e^2}\sum_{j=1}^N
\left(\frac{\partial^2 W}{\partial x^2}\right)({\bf r}_j)+
\frac{\hbar^2e^2}{m_e^3}\sum_{j=1}^N[B_{{\rm P},z}({\bf r}_j)]^2
+\frac{2\hbar^2e^2B}{m_e^3}\sum_{j=1}^N B_{{\rm P},z}({\bf r}_j)\ret
&+&\frac{\hbar^2e}{2m_e^2}\sum_{j=1}^N 
\left[v_{y,j}^{(0)}(\mbox{\boldmath $\phi$})
\left(\frac{\partial B_{{\rm P},z}}{\partial x}\right)({\bf r}_j)
+\left(\frac{\partial B_{{\rm P},z}}{\partial x}\right)({\bf r}_j)
v_{y,j}^{(0)}(\mbox{\boldmath $\phi$})\right].
\label{dcHvx}
\end{eqnarray}
Similarly we have
\begin{eqnarray}
& &\left[v_{{\rm tot},y}^{(0)}(\mbox{\boldmath $\phi$}),
[H_0^{(N)}(\mbox{\boldmath $\phi$}),v_{{\rm tot},y}^{(0)}
(\mbox{\boldmath $\phi$})]\right]\ret
&=&\frac{\hbar^2e^2B^2}{m_e^3}N+\frac{\hbar^2}{m_e^2}\sum_{j=1}^N
\left(\frac{\partial^2 W}{\partial y^2}\right)({\bf r}_j)+
\frac{\hbar^2e^2}{m_e^3}\sum_{j=1}^N[B_{{\rm P},z}({\bf r}_j)]^2
+\frac{2\hbar^2e^2B}{m_e^3}\sum_{j=1}^N B_{{\rm P},z}({\bf r}_j)\ret
&-&\frac{\hbar^2e}{2m_e^2}\sum_{j=1}^N 
\left[v_{x,j}^{(0)}(\mbox{\boldmath $\phi$})
\left(\frac{\partial B_{{\rm P},z}}{\partial y}\right)({\bf r}_j)
+\left(\frac{\partial B_{{\rm P},z}}{\partial y}\right)({\bf r}_j)
v_{x,j}^{(0)}(\mbox{\boldmath $\phi$})\right].
\label{dcHvy}
\end{eqnarray}
by using the commutation relations (\ref{vxvyH0commu}) and (\ref{vxvycommu}).

{From} the expressions (\ref{dcHvx}) and (\ref{dcHvy}) for 
the double commutators, it is sufficient to evaluate the ground state 
expectation values of the last sums in both the right-hand sides. 
Since all the terms in the summands can be treated in the same way, 
we consider only the second term in the summand of the last sum in 
(\ref{dcHvy}). Using the Schwarz inequality, we have 
\begin{eqnarray}
& &\left|\sum_{j=1}^N \left\langle\Phi_{0,\mu}^{(N)}(\mbox{\boldmath $\phi$}), 
\left(\frac{\partial B_{{\rm P},z}}{\partial y}
\right)({\bf r}_j)v_{x,j}^{(0)}(\mbox{\boldmath $\phi$})
\Phi_{0,\mu}^{(N)}(\mbox{\boldmath $\phi$})
\right\rangle\right|\ret
&\le&\left[\sum_{i=1}^N \left\langle\Phi_{0,\mu}^{(N)}(\mbox{\boldmath $\phi$}),
\left|\left(\frac{\partial B_{{\rm P},z}}{\partial y}
\right)({\bf r}_i)\right|^2\Phi_{0,\mu}^{(N)}(\mbox{\boldmath $\phi$})
\right\rangle\right]^{1/2}
\left[\sum_{j=1}^N \left\langle\Phi_{0,\mu}^{(N)}(\mbox{\boldmath $\phi$}),
\left[v_{x,j}^{(0)}(\mbox{\boldmath $\phi$})\right]^2
\Phi_{0,\mu}^{(N)}(\mbox{\boldmath $\phi$})
\right\rangle\right]^{1/2}\ret
&\le&\left(\frac{2}{m_e}\right)^{1/2}N^{1/2}
\left\Vert\frac{\partial B_{{\rm P},z}}{\partial y}\right\Vert_\infty
\left[\left\langle\Phi_{0,\mu}^{(N)}(\mbox{\boldmath $\phi$}),
H_0^{(N)}(\mbox{\boldmath $\phi$})
\Phi_{0,\mu}^{(N)}(\mbox{\boldmath $\phi$})\right\rangle+N\left\Vert W\right\Vert_\infty
\right]^{1/2}\ret
&=&\left(\frac{2}{m_e}\right)^{1/2}N^{1/2}
\left\Vert\frac{\partial B_{{\rm P},z}}{\partial y}\right\Vert_\infty
\left[E_{0,\mu}^{(N)}(\mbox{\boldmath $\phi$})+N\left\Vert W\right\Vert_\infty
\right]^{1/2}.
\label{expgsDBv}
\end{eqnarray}
For getting the second inequality, we have used the inequality, 
\begin{equation}
\sum_{j=1}^N \frac{m_e}{2}\left[v_{x,j}^{(0)}(\mbox{\boldmath $\phi$})\right]^2
\le H_0^{(N)}(\mbox{\boldmath $\phi$})+N\left\Vert W\right\Vert_\infty
\end{equation}
which follows from the assumption $W^{(2)}\ge 0$. 
By relying on the decay assumption (\ref{decayconditionW2}) 
for the interaction $W^{(2)}$, we can prove \cite{Koma1} 
that the energy eigenvalue $E_{0,\mu}^{(N)}(\mbox{\boldmath $\phi$})$ is of order of $N$. 
See Appendix~\ref{estEU} for the detail. 
Consequently the right-hand side of the last line in (\ref{expgsDBv}) 
is of order of $N$. 

%%%%%%%%%%%%%%%%%%%%%%%%%%%%%%%%%%%%%%%%%%%%%%%%%%%%%%%%%
\section[Proofs of Theorems 7.6, 7.8 and 7.12]{Proofs of 
Theorems~\ref{1dtranslationinv}, \ref{fractionTheorem1} 
and \ref{fractionTheorem2}}
\label{fractionalproofs}
\setcounter{equation}{0}
\setcounter{theorem}{0}

We begin with rewriting the Hall conductance $\sigma_{xy}(\mbox{\boldmath $\phi$})$ of 
(\ref{sigmaxyphinew}).  
Using the commutation relation (\ref{vxvyH0commu}), the summand in 
(\ref{sigmaxyphinew}) can be written as 
\begin{eqnarray}
& &\left\langle\Phi_{0,\mu}^{(N)}(\mbox{\boldmath $\phi$}),
v_{{\rm tot},y}^{(0)}(\mbox{\boldmath $\phi$})
\frac{1-Q(E_0^{(N)}(\mbox{\boldmath $\phi$}))}{[E_{0,\mu}^{(N)}(\mbox{\boldmath $\phi$})
-H_0^{(N)}(\mbox{\boldmath $\phi$})]^2}
v_{{\rm tot},x}^{(0)}(\mbox{\boldmath $\phi$})
\Phi_{0,\mu}^{(N)}(\mbox{\boldmath $\phi$})\right\rangle
-{\rm c.c.}\ret
&=&-\frac{2im_e}{\hbar eB}\left\langle\Phi_{0,\mu}^{(N)}(\mbox{\boldmath $\phi$}),
v_{{\rm tot},y}^{(0)}(\mbox{\boldmath $\phi$})
\frac{1-Q(E_0^{(N)}(\mbox{\boldmath $\phi$}))}{E_{0,\mu}^{(N)}(\mbox{\boldmath $\phi$})
-H_0^{(N)}(\mbox{\boldmath $\phi$})}
v_{{\rm tot},y}^{(0)}(\mbox{\boldmath $\phi$})
\Phi_{0,\mu}^{(N)}(\mbox{\boldmath $\phi$})\right\rangle\ret
&+&\left[\left\langle\Phi_{0,\mu}^{(N)}(\mbox{\boldmath $\phi$}),
v_{{\rm tot},y}^{(0)}(\mbox{\boldmath $\phi$})
\frac{1-Q(E_0^{(N)}(\mbox{\boldmath $\phi$}))}{[E_{0,\mu}^{(N)}(\mbox{\boldmath $\phi$})
-H_0^{(N)}(\mbox{\boldmath $\phi$})]^2}
I_x(\mbox{\boldmath $\phi$})\Phi_{0,\mu}^{(N)}(\mbox{\boldmath $\phi$})
\right\rangle-{\rm c.c.}\right]
\label{F1}
\end{eqnarray}
in the same way as in (\ref{commuteidentity}), where c.c. stands for the complex conjugate.

First we prove Theorem~\ref{1dtranslationinv} by using the above expression (\ref{F1}). 
We assume ${\bf A}_{\rm P}=0$, i.e., $B_{{\rm P},z}=0$.  
We treat only the case that the electrostatic potential $W$ is a function 
of the single variable $x$ only because we can treat the other case that $W$ is a function 
of the single variable $y$ only in the same way. 
{From} the assumptions and the expression (\ref{defIx}) of $I_x(\mbox{\boldmath $\phi$})$,  
one has $I_x(\mbox{\boldmath $\phi$})=0$. Further we have 
\begin{eqnarray}
& &2\sum_{\mu=1}^q \left\langle\Phi_{0,\mu}^{(N)}(\mbox{\boldmath $\phi$}),
v_{{\rm tot},y}^{(0)}(\mbox{\boldmath $\phi$})\frac{1-Q(E_0^{(N)}(\mbox{\boldmath $\phi$}))}
{E_{0,\mu}^{(N)}(\mbox{\boldmath $\phi$})-H_0^{(N)}(\mbox{\boldmath $\phi$})}
v_{{\rm tot},y}^{(0)}(\mbox{\boldmath $\phi$})
\Phi_{0,\mu}^{(N)}(\mbox{\boldmath $\phi$})\right\rangle\ret 
&=&\frac{\partial}{\partial \phi_y}{\rm Tr}\ v_{{\rm tot},y}^{(0)}(\mbox{\boldmath $\phi$})
Q(E_0^{(N)}(\mbox{\boldmath $\phi$}))-\frac{Nq}{m_e}
\label{F3}
\end{eqnarray}
in the same way as in Sec.~\ref{Sec:QAHC}. Combining these observations, 
(\ref{sigmaxyphinew}) and (\ref{F1}), we obtain the desired result, 
\begin{equation}
\overline{\sigma_{xy}(\mbox{\boldmath $\phi$})}
=-\frac{e^2}{h}\frac{N}{M}=-\frac{e^2}{h}\nu.
\end{equation}

Now let us return to the general setting. We rewrite the right-hand side 
of (\ref{F1}) further. In the same way, we have  
\begin{eqnarray}
& &\left\langle\Phi_{0,\mu}^{(N)}(\mbox{\boldmath $\phi$}),
v_{{\rm tot},y}^{(0)}(\mbox{\boldmath $\phi$})
\frac{1-Q(E_0^{(N)}(\mbox{\boldmath $\phi$}))}{[E_{0,\mu}^{(N)}(\mbox{\boldmath $\phi$})
-H_0^{(N)}(\mbox{\boldmath $\phi$})]^2}
I_x(\phi)\Phi_{0,\mu}^{(N)}(\mbox{\boldmath $\phi$})\right\rangle-{\rm c.c.}\ret
&=&-\frac{im_e}{\hbar eB}\left[\left\langle\Phi_{0,\mu}^{(N)}(\mbox{\boldmath $\phi$}),
v_{{\rm tot},x}^{(0)}(\mbox{\boldmath $\phi$})
\frac{1-Q(E_0^{(N)}(\mbox{\boldmath $\phi$}))}{E_{0,\mu}^{(N)}(\mbox{\boldmath $\phi$})
-H_0^{(N)}(\mbox{\boldmath $\phi$})}I_x(\mbox{\boldmath $\phi$})
\Phi_{0,\mu}^{(N)}(\mbox{\boldmath $\phi$})
\right\rangle+{\rm c.c.}\right]\ret
&+&\left[\left\langle\Phi_{0,\mu}^{(N)}(\mbox{\boldmath $\phi$}),
I_x(\mbox{\boldmath $\phi$})
\frac{1-Q(E_0^{(N)}(\mbox{\boldmath $\phi$}))}{[E_{0,\mu}^{(N)}(\mbox{\boldmath $\phi$})
-H_0^{(N)}(\mbox{\boldmath $\phi$})]^2}
I_y(\mbox{\boldmath $\phi$})\Phi_{0,\mu}^{(N)}(\mbox{\boldmath $\phi$})
\right\rangle-{\rm c.c.}\right],
\label{F2}
\end{eqnarray}
and 
\begin{eqnarray}
& &\sum_{\mu=1}^q\left\langle\Phi_{0,\mu}^{(N)}(\mbox{\boldmath $\phi$}),
v_{{\rm tot},x}^{(0)}(\mbox{\boldmath $\phi$})
\frac{1-Q(E_0^{(N)}(\mbox{\boldmath $\phi$}))}{E_{0,\mu}^{(N)}(\mbox{\boldmath $\phi$})
-H_0^{(N)}(\mbox{\boldmath $\phi$})}I_x(\mbox{\boldmath $\phi$})
\Phi_{0,\mu}^{(N)}(\mbox{\boldmath $\phi$})
\right\rangle+{\rm c.c.}\ret
&=&\frac{\partial}{\partial \phi_x}{\rm Tr}\ I_x(\mbox{\boldmath $\phi$})
Q(E_0^{(N)}(\mbox{\boldmath $\phi$}))
+\frac{1}{m_eB}\sum_{j=1}^N{\rm Tr}\ B_{{\rm P},z}({\bf r}_j)
Q(E_0^{(N)}(\mbox{\boldmath $\phi$})). 
\label{F4}
\end{eqnarray}
By combining (\ref{sigmaxyphinew}), (\ref{F1}), (\ref{F3}), (\ref{F2}) and 
(\ref{F4}), the averaged Hall conductance can be written as 
\begin{equation}
\overline{\sigma_{xy}(\mbox{\boldmath $\phi$})}=
-\frac{e^2}{h}\frac{N}{M}+\frac{e^2}{h}
\frac{1}{MBq}\sum_{j=1}^N 
\overline{{\rm Tr}\ B_{{\rm P},z}({\bf r}_j)Q(E_0^{(N)}(\mbox{\boldmath $\phi$}))}+
\overline{\Delta\sigma_{xy}(\mbox{\boldmath $\phi$})},
\label{avsigmaxyphi}
\end{equation}
where
\begin{equation}
\Delta\sigma_{xy}(\mbox{\boldmath $\phi$})=
\frac{i\hbar e^2}{L_xL_y}
\frac{1}{q}\sum_{\mu=1}^q
\left[\left\langle\Phi_{0,\mu}^{(N)}(\mbox{\boldmath $\phi$}),I_x(\mbox{\boldmath $\phi$})
\frac{1-Q(E_0^{(N)}(\mbox{\boldmath $\phi$}))}{[E_{0,\mu}^{(N)}(\mbox{\boldmath $\phi$})
-H_0^{(N)}(\mbox{\boldmath $\phi$})]^2}
I_y(\mbox{\boldmath $\phi$})\Phi_{0,\mu}^{(N)}(\mbox{\boldmath $\phi$})
\right\rangle-{\rm c.c.}\right].
\label{csigmaxyphi}
\end{equation}
The sum in the right-hand side of (\ref{avsigmaxyphi}) is easily evaluated as 
\begin{equation}
\left|\frac{e^2}{h}
\frac{1}{MBq}\sum_{j=1}^N 
\overline{{\rm Tr}\ B_{{\rm P},z}({\bf r}_j)Q(E_0^{(N)}(\mbox{\boldmath $\phi$}))}\right|
\le \frac{e^2}{h}\frac{N}{M}
\frac{\left\Vert B_{{\rm P},z}\right\Vert_\infty}{B}. 
\label{Bbound}
\end{equation}
Next we estimate the right-hand side of (\ref{csigmaxyphi}). 
In the same way as in the preceding Appendix~\ref{Boundsigmaxy}, we have 
\begin{equation}
\left|\Delta\sigma_{xy}(\mbox{\boldmath $\phi$})\right|\le 
\left(\frac{\hbar \omega_c}{\Delta E}\right)^3
\frac{e^2}{h}\frac{N}{M}\delta'(\mbox{\boldmath $\phi$})
\label{estDeltasigmaxyphi}
\end{equation}
with
\begin{equation}
\delta'(\mbox{\boldmath $\phi$})=\frac{m_e}{(\hbar\omega_c)^2N}\prod_{s=x,y}
\sqrt{\frac{1}{q}\sum_{\mu=1}^q\left\langle\Phi_{0,\mu}^{(N)}(\mbox{\boldmath $\phi$}),
\left[I_s(\mbox{\boldmath $\phi$}),[H_0^{(N)}(\mbox{\boldmath $\phi$}),
I_s(\mbox{\boldmath $\phi$})]\right]\Phi_{0,\mu}^{(N)}
(\mbox{\boldmath $\phi$})\right\rangle}. 
\label{deltaprime}
\end{equation}
In the rest of this Appendix, we will show 
\begin{equation}
\delta'(\mbox{\boldmath $\phi$})\le {\tilde \delta'},
\label{deltaprimebound}
\end{equation}
where ${\tilde \delta'}$ is independent of $\mbox{\boldmath $\phi$}$, 
the number $N$ of the electrons and of the sizes $L_x,L_y$ of the system. 
But ${\tilde \delta'}$ is a continuous function of 
the norms $\Vert D^{(m,n)}B_{{\rm P},z}\Vert_\infty$, 
$\Vert D^{(m,n)}W\Vert_\infty$ and satisfies 
${\tilde \delta'}=0$ in the special point with ${\bf A}_{\rm P}=0$ and $W=0$. 
Therefore ${\tilde \delta'}$ becomes small for the weak potentials 
${\bf A}_{\rm P}$ and $W$. 
Combining (\ref{avsigmaxyphi}), (\ref{Bbound}), (\ref{estDeltasigmaxyphi}) 
and (\ref{deltaprimebound}), we have the desired result, 
\begin{equation}
-\frac{e^2}{h}\nu(1+\delta)\le \overline{\sigma_{xy}(\mbox{\boldmath $\phi$})}
\le -\frac{e^2}{h}\nu(1-\delta),
\label{E12} 
\end{equation}
with
\begin{equation}
\delta=\frac{\Vert B_{{\rm P},z}\Vert_\infty}{B}
+\left(\frac{\hbar\omega_c}{\Delta E}\right)^3{\tilde \delta'}.
\end{equation}
In passing, we remark that we can obtain a similar bound for 
the non-averaged Hall conductance $\sigma_{xy}(\mbox{\boldmath $\phi$})$ 
to (\ref{E12}) by using (\ref{vyvxH0commu}) and (\ref{vxvycommu})  
for the first term in the right-hand side of (\ref{F1}).

In order to prove the bound (\ref{deltaprimebound}), 
it is sufficient to estimate the ground 
state expectation values of the double commutators 
$\left[I_s(\mbox{\boldmath $\phi$}),[H_0^{(N)}(\mbox{\boldmath $\phi$}),
I_s(\mbox{\boldmath $\phi$})]\right]$ in (\ref{deltaprime}). 
To this end, let us calculate the double commutators. 
In the following, we will consider only the case with $s=x$ because 
we can treat the case with $s=y$ exactly in the same way. Note that 
\begin{eqnarray}
[H_0^{(N)}(\mbox{\boldmath $\phi$}),I_x(\mbox{\boldmath $\phi$})]
&=&\frac{m_e}{2eB}\sum_{j=1}^N 
\left[\left(v_{x,j}^{(0)}(\mbox{\boldmath $\phi$})\right)^2+
\left(v_{y,j}^{(0)}(\mbox{\boldmath $\phi$})\right)^2,
\left(\frac{\partial W}{\partial y}\right)({\bf r}_j)\right]\ret
&-&\frac{m_e}{4B}\sum_{j=1}^N 
\left[\left(v_{x,j}^{(0)}(\mbox{\boldmath $\phi$})\right)^2+
\left(v_{y,j}^{(0)}(\mbox{\boldmath $\phi$})\right)^2,
B_{{\rm P},z}({\bf r}_j)v_{x,j}^{(0)}(\mbox{\boldmath $\phi$})
+v_{x,j}^{(0)}(\mbox{\boldmath $\phi$})B_{{\rm P},z}({\bf r}_j)\right]\ret
&-&\frac{1}{2B}\sum_{j=1}^N 
\left[W({\bf r}_j),
B_{{\rm P},z}({\bf r}_j)v_{x,j}^{(0)}(\mbox{\boldmath $\phi$})
+v_{x,j}^{(0)}(\mbox{\boldmath $\phi$})B_{{\rm P},z}({\bf r}_j)\right]\ret
&-&\frac{1}{2B}\sum_{i,j}\sum_{\ell=1}^N\left[W^{(2)}({\bf r}_i-{\bf r}_j),
B_{{\rm P},z}({\bf r}_\ell)v_{x,\ell}^{(0)}(\mbox{\boldmath $\phi$})
+v_{x,\ell}^{(0)}(\mbox{\boldmath $\phi$})B_{{\rm P},z}({\bf r}_\ell)\right]\ret
&=&\sum_{s=1}^3J_x^{(0,s)}+J_x^{(x)}(\mbox{\boldmath $\phi$})
+\sum_{s=1}^3J_x^{(y,s)}(\mbox{\boldmath $\phi$})+J_x^{(xx)}(\mbox{\boldmath $\phi$})
+J_x^{(xy)}(\mbox{\boldmath $\phi$}),
\end{eqnarray}
where the operators in the last line are given by 
\begin{equation}
J_x^{(0,1)}=-\frac{i\hbar}{m_eB}\sum_{j=1}^N B_{{\rm P},z}({\bf r}_j)
\left(\frac{\partial W}{\partial x}\right)({\bf r}_j),
\end{equation}
\begin{equation}
J_x^{(0,2)}=-\frac{i\hbar^3}{4m_e^2B}\sum_{j=1}^N \left[
\left(\frac{\partial^3 B_{{\rm P},z}}{\partial x^3}\right)
({\bf r}_j)+\left(\frac{\partial^3 B_{{\rm P},z}}{\partial x\partial y^2}
\right)({\bf r}_j)\right],
\end{equation}
\begin{equation}
J_x^{(0,3)}=-\frac{i\hbar}{m_eB}\sum_{i,j}[B_{{\rm P},z}({\bf r}_i)-
B_{{\rm P},z}({\bf r}_j)]\left(\frac{\partial W^{(2)}}{\partial x}\right)
({\bf r}_i-{\bf r}_j),
\end{equation}
\begin{equation}
J_x^{(x)}(\mbox{\boldmath $\phi$})=-\frac{i\hbar}{2eB}\sum_{j=1}^N 
\left[v_{x,j}^{(0)}(\mbox{\boldmath $\phi$})
\left(\frac{\partial^2 W}{\partial x \partial y}\right)({\bf r}_j)+
\left(\frac{\partial^2 W}{\partial x \partial y}\right)({\bf r}_j)
v_{x,j}^{(0)}(\mbox{\boldmath $\phi$})\right],
\end{equation}
\begin{equation}
J_x^{(y,1)}(\mbox{\boldmath $\phi$})=
-\frac{i\hbar}{2eB}\sum_{j=1}^N \left[v_{y,j}^{(0)}(\mbox{\boldmath $\phi$})
\left(\frac{\partial^2 W}{\partial y^2}\right)({\bf r}_j)+
\left(\frac{\partial^2 W}{\partial y^2}\right)({\bf r}_j)
v_{y,j}^{(0)}(\mbox{\boldmath $\phi$})\right],
\end{equation}
\begin{equation}
J_x^{(y,2)}(\mbox{\boldmath $\phi$})=-\frac{i\hbar e}{2m_e}\sum_{j=1}^N 
\left[v_{y,j}^{(0)}(\mbox{\boldmath $\phi$})B_{{\rm P},z}({\bf r}_j)
+B_{{\rm P},z}({\bf r}_j)v_{y,j}^{(0)}(\mbox{\boldmath $\phi$})\right],
\end{equation}
\begin{equation}
J_x^{(y,3)}(\mbox{\boldmath $\phi$})=-\frac{i\hbar e}{2m_eB}\sum_{j=1}^N 
\left\{v_{y,j}^{(0)}(\mbox{\boldmath $\phi$})\left[B_{{\rm P},z}({\bf r}_j)\right]^2
+\left[B_{{\rm P},z}({\bf r}_j)\right]^2v_{y,j}^{(0)}(\mbox{\boldmath $\phi$})\right\},
\end{equation}
\begin{equation}
J_x^{(xx)}(\mbox{\boldmath $\phi$})=\frac{i\hbar}{B}\sum_{j=1}^N 
v_{x,j}^{(0)}(\mbox{\boldmath $\phi$})\left(\frac{\partial B_{{\rm P},z}}{\partial x}
\right)({\bf r}_j)v_{x,j}^{(0)}(\mbox{\boldmath $\phi$}),
\end{equation}
and 
\begin{equation}
J_x^{(xy)}(\mbox{\boldmath $\phi$})=\frac{i\hbar}{2B}\sum_{j=1}^N \left[
v_{x,j}^{(0)}(\mbox{\boldmath $\phi$})\left(\frac{\partial B_{{\rm P},z}}{\partial y}
\right)({\bf r}_j)v_{y,j}^{(0)}(\mbox{\boldmath $\phi$})+
v_{y,j}^{(0)}(\mbox{\boldmath $\phi$})\left(\frac{\partial B_{{\rm P},z}}{\partial y}
\right)({\bf r}_j)v_{x,j}^{(0)}(\mbox{\boldmath $\phi$})\right].
\end{equation}
Hence the double commutator becomes 
\begin{eqnarray}
\left[I_x(\mbox{\boldmath $\phi$}),[H_0^{(N)}(\mbox{\boldmath $\phi$}),
I_x(\mbox{\boldmath $\phi$})]\right]
&=&\sum_{s=1}^3 [I_x(\mbox{\boldmath $\phi$}),J_x^{(0,s)}]
+[I_x(\mbox{\boldmath $\phi$}),J_x^{(x)}(\mbox{\boldmath $\phi$})]
+\sum_{s=1}^3 [I_x(\mbox{\boldmath $\phi$}),J_x^{(y,s)}(\mbox{\boldmath $\phi$})]\ret
&+&[I_x(\mbox{\boldmath $\phi$}),J_x^{(xx)}(\mbox{\boldmath $\phi$})]+
[I_x(\mbox{\boldmath $\phi$}),J_x^{(xy)}]. 
\label{doublecommuIxH0Ix}
\end{eqnarray}
In order to prove the boundedness of $\delta'(\mbox{\boldmath $\phi$})$ of 
(\ref{deltaprime}), we shall show that the ground state expectation values 
for all the commutators in this right-hand side are of order $N$. 

The three commutators in the first sum in the right-hand side 
of (\ref{doublecommuIxH0Ix}) are calculated as 
\begin{eqnarray}
& &[I_x(\mbox{\boldmath $\phi$}),J_x^{(0,1)}]\ret&=&
\frac{\hbar^2}{m_e^2B^2}\sum_{j=1}^N 
\left\{\left[B_{{\rm P},z}({\bf r}_j)\right]^2
\left(\frac{\partial^2 W}{\partial x^2}\right)({\bf r}_j)
+B_{{\rm P},z}({\bf r}_j)\left(\frac{\partial B_{{\rm P},z}}{\partial x}
\right)({\bf r}_j)\left(\frac{\partial W}{\partial x}\right)({\bf r}_j)
\right\},\ret
\end{eqnarray}
\begin{equation}
[I_x(\mbox{\boldmath $\phi$}),J_x^{(0,2)}]=\frac{\hbar^4}{4m_e^3B^2}\sum_{j=1}^N
B_{{\rm P},z}({\bf r}_j)\left[\left(
\frac{\partial^4B_{{\rm P},z}}{\partial x^4}\right)({\bf r}_j)
+\left(\frac{\partial^4B_{{\rm P},z}}{\partial x^2\partial y^2}\right)
({\bf r}_j)\right],
\end{equation}
and 
\begin{eqnarray}
& &[I_x(\mbox{\boldmath $\phi$}),J_x^{(0,3)}]\ret
&=&\frac{\hbar^2}{m_e^2B^2}\sum_{i,j}\left[B_{{\rm P},z}({\bf r}_i)-
B_{{\rm P},z}({\bf r}_j)\right]^2\left(\frac{\partial^2W^{(2)}}{\partial x^2}
\right)({\bf r}_i-{\bf r}_j)\ret
&+&\frac{\hbar^2}{m_e^2B^2}\sum_{i,j}
\left[B_{{\rm P},z}({\bf r}_i)
\left(\frac{\partial B_{{\rm P},z}}{\partial x}\right)({\bf r}_i)
-B_{{\rm P},z}({\bf r}_j)
\left(\frac{\partial B_{{\rm P},z}}{\partial x}\right)({\bf r}_j)\right]
\left(\frac{\partial W^{(2)}}{\partial x}\right)({\bf r}_i-{\bf r}_j).\ret
\end{eqnarray}
Clearly the first two ones are of order $N$ from the assumptions. 
The ground state expectation value of the rest one is evaluated as 
\begin{eqnarray}
& &\left|\left\langle \Phi_{0,\mu}^{(N)}(\mbox{\boldmath $\phi$}),
[I_x(\mbox{\boldmath $\phi$}),J_x^{(0,3)}]
\Phi_{0,\mu}^{(N)}(\mbox{\boldmath $\phi$})\right\rangle\right|\ret
&\le& \frac{4\hbar^2}{m_e^2}
\left(\frac{\left\Vert B_{{\rm P},z}\right\Vert_\infty}{B}\right)^2
\sum_{i,j}\left\langle \Phi_{0,\mu}^{(N)}(\mbox{\boldmath $\phi$}),
\left|\left(\frac{\partial^2W^{(2)}}{\partial x^2}
\right)({\bf r}_i-{\bf r}_j)\right|\Phi_{0,\mu}^{(N)}(\mbox{\boldmath $\phi$})
\right\rangle\ret
&+&\frac{2\hbar^2}{m_e^2B}\frac{\left\Vert B_{{\rm P},z}\right\Vert_\infty}{B}
\left\Vert \frac{\partial B_{{\rm P},z}}{\partial x}\right\Vert_\infty
\sum_{i,j}\left\langle \Phi_{0,\mu}^{(N)}(\mbox{\boldmath $\phi$}),
\left|\left(\frac{\partial W^{(2)}}{\partial x}
\right)({\bf r}_i-{\bf r}_j)\right|\Phi_{0,\mu}^{(N)}(\mbox{\boldmath $\phi$})
\right\rangle\ret
&\le& \frac{4\alpha_{\rm int}\hbar\omega_c}{m_e}
\frac{\left\Vert B_{{\rm P},z}\right\Vert_\infty}{B}
\left(\frac{\left\Vert B_{{\rm P},z}\right\Vert_\infty}{B}
+\frac{\ell_B}{2B}\left\Vert \frac{\partial B_{{\rm P},z}}{\partial x}
\right\Vert_\infty\right)\ret
& &\hspace{3cm}\times\sum_{i,j}\left\langle \Phi_{0,\mu}^{(N)}(\mbox{\boldmath $\phi$}),
W^{(2)}({\bf r}_i-{\bf r}_j)\Phi_{0,\mu}^{(N)}(\mbox{\boldmath $\phi$})\right\rangle, 
\end{eqnarray}
where we have used the assumptions (\ref{assumW21}) and (\ref{assumW22}). 
Further, by using 
the inequality 
\begin{eqnarray}
\sum_{i,j}\left\langle \Phi_{0,\mu}^{(N)}(\mbox{\boldmath $\phi$}),
W^{(2)}({\bf r}_i-{\bf r}_j)\Phi_{0,\mu}^{(N)}(\mbox{\boldmath $\phi$})\right\rangle
&\le& \left\langle \Phi_{0,\mu}^{(N)}(\mbox{\boldmath $\phi$}),
H_0^{(N)}(\mbox{\boldmath $\phi$})\Phi_{0,\mu}^{(N)}(\mbox{\boldmath $\phi$})\right\rangle
+N\left\Vert W\right\Vert\ret
&=&E_{0,\mu}^{(N)}(\mbox{\boldmath $\phi$})+N\left\Vert W\right\Vert, 
\end{eqnarray}
we have 
\begin{eqnarray}
\left|\left\langle \Phi_{0,\mu}^{(N)}(\mbox{\boldmath $\phi$}),
[I_x(\mbox{\boldmath $\phi$}),J_x^{(0,3)}]
\Phi_{0,\mu}^{(N)}(\mbox{\boldmath $\phi$})\right\rangle\right|&\le& 
\frac{4\alpha_{\rm int}\hbar\omega_c}{m_e}
\frac{\left\Vert B_{{\rm P},z}\right\Vert_\infty}{B}
\left(\frac{\left\Vert B_{{\rm P},z}\right\Vert_\infty}{B}
+\frac{\ell_B}{2B}\left\Vert \frac{\partial B_{{\rm P},z}}{\partial x}
\right\Vert_\infty\right)\ret
&\times&\left(E_{0,\mu}^{(N)}(\mbox{\boldmath $\phi$})+N\left\Vert W\right\Vert\right). 
\end{eqnarray}
Since the ground state energy $E_{0,\mu}^{(N)}(\mbox{\boldmath $\phi$})$ is of order $N$ 
as shown in Appendix~\ref{estEU}, this right-hand side is 
of order $N$. We remark that, when ${\bf A}_{\rm P}=0$, we do not 
need the assumptions (\ref{assumW21}) and (\ref{assumW22}) 
because all the operators $J_x^{(0,s)}$ are vanishing. 

Note that 
\begin{eqnarray}
[I_x(\mbox{\boldmath $\phi$}),J_x^{(x)}(\mbox{\boldmath $\phi$})]
&=&\frac{\hbar^2}{m_ee^2B^2}\sum_{j=1}^N
\left[\left(\frac{\partial^2 W}{\partial x\partial y}\right)
({\bf r}_j)\right]^2\ret
&+&\frac{\hbar^2}{2m_eeB^2}\sum_{j=1}^N
\left\{v_{x,j}^{(0)}(\mbox{\boldmath $\phi$}),B_{{\rm P},z}({\bf r}_j)
\left(\frac{\partial^3 W}{\partial x^2\partial y}\right)({\bf r}_j)\right\}
\ret
&-&\frac{\hbar^2}{2m_eeB^2}\sum_{j=1}^N
\left\{v_{x,j}^{(0)}(\mbox{\boldmath $\phi$}),
\left(\frac{\partial B_{{\rm P},z}}{\partial x}\right)({\bf r}_j)
\left(\frac{\partial^2 W}{\partial x\partial y}\right)({\bf r}_j)\right\},
\label{commuIxJx}
\end{eqnarray}
\begin{eqnarray}
[I_x(\mbox{\boldmath $\phi$}),J_x^{(y,1)}(\mbox{\boldmath $\phi$})]
&=&\frac{\hbar^2}{m_ee^2B^2}\sum_{j=1}^N
\left[\left(\frac{\partial^2 W}{\partial y^2}\right)
({\bf r}_j)\right]^2\ret
&+&\frac{\hbar^2}{2m_eeB^2}\sum_{j=1}^N
\left\{v_{y,j}^{(0)}(\mbox{\boldmath $\phi$}),B_{{\rm P},z}({\bf r}_j)
\left(\frac{\partial^3 W}{\partial x\partial y^2}\right)({\bf r}_j)\right\}
\ret
&-&\frac{\hbar^2}{2m_eeB^2}\sum_{j=1}^N
\left\{v_{x,j}^{(0)}(\mbox{\boldmath $\phi$}),
\left(\frac{\partial B_{{\rm P},z}}{\partial y}\right)({\bf r}_j)
\left(\frac{\partial^2 W}{\partial y^2}\right)({\bf r}_j)\right\}\ret
&+&\frac{\hbar^2}{m_e^2B}\sum_{j=1}^N B_{{\rm P},z}({\bf r}_j)
\left(\frac{\partial^2 W}{\partial y^2}\right)({\bf r}_j)
+\frac{\hbar^2}{m_e^2B^2}\sum_{j=1}^N [B_{{\rm P},z}({\bf r}_j)]^2
\left(\frac{\partial^2 W}{\partial y^2}\right)({\bf r}_j),\ret
\label{commuIxJy1}
\end{eqnarray}
\begin{eqnarray}
[I_x(\mbox{\boldmath $\phi$}),J_x^{(y,2)}(\mbox{\boldmath $\phi$})]
&=&\frac{\hbar^2}{m_e^2B}\sum_{j=1}^N 
B_{{\rm P},z}({\bf r}_j)\left(\frac{\partial^2 W}{\partial y^2}\right)
({\bf r}_j)\ret
&+&\frac{\hbar^2e}{2m_e^2B}\sum_{j=1}^N 
\left\{v_{y,j}^{(0)}(\mbox{\boldmath $\phi$}),B_{{\rm P},z}({\bf r}_j)
\left(\frac{\partial B_{{\rm P},z}}{\partial x}\right)({\bf r}_j)\right\}\ret
&-&\frac{\hbar^2e}{2m_e^2B}\sum_{j=1}^N 
\left\{v_{x,j}^{(0)}(\mbox{\boldmath $\phi$}),B_{{\rm P},z}({\bf r}_j)
\left(\frac{\partial B_{{\rm P},z}}{\partial y}\right)({\bf r}_j)\right\}\ret
&+&\frac{\hbar^2e^2}{m_e^3}\sum_{j=1}^N [B_{{\rm P},z}({\bf r}_j)]^2+
\frac{\hbar^2e^2}{m_e^3B}\sum_{j=1}^N [B_{{\rm P},z}({\bf r}_j)]^3,
\end{eqnarray}
and 
\begin{eqnarray}
[I_x(\mbox{\boldmath $\phi$}),J_x^{(y,3)}(\mbox{\boldmath $\phi$})]
&=&\frac{\hbar^2}{m_e^2B^2}\sum_{j=1}^N 
[B_{{\rm P},z}({\bf r}_j)]^2\left(\frac{\partial^2 W}{\partial y^2}\right)
({\bf r}_j)\ret
&+&\frac{\hbar^2e}{m_e^2B^2}\sum_{j=1}^N 
\left\{v_{y,j}^{(0)}(\mbox{\boldmath $\phi$}),[B_{{\rm P},z}({\bf r}_j)]^2
\left(\frac{\partial B_{{\rm P},z}}{\partial x}\right)({\bf r}_j)\right\}\ret
&-&\frac{\hbar^2e}{2m_e^2B^2}\sum_{j=1}^N 
\left\{v_{x,j}^{(0)}(\mbox{\boldmath $\phi$}),[B_{{\rm P},z}({\bf r}_j)]^2
\left(\frac{\partial B_{{\rm P},z}}{\partial y}\right)({\bf r}_j)\right\}\ret
&+&\frac{\hbar^2e^2}{m_e^3B}\sum_{j=1}^N [B_{{\rm P},z}({\bf r}_j)]^3+
\frac{\hbar^2e^2}{m_e^3B^2}\sum_{j=1}^N [B_{{\rm P},z}({\bf r}_j)]^4,
\end{eqnarray}
where $\{X,Y\}=XY+YX$ for operators $X,Y$. Since the terms including 
the velocity operators $v_{s,j}^{(0)}(\mbox{\boldmath $\phi$})$ in these 
right-hand sides can be estimated in the same 
way as in the previous Appendix~\ref{Boundsigmaxy}, 
we can get the desired estimates of order $N$ for their ground state 
expectation values. 

Note that 
\begin{eqnarray}
& &[I_x(\mbox{\boldmath $\phi$}),J_x^{(xx)}(\mbox{\boldmath $\phi$})]\ret
&=&-\frac{\hbar^2}{m_eeB^2}
\sum_{j=1}^N \left\{v_{x,j}^{(0)}(\mbox{\boldmath $\phi$}),
\left(\frac{\partial B_{{\rm P},z}}{\partial x}\right)({\bf r}_j)
\left(\frac{\partial^2 W}{\partial x\partial y}\right)({\bf r}_j)\right\}\ret
&+&\frac{\hbar^2}{m_eB^2}\sum_{j=1}^N v_{x,j}^{(0)}(\mbox{\boldmath $\phi$})\left[
2\left|\left(\frac{\partial B_{{\rm P},z}}{\partial x}\right)({\bf r}_j)
\right|^2-B_{{\rm P},z}({\bf r}_j)
\left(\frac{\partial^2 B_{{\rm P},z}}{\partial x^2}\right)({\bf r}_j)
\right]v_{x,j}^{(0)}(\mbox{\boldmath $\phi$})\ret
&-&\frac{\hbar^4}{2m_e^3B^2}\sum_{j=1}^N 
\left[
\left|\left(\frac{\partial^2 B_{{\rm P},z}}{\partial x^2}\right)
({\bf r}_j)\right|^2+
\left(\frac{\partial B_{{\rm P},z}}{\partial x}\right)({\bf r}_j)
\left(\frac{\partial^3 B_{{\rm P},z}}{\partial x^3}\right)({\bf r}_j)
\right].
\end{eqnarray}
Clearly the ground state expectation values for the first and third sums 
can be evaluated in the same way as in the above. 
For the second sum, we have 
$$
\left|\sum_{j=1}^N \left\langle\Phi_{0,\mu}^{(N)}(\mbox{\boldmath $\phi$}),
v_{x,j}^{(0)}(\mbox{\boldmath $\phi$})\left[
2\left|\left(\frac{\partial B_{{\rm P},z}}{\partial x}\right)({\bf r}_j)
\right|^2-B_{{\rm P},z}({\bf r}_j)
\left(\frac{\partial^2 B_{{\rm P},z}}{\partial x^2}\right)({\bf r}_j)
\right]v_{x,j}^{(0)}(\mbox{\boldmath $\phi$})
\Phi_{0,\mu}^{(N)}(\mbox{\boldmath $\phi$})\right\rangle\right|
$$
\begin{eqnarray}
&\le&\left(2\left\Vert\frac{\partial B_{{\rm P},z}}{\partial x}
\right\Vert_\infty^2+\left\Vert B_{{\rm P},z}\right\Vert_\infty
\left\Vert\frac{\partial^2 B_{{\rm P},z}}{\partial x^2}\right\Vert_\infty
\right)\sum_{j=1}^N \left\langle\Phi_{0,\mu}^{(N)}(\phi),
\left(v_{x,j}^{(0)}(\mbox{\boldmath $\phi$})\right)^2
\Phi_{0,\mu}^{(N)}(\mbox{\boldmath $\phi$})\right\rangle\ret
&\le&\frac{2}{m_e}\left(2\left\Vert\frac{\partial B_{{\rm P},z}}{\partial x}
\right\Vert_\infty^2+\left\Vert B_{{\rm P},z}\right\Vert_\infty
\left\Vert\frac{\partial^2 B_{{\rm P},z}}{\partial x^2}\right\Vert_\infty
\right)\left[\left\langle\Phi_{0,\mu}^{(N)}(\mbox{\boldmath $\phi$}),
H_0^{(N)}(\mbox{\boldmath $\phi$})\Phi_{0,\mu}^{(N)}(\mbox{\boldmath $\phi$})
\right\rangle+N\left\Vert W\right\Vert_\infty\right]
\ret
&=&\frac{2}{m_e}\left(2\left\Vert\frac{\partial B_{{\rm P},z}}{\partial x}
\right\Vert_\infty^2+\left\Vert B_{{\rm P},z}\right\Vert_\infty
\left\Vert\frac{\partial^2 B_{{\rm P},z}}{\partial x^2}\right\Vert_\infty
\right)\left[E_{0,\mu}^{(N)}(\mbox{\boldmath $\phi$})
+N\left\Vert W\right\Vert_\infty\right].
\end{eqnarray}
Thus the corresponding contribution is of order $N$. 

Finally let us compute the ground state expectation value of 
$[I_x(\mbox{\boldmath $\phi$}),J_x^{(xy)}(\mbox{\boldmath $\phi$})]$. 
In order to make this task easier, we decompose the operator 
$I_x(\mbox{\boldmath $\phi$})$ into two parts as 
\begin{equation}
I_x(\mbox{\boldmath $\phi$})=I_x^{(1)}+I_x^{(2)}(\mbox{\boldmath $\phi$})
\end{equation}
with 
\begin{equation}
I_x^{(1)}=\frac{1}{eB}\sum_{j=1}^N 
\left(\frac{\partial W}{\partial y}\right)({\bf r}_j),
\end{equation}
and 
\begin{equation}
I_x^{(2)}(\mbox{\boldmath $\phi$})=-\frac{1}{2B}\sum_{j=1}^N 
\left[B_{{\rm P},z}({\bf r}_j)v_{x,j}^{(0)}(\mbox{\boldmath $\phi$})+
v_{x,j}^{(0)}(\mbox{\boldmath $\phi$})
B_{{\rm P},z}({\bf r}_j)\right].
\end{equation}
Note that 
\begin{eqnarray}
[I_x^{(1)},J_x^{(xy)}(\mbox{\boldmath $\phi$})]&=&-\frac{\hbar^2}{2m_eeB^2}
\sum_{j=1}^N \left\{v_{x,j}^{(0)}(\mbox{\boldmath $\phi$}),
\left(\frac{\partial B_{{\rm P},z}}{\partial y}\right)({\bf r}_j)
\left(\frac{\partial^2W}{\partial y^2}\right)({\bf r}_j)\right\}\ret
&-&\frac{\hbar^2}{2m_eeB^2}
\sum_{j=1}^N \left\{v_{y,j}^{(0)}(\mbox{\boldmath $\phi$}),
\left(\frac{\partial B_{{\rm P},z}}{\partial y}\right)({\bf r}_j)
\left(\frac{\partial^2W}{\partial x\partial y}\right)({\bf r}_j)\right\}, 
\end{eqnarray}
and 
\begin{equation}
[I_x^{(2)}(\mbox{\boldmath $\phi$}),J_x^{(xy)}(\mbox{\boldmath $\phi$})]=
K_x^{(xx)}(\mbox{\boldmath $\phi$})+K_x^{(xy)}(\mbox{\boldmath $\phi$})
+K_x^{(x)}(\mbox{\boldmath $\phi$})+K_x^{(0)}
\end{equation}
with 
\begin{equation}
K_x^{(xx)}(\mbox{\boldmath $\phi$})=\frac{\hbar^2}{m_eB^2}\sum_{j=1}^N 
v_{x,j}^{(0)}(\mbox{\boldmath $\phi$})
\left[\left(\frac{\partial B_{{\rm P},z}}{\partial y}
\right)({\bf r}_j)\right]^2
v_{x,j}^{(0)}(\mbox{\boldmath $\phi$}),
\end{equation}
\begin{eqnarray}
K_x^{(xy)}(\mbox{\boldmath $\phi$})&=&
-\frac{\hbar^2}{2m_eB^2}\sum_{j=1}^N 
\left[v_{x,j}^{(0)}(\mbox{\boldmath $\phi$})B_{{\rm P},z}
({\bf r}_j)\left(\frac{\partial^2 B_{{\rm P},z}}{\partial x\partial y}
\right)({\bf r}_j)v_{y,j}^{(0)}(\mbox{\boldmath $\phi$})+(x\leftrightarrow y)\right]\ret
&+&\frac{\hbar^2}{2m_eB^2}\sum_{j=1}^N \left[v_{x,j}^{(0)}(\mbox{\boldmath $\phi$})
\left(\frac{\partial B_{{\rm P},z}}{\partial x}\right)({\bf r}_j)
\left(\frac{\partial B_{{\rm P},z}}{\partial y}\right)({\bf r}_j)
v_{y,j}^{(0)}(\mbox{\boldmath $\phi$})+(x\leftrightarrow y)\right],\ret
\label{Kxy}
\end{eqnarray}
\begin{eqnarray}
K_x^{(x)}(\mbox{\boldmath $\phi$})&=&-\frac{\hbar^2e}{2m_e^2B}\sum_{j=1}^N 
\left\{v_{x,j}^{(0)}(\mbox{\boldmath $\phi$}),B_{{\rm P},z}({\bf r}_j)
\left(\frac{\partial B_{{\rm P},z}}{\partial y}\right)({\bf r}_j)\right\}\ret
&-&\frac{\hbar^2e}{2m_e^2B^2}\sum_{j=1}^N 
\left\{v_{x,j}^{(0)}(\mbox{\boldmath $\phi$}),\left[B_{{\rm P},z}({\bf r}_j)\right]^2
\left(\frac{\partial B_{{\rm P},z}}{\partial y}\right)({\bf r}_j)\right\},
\end{eqnarray}
and
\begin{eqnarray}
K_x^{(0)}&=&-\frac{\hbar^4}{4m_e^3B^2}\sum_{j=1}^N \left[
2\left(\frac{\partial B_{{\rm P},z}}{\partial y}\right)({\bf r}_j)
\left(\frac{\partial^3 B_{{\rm P},z}}{\partial x^2\partial y}\right)({\bf r}_j)
+\left(\frac{\partial^2 B_{{\rm P},z}}{\partial x^2}\right)({\bf r}_j)
\left(\frac{\partial^2 B_{{\rm P},z}}{\partial y^2}\right)({\bf r}_j)\right]
\ret
&-&\frac{\hbar^4}{4m_e^3B^2}\sum_{j=1}^N 
\left[\left(\frac{\partial^2B_{{\rm P},z}}{\partial x\partial y}\right)
({\bf r}_j)\right]^2.
\end{eqnarray}
Hence all the contributions except for that for $K_x^{(xy)}(\mbox{\boldmath $\phi$})$ 
can be evaluated in the same way as in the above. 
We shall show that the ground state expectation value 
of $K_x^{(xy)}(\mbox{\boldmath $\phi$})$ 
is of order $N$, too. Using the Schwarz inequality, we have 
\begin{eqnarray}
& &\left|\sum_{j=1}^N \left\langle\Phi_{0,\mu}^{(N)}(\mbox{\boldmath $\phi$}),
v_{x,j}^{(0)}(\mbox{\boldmath $\phi$})B_{{\rm P},z}
({\bf r}_j)\left(\frac{\partial^2 B_{{\rm P},z}}{\partial x\partial y}
\right)({\bf r}_j)v_{y,j}^{(0)}(\mbox{\boldmath $\phi$})
\Phi_{0,\mu}^{(N)}(\mbox{\boldmath $\phi$})\right\rangle
\right|\ret
&\le&\sqrt{\sum_{j=1}^N \left\langle\Phi_{0,\mu}^{(N)}(\mbox{\boldmath $\phi$}),
v_{x,j}^{(0)}(\mbox{\boldmath $\phi$})
\left|B_{{\rm P},z}({\bf r}_j)\right|^2v_{x,j}^{(0)}(\mbox{\boldmath $\phi$})
\Phi_{0,\mu}^{(N)}(\mbox{\boldmath $\phi$})\right\rangle}\ret
&\times&\sqrt{\sum_{j=1}^N \left\langle\Phi_{0,\mu}^{(N)}(\mbox{\boldmath $\phi$}),
v_{y,j}^{(0)}(\mbox{\boldmath $\phi$})\left|
\left(\frac{\partial^2 B_{{\rm P},z}}{\partial x\partial y}
\right)({\bf r}_j)\right|^2
v_{y,j}^{(0)}(\mbox{\boldmath $\phi$})
\Phi_{0,\mu}^{(N)}(\mbox{\boldmath $\phi$})\right\rangle}\ret
&\le&\left\Vert B_{{\rm P},z}\right\Vert_\infty
\left\Vert\frac{\partial^2 B_{{\rm P},z}}{\partial x\partial y}
\right\Vert_\infty\sqrt{\sum_{j=1}^N 
\left\langle\Phi_{0,\mu}^{(N)}(\mbox{\boldmath $\phi$}),
\left(v_{x,j}^{(0)}(\mbox{\boldmath $\phi$})\right)^2
\Phi_{0,\mu}^{(N)}(\mbox{\boldmath $\phi$})\right\rangle}\ret
& &\qquad\qquad\times 
\sqrt{\sum_{j=1}^N \left\langle\Phi_{0,\mu}^{(N)}(\mbox{\boldmath $\phi$}),
\left(v_{y,j}^{(0)}(\mbox{\boldmath $\phi$})\right)^2
\Phi_{0,\mu}^{(N)}(\mbox{\boldmath $\phi$})\right\rangle}.
\end{eqnarray}
This right-hand side is of order $N$ in the same method as in the above. 
All the rest of the contributions in the right-hand side of (\ref{Kxy}) 
can be treated in the same way. Consequently $\delta'(\mbox{\boldmath $\phi$})$ of 
(\ref{deltaprime}) has been proved to be bounded uniformly in 
the number $N$ of the electrons. 

Consider the special case with ${\bf A}_{\rm P}=0$ which corresponds 
to the situation of Theorem~\ref{fractionTheorem1}. 
Then all the contributions except for the first sum in (\ref{commuIxJx}) 
and for the first sum in (\ref{commuIxJy1}) are vanishing, 
and a similar result can be obtained for the double commutator 
$[I_y(\mbox{\boldmath $\phi$}),[H_0^{(N)}(\mbox{\boldmath $\phi$}),
I_y(\mbox{\boldmath $\phi$})]]$. As a result, we have the bound, 
\begin{equation}
\delta'(\mbox{\boldmath $\phi$})\le\frac{2\ell_B^4}{(\hbar\omega_c)^2}\max_{m+n=2}
\left\Vert D^{(m,n)}W\right\Vert_\infty^2. 
\end{equation}
Combining this with (\ref{avsigmaxyphi}) and (\ref{estDeltasigmaxyphi}), 
we obtain the desired result (\ref{fractionalbound2}) with (\ref{defdelta}). 

%%%%%%%%%%%%%%%%%%%%%%%%%%%%%%%%%%%%%%%%%%%%
\Section{Estimate of the ground state energies $E_{0,\mu}^{(N)}(\mbox{\boldmath $\phi$})$}
\label{estEU}

In this appendix, we show that all of 
the ground state energies $E_{0,\mu}^{(N)}(\mbox{\boldmath $\phi$})$ are of order 
of $N$. 

Let $\Psi_0^{(N)}(\mbox{\boldmath $\phi$})$ be a ground state vector of the Landau 
Hamiltonian, 
\begin{equation}
H_{\rm L}^{(N)}(\mbox{\boldmath $\phi$})=\sum_{j=1}^N \frac{1}{2m_e}
\left[(p_{x,j}-eBy_j+\phi_x)^2+(p_{y,j}+\phi_y)^2\right], 
\end{equation}
for the non-interacting $N$ electrons. Then one has 
\begin{equation}
E_{0,\mu}^{(N)}(\mbox{\boldmath $\phi$})\le 
\left\langle\Psi_0^{(N)}(\mbox{\boldmath $\phi$}),H_0^{(N)}(\mbox{\boldmath $\phi$})
\Psi_0^{(N)}(\mbox{\boldmath $\phi$})\right\rangle+\Delta{\cal E}(\mbox{\boldmath $\phi$})
\quad \mbox{for } \mu=1,2,\ldots,q, 
\end{equation}
where $H_0^{(N)}(\mbox{\boldmath $\phi$})$ is the Hamiltonian (\ref{Ham0phi}) of 
the present system, and $\Delta{\cal E}(\mbox{\boldmath $\phi$})$ is given by 
\begin{equation}
\Delta{\cal E}(\mbox{\boldmath $\phi$})=\max_{\mu,\mu'}
\left|E_{0,\mu}^{(N)}(\mbox{\boldmath $\phi$})
-E_{0,\mu'}^{(N)}(\mbox{\boldmath $\phi$})\right|. 
\end{equation}
Using the eigenvectors $\varphi_{n,k}^{\rm P}(\mbox{\boldmath $\phi$})$ of 
(\ref{varphiP}) for the single electron Hamiltonian 
${\cal H}_0(\mbox{\boldmath $\phi$})$ of (\ref{hamsingleLandau}), 
the ground state expectation value for $H_0^{(N)}(\mbox{\boldmath $\phi$})$ 
can be written as 
\begin{eqnarray}
& &\left\langle\Psi_0^{(N)}(\mbox{\boldmath $\phi$}),H_0^{(N)}(\mbox{\boldmath $\phi$})
\Psi_0^{(N)}(\mbox{\boldmath $\phi$})\right\rangle\ret
&=&\sum_{n,k} 
\left\langle\varphi_{n,k}^{\rm P}(\mbox{\boldmath $\phi$}),{\cal H}(\mbox{\boldmath $\phi$})
\varphi_{n,k}^{\rm P}(\mbox{\boldmath $\phi$})\right\rangle
+\sum_{1\le i<j\le N}\left\langle\Psi_0^{(N)}(\mbox{\boldmath $\phi$}),
W^{(2)}({\bf r}_i-{\bf r}_j)\Psi_0^{(N)}(\mbox{\boldmath $\phi$})\right\rangle\ret
&\le& \sum_{n,k}\left[{\cal E}_{n,k}+\frac{\sqrt{2}e}{\sqrt{m_e}}
\Vert |{\bf A}_{\rm P}|\Vert_\infty
\sqrt{{\cal E}_{n,k}}+\frac{e^2}{2m_e}\left(\Vert |{\bf A}_{\rm P}|\Vert_\infty\right)^2
+\left\Vert W^+\right\Vert_\infty\right]\ret
&+&\sum_{1\le i<j\le N}\left\langle\Psi_0^{(N)}(\mbox{\boldmath $\phi$}),
W^{(2)}({\bf r}_i-{\bf r}_j)\Psi_0^{(N)}(\mbox{\boldmath $\phi$})\right\rangle,
\end{eqnarray}
where ${\cal E}_{n,k}=(n+1/2)\hbar\omega_c$, 
the Hamiltonian ${\cal H}(\mbox{\boldmath $\phi$})$ is given by (\ref{hamphiAW}), 
and we have used the inequality (\ref{calHphibound}). 
Clearly the first sum in the right-hand side of the inequality is of order 
$N$, and so it is enough to estimate the ground state expectation value  
of the electron-electron interaction energy. But this quantity 
is of order $N$ in the same way as in Appendix~G of ref.~\cite{Koma2}.

%%%%%%%%%%%%%%%%%%%%%%%%%%%%%%%%%%%%%%%%%%%%%%%%%%%%%%%%%%%%%
%newpage

\end{document}